\documentclass[12pt]{iopart}

\newcommand{\dd}{\,\mathrm{d}}

\usepackage{bm}
\usepackage{iopams}
\usepackage{braket}
\usepackage{setstack}
\usepackage{citesort}
\usepackage{cite}
\usepackage{longtable}
\usepackage{caption}
\usepackage{multirow}
\usepackage{graphicx}
\usepackage{ntheorem}
\usepackage{stackrel}
\usepackage{subfigure}
\usepackage{hhline}
\usepackage{epstopdf}
\usepackage[fleqn]{cases} 

\newtheorem{proposition}{Proposition}

\begin{document}

\title{Kurtosis of von Neumann entanglement entropy}
\author{Youyi Huang, Lu Wei, and Bjordis Collaku }
\address{Department of Electrical and Computer Engineering, University of Michigan - Dearborn, Michigan 48128, USA}
\ead{\{youyih, luwe, bcollaku\}@umich.edu}
\vspace{10pt}
\begin{indented}
\item[]July 2021
\end{indented}

\begin{abstract}
In this work, we study the statistical behavior of entanglement in quantum bipartite systems under the Hilbert-Schmidt ensemble as assessed by the standard measure - the von Neumann entropy.  Expressions of the first three exact cumulants of von Neumann entropy are known in the literature. The main contribution of the present work is the exact formula of the corresponding fourth cumulant that controls the tail behavior of the distribution. As a key ingredient in deriving the result, we make use of newly observed unsimplifiable summation bases that lead to a complete cancellation. In addition to providing further evidence of the conjectured Gaussian limit of the von Neumann entropy, the obtained formula also provides an improved finite-size approximation to the distribution.
\end{abstract}

\vspace{2pc}

\maketitle
\section{Introduction and the main result}\label{sec:intr}
The main focus of quantum information theory is the theoretical underpinnings of quantum technologies such as quantum computing and quantum communications. The concept of entanglement has played a crucial role in the development of quantum mechanics. The phenomenon of quantum entanglement is also the resource and medium that enable quantum technologies.

We consider the quantum bipartite model proposed by Page \cite{Page93} in the year 1993, which is a useful model in describing the interaction of a physical object and its environment. In particular, we aim to understand the degree of entanglement of such a model as measured by von Neumann entropy, the statistical behavior of which is encoded from its cumulants/moments. In principle, the knowledge of all moments determines uniquely the distribution of von Neumann entropy due to its compact support. In practice, finite-size approximations to the distribution of the entropy can be constructed from the first a few cumulants. In the literature, the first three cumulants of von Neumann entropy have been studied in~\cite{Page93,Foong94,Ruiz95,VPO16,Wei17,Bianchi19,Wei20} among others. The first two cumulants are relevant to the average and fluctuation behavior, whereas the third cumulant, i.e., the skewness, describes the degree of asymmetry of the distribution. As a further step towards understanding its full statistical distribution, we study the (excess) kurtosis of von Neumann entropy that controls the tails of the distribution. The computation of kurtosis requires additionally the corresponding fourth cumulant.
 
The bipartite model is formulated as follows. Consider a composite quantum system that consists of two subsystems $A$ and $B$ of Hilbert space dimensions $m$ and $n$, respectively. The Hilbert space $\mathcal{H}_{A+B}$ of the composite system is given by the tensor product of the Hilbert spaces of the subsystems, $\mathcal{H}_{A+B}=\mathcal{H}_{A}\otimes\mathcal{H}_{B}$. A random pure state of the composite system is defined as a linear combination of the random coefficients $x_{i,j}$ and the complete basis $\left\{\Ket{i^{A}}\right\}$ and $\left\{\Ket{j^{B}}\right\}$ of $\mathcal{H}_{A}$ and $\mathcal{H}_{B}$,
\begin{equation}
\Ket{\psi}=\sum_{i=1}^{m}\sum_{j=1}^{n}x_{i,j}\Ket{i^{A}}\otimes\Ket{j^{B}}.
\end{equation}
The corresponding density matrix in the random pure state is
\begin{equation}\label{eq:rho}
\rho=\Ket{\psi}\Bra{\psi},
\end{equation}
which has the natural constraint $\tr(\rho)=1$. This implies that the $m\times n$ random coefficient matrix $\mathbf{X}=(x_{i,j})$ satisfies \begin{equation}\label{eq:pcv}
\tr\left(\mathbf{XX}^{\dag}\right)=1.
\end{equation}
Without loss of generality, we assume that $m\leq n$. The reduced density matrix $\rho_{A}$ of the smaller subsystem $A$ is computed by partial tracing of the full density matrix~(\ref{eq:rho}) over the other subsystem $B$ as
\begin{equation}
\rho_{A}=\tr_{B}(\rho).
\end{equation}
The Schmidt decomposition of $\rho_{A}$ is given by
\begin{equation}
\rho_{A}=\sum_{i=1}^{m}\lambda_{i}\Ket{\phi_{i}^{A}}\Bra{\phi_{i}^{A}},
\end{equation}
where $0<\lambda_{i}<1$ is the $i$-th eigenvalue of the $m\times m$ Hermitian matrix $\mathbf{W}=\mathbf{XX}^{\dag}$ with $\Ket{\phi_{i}^{A}}$ being the corresponding eigenvector. The conservation of probability~(\ref{eq:pcv}) now leads to the fixed-trace constraint
\begin{equation}\label{eq:lamcon}
\sum_{i=1}^{m}\lambda_{i}=1.
\end{equation}
 The corresponding eigenvalue density of $\mathbf{W}$ is well-known (see, e.g.~\cite{Page93})
\begin{equation}\label{eq:fte}
f\left(\bm{\lambda}\right)=\frac{\Gamma(mn)}{C}~\delta\left(1-\sum_{i=1}^{m}\lambda_{i}\right)\prod_{1\leq i<j\leq m}\left(\lambda_{i}-\lambda_{j}\right)^{2}\prod_{i=1}^{m}\lambda_{i}^{n-m},
\end{equation}
where $\delta(\cdot)$ is the Dirac delta function and the constant
\begin{equation}\label{eq:con}
C=\prod_{i=1}^{m}\Gamma(n-i+1)\Gamma(i).
\end{equation}
This fixed-trace ensemble~(\ref{eq:fte}) is also known as the (unitary) Hilbert-Schmidt measure. The above discussed bipartite model is useful in describing the interaction of various real-world quantum systems. For example, the subsystem $A$ is the black hole and the subsystem $B$ is the associated radiation field~\cite{Page93}. The degree of entanglement of subsystems can be measured by entanglement entropies, which are functions of eigenvalues of $\mathbf{W}$. We consider the standard measure of von Neumann entropy of the subsystem with density matrix $\rho_{A}$,
\begin{equation}\label{eq:vN}
S=-\tr\left(\rho_{A}\ln\rho_{A}\right)=-\sum_{i=1}^{m}\lambda_{i}\ln\lambda_{i},~~~~~~S\in\left[0, \ln{m}\right],
\end{equation}
which attains the separable state ($S=0$) when $\lambda_{1}=1$, $\lambda_{2}=\dots=\lambda_{m}=0$ and the maximally-entangled state ($S=\ln{m}$) when $\lambda_{1}=\dots\lambda_{m}=1/m$. Statistical information of the von Neumann entropy is encoded through its moments
\begin{equation}
\mathbb{E}_{f}\!\left[S^{k}\right],~~~~~~k=1,2,3,\dots,
\end{equation}
where the expectation is taken over the Hilbert-Schmidt measure~(\ref{eq:fte}). For a comprehensive treatment of the density matrix formulism including other measures and entropies, for example, the Bures-Hall measure and the corresponding quantum purity, von Neumann entropy, and Tsallis entropy, we refer readers to the book~\cite{BZ06} and references therein.

Generally, the moment sequence, $m_{1},m_{2},m_{3},\dots$, and the cumulant sequence, $\kappa_{1},\kappa_{2},\kappa_{3},\dots$, for a random variable are related: the $i$-th moment is an $i$-th degree polynomial in the first $i$ cumulants and vice versa. In particular, the relation pairs up to $i=4$ are 
\begin{equation}\label{eq:1km} 
 \qquad \!\!\!\!\! m_{1} = \kappa_{1} 
\end{equation}
\numparts
\begin{numcases}{}
 m_{2} = \kappa_{2}+\kappa_{1}^{2}\\
\kappa_{2} = m_{2}-m_{1}^{2} \label{eq:2mk}
\end{numcases}
\endnumparts
\numparts
\begin{numcases}{}
   m_{3} = \kappa_{3}+3\kappa_{2}\kappa_{1}+\kappa^{3}_{1}\\
\kappa_{3} =  m_{3}-3m_{2}m_{1}+2m_{1}^{3}
\end{numcases}
\endnumparts
\numparts
\begin{numcases}{}
m_{4} = \kappa_{4}+4\kappa_{3}\kappa_{1}+3\kappa_{1}^2+6\kappa_{2}\kappa_{1}^2+\kappa_{1}^{4}\\
\kappa_{4} = m_{4}-4m_{3}m_{1}-3m_{2}^{2}+12m_{2}m_{1}^2-6m_{1}^4. \label{eq:4mk}
\end{numcases}
\endnumparts
It turns out that the cumulants of von Neumann entropy can be expressed through polygamma functions\cite{Page93, Wei17, Wei20}. The $i$-th order polygamma function is defined as
\begin{equation}\label{eq:pgd}
\psi_{i}(z)=\frac{\partial^{i+1}\ln\Gamma(z)}{\partial z^{i+1}}=(-1)^{i+1}i!\sum_{k=0}^{\infty}\frac{1}{(k+z)^{i+1}},
\end{equation}
which also admits a finite sum expression for positive integer arguments. In particular, the first four orders of polygamma functions are involved in the present work, the finite sum forms of which are 
\begin{equation}\label{eq:p0}
\psi_{0}(l)=-\gamma+\sum_{k=1}^{l-1}\frac{1}{k}
\end{equation}
\begin{equation}\label{eq:p1}
\psi_{1}(l)=\frac{\pi^{2}}{6}-\sum_{k=1}^{l-1}\frac{1}{k^{2}}
\end{equation}
\begin{equation}\label{eq:p2}
\psi_{2}(l)=-2\zeta(3)+2\sum_{k=1}^{l-1}\frac{1}{k^{3}}
\end{equation}
\begin{equation}\label{eq:p3}
\psi_{3}(l)=\frac{\pi^{4}}{15}-6\sum_{k=1}^{l-1}\frac{1}{k^{4}},
\end{equation}
where $\gamma\approx0.5772$ is the Euler's constant and $\zeta(3)\approx1.20206$ is the Ap\'{e}ry's constant with

\begin{equation}\label{eq:z}
\zeta(s)=\sum_{k=1}^{\infty}\frac{1}{k^{s}}
\end{equation}
being the Riemann zeta function. 

In the present paper, we focus on the kurtosis of von Neumann entropy defined as 
\begin{equation}\label{eq:skew}
\gamma_{2}=\mathbb{E}_{f}\!\left[X^4\right]-3=\frac{\kappa_{4}}{\kappa_{2}^{2}},
\end{equation}
where $X$ denotes the standardized von Neumann entropy
\begin{equation}\label{eq:X}
X=\frac{S-\kappa_{1}}{\sqrt{\kappa_{2}}}.
\end{equation}
 The kurtosis describes the shape of the distribution in terms of the  tails. As seen from the definition~(\ref{eq:skew}), calculating the kurtosis requires the additional knowledge of the fourth cumulant of $S$. The exact formula of the fourth cumulant is given in the following proposition, which is the main contribution of this work.
\begin{proposition}\label{prop1}
For a bipartite system of subsystem dimensions $m$ and $n$ $(m\leq n)$, the exact fourth cumulant of von Neumann entropy~(\ref{eq:vN}) under the Hilbert-Schmidt ensemble~(\ref{eq:fte}) is given by
\begin{eqnarray}\label{eq:k4}
\kappa_{4}=&\!d_1\psi _3(m n+1)+d_2\psi _3(n)+d_3\psi _2(n)+d_4 \psi _1^2(n)+d_5 \psi _1(n)+d_6,
\end{eqnarray}
where the coefficients $d_1$, $d_2$, $\dots$, $d_6$ are listed in~table 1.
\end{proposition}

Detailed derivation of proposition \ref{prop1} is presented in the next section. For convenience, we summarize the first four cumulants of von Neumann entropy under the Hilbert-Schmidt ensemble in~table 1.

\renewcommand\thetable{1(a)} 
\begin{longtable}[h!]{l c}
 \bs \bs
\caption{First four cumulants of von Neumann entropy}\label{t:k1to4}\\

\hline\hline

Cumulants &  References    \\ \bs

\bs\multirow{1}{*}{$\kappa_{1}=a_1\psi _0(m n+1)+a_2\psi _0(n)+a_3$} & \quad \cite{Page93,Foong94,Ruiz95}  \\ \bs
\bs\multirow{1}{*}{$\kappa_{2}=b_1\psi _1(m n+1)+b_2\psi _1(n)+b_3$} & \quad \cite{VPO16,Wei17,Bianchi19}  \\ \bs
\bs\multirow{1}{*}{$\kappa_{3}=c_1\psi _2(m n+1)+c_2\psi _2(n)+c_3\psi _1(n)+c_4$} & \quad \cite{Bianchi19,Wei20}  \\ \bs
\bs\multirow{1}{*}{$\kappa_{4}=d_1\psi _3(m n+1)+d_2\psi _3(n)+d_3\psi _2(n)+d_4 \psi _1^2(n)+d_5 \psi _1(n)+d_6$} & \quad present work  \\ \bs
\hline\hline
\end{longtable}

\begin{table}[h]
\centering 
\normalsize

\renewcommand\thetable{1(b)} 

\caption{Coefficients of first four cumulants of von Neumann entropy}

\begin{subtable}[\label{table:1a} Coefficients~of~$\kappa_1$]
{
\begin{tabular}{l  p{2cm}}
\hhline{==}
\bs\multirow{1}{*}{$a_1=$} & $1$  \\
\bs\multirow{1}{*}{$a_2=$} & $-1$  \\
\bs\bs\multirow{1}{*}{$a_3=$} & $\displaystyle -\frac{m+1}{2n}$  \\
\end{tabular}}
\end{subtable}
\qquad\qquad\qquad\qquad\qquad~~~~~~~~
\subtable[\label{table:1b}Coefficients of $\kappa_2$]
{\begin{tabular}{l l}
\hhline{==}
\bs\multirow{1}{*}{$b_1=$} & $-1$  \\ 
\bs\multirow{1}{*}{$b_2=$} & $\displaystyle \frac{m+n}{mn+1}$  \\
\bs\multirow{1}{*}{$b_3=$} & $-\displaystyle \frac{(m+1)(m+2n+1)}{4n^{2}(mn+1)}$  \\ 
\end{tabular}}
\end{table}

\begin{table}[h]
\centering
\normalsize
\ContinuedFloat
\subtable[\label{table:1c}Coefficients of $\kappa_3$]
{\begin{tabular}{l  p{13.8cm}}
\hhline{==}
\bs\multirow{1}{*}{$~c_1=$} & $1$  \\ \bs
\bs\multirow{1}{*}{$~c_2=$} & $\displaystyle -\frac{\left(m^2+3 m n+n^2+1\right)}{(m n+1) (m n+2)}$  \\ \bs
\bs\multirow{1}{*}{$~c_3=$} & $\displaystyle \frac{\left(m^2-1\right) \left(m n-3 n^2+1\right)}{n (m n+1)^2 (m n+2)}$  \\ \bs
\bs\multirow{1}{*}{$~c_4=$} & $\displaystyle -\frac{(m+1)}{4 n^3 (m n+1)^2 (m n+2)}  \left(3 m^2 n^2+2 m^3 n+4 m^2 n+2 m^2+4 m n^3+3 m n^2+\right.$ \\ \bs
 & $8 m n+4 m+\left.10 n^2+6 n+2\right)$\\ \bs \bs
\end{tabular}}
\end{table}

\begin{table}[h]
\centering
\normalsize
\ContinuedFloat
\subtable[\label{t:k4S}Coefficients of $\kappa_4$]
{\begin{tabular}{l  p{13.8cm}}
\hhline{==}
\bs\multirow{1}{*}{$d_1=$} &$-1$  \\ \bs
\bs\multirow{1}{*}{$d_2=$} & $\displaystyle\frac{(m+n) \left(m^2+5 m n+n^2+5\right) }{(m n+1) (m n+2) (m n+3)}$\\ \bs
\bs\multirow{1}{*}{$d_3=$} &  $\displaystyle\frac{(m-1) (m+1)}{n (m n+1)^2(m n+2)^2 (m n+3)}\left(6 m^2 n^3-3 m^3 n^2-9 m^2 n+12 m n^4+~~~\right.$  \\ \bs
& $\!\left.6 m n^2-6 m+20 n^3-8 n\right)$\\\bs
\bs\multirow{1}{*}{$d_4=$} & $\displaystyle\frac{6 \left(m^2-1\right) \left(n^2-1\right)}{(m n+1)^2 (m n+2) (m n+3)}$\\ \bs
\bs\multirow{1}{*}{$d_5=$} &$\displaystyle\frac{m^2-1}{n^2 (m n+1)^3 (m n+2)^2 (m n+3)}\left(3 m^4 n^3-9 m^3 n^4+15 m^3 n^2-6 m^2 n^4-\right.$\\ \bs
& $21 m^2 n^3+6 m^2 n^2+24 m^2 n-36 m n^4-18 m n^3-4 m n^2+18 m n+12 m-$ \\ \bs
& $60 n^3-\!\left.12 n^2+8 n+12\right)$ \\ \bs
\bs\multirow{1}{*}{$d_{6}=$} & $-\displaystyle\frac{m+1}{8 n^4 (m n+1)^3 (m n+2)^2 (m n+3)}\left(15 m^6 n^3+20 m^5 n^4+45 m^5 n^3+
\right.$ \\ \bs
& $63 m^5 n^2+24 m^4 n^5+40 m^4 n^4+185 m^4 n^3+189 m^4 n^2+24 m^3 n^6+24 m^3 n^5+$ \\ \bs
& $200 m^3 n^4+295 m^3 n^3+453 m^3 n^2+192 m^2 n^5+180 m^2 n^4+560 m^2 n^3+$ \\ \bs
& $591 m^2 n^2+84 m^4 n+252 m^3 n+396 m^2 n+36 m^3+108 m^2+520 m n^4+$ \\ \bs
& $420 m n^3+576 m n^2+372 m n+108 m+\!\left.448 n^3+312 n^2+144 n+36\right)$ \\ \bs
\bs
\hhline{==}

\end{tabular}}
\end{table}
\pagebreak

By definition, the first two cumulants of the standardized von Neumann entropy~\eref{eq:X}  are
\begin{equation}
\kappa_{1}^{(X)}=0,\qquad\kappa_{2}^{(X)}=1.
\end{equation}
The higher order cumulants of $S$ and $X$ are related by
\begin{equation}\label{eq:SXc}
\kappa_{j}^{(X)}=\frac{\kappa_{j}}{\kappa_{2}^{j/2}},\qquad j\geq3.
\end{equation}
It is conjectured that the standardized von Neumann entropy~\eref{eq:X} converges to a Gaussian distribution under the asymptotic regime~\cite{Wei20}
\begin{equation}\label{eq:lim}
m\to\infty,~~~~n\to\infty,~~~~\frac{m}{n}=c\in(0,1].
\end{equation}
Proving the above conjecture requires showing that all the higher cumulants~\eref{eq:SXc} vanish in the regime~\eref{eq:lim}. It is shown in~\cite{Wei20} that the third cumulant of $X$, which can be interpreted as the skewness of $S$, vanishes under the limit~\eref{eq:lim}. 
With the exact formula~\eref{eq:k4} obtained in the present work, a further evidence to the conjecture is observed by taking the limit~{\eref{eq:lim},

\begin{equation}
\gamma_2=\kappa_{4}^{(X)}=\frac{\kappa_{4}}{\kappa_{2}^{2}}\rightarrow 0.\label{eq:ev2}
\end{equation}
Namely, the fourth cumulant of $X$ (or the kurtosis of $S$) has been shown to vanish under the asymptotic regime~\eref{eq:lim}.
\begin{figure}[!h]
\centering
\includegraphics[width=0.85\linewidth]{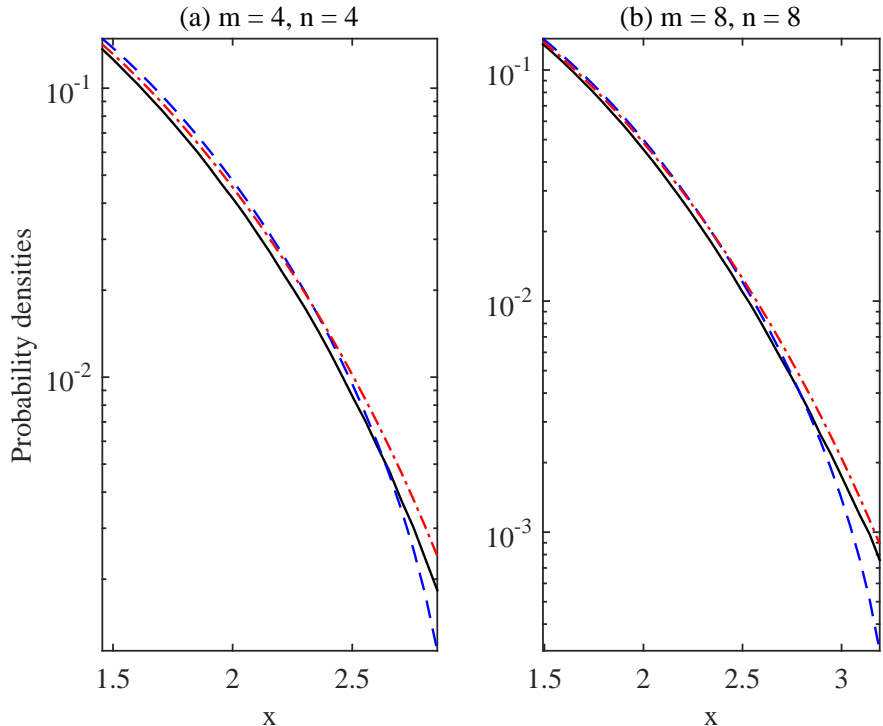}
\caption{Log-linear plot of probability densities of standardized von Neumann entropy~(\ref{eq:X}) of subsystem dimensions $m=4$, $n=4$ in subplot (a) and $m=8$, $n=8$ in subplot (b) : A comparison of Gaussian approximation with the $\kappa_{3}$ correction term~\cite{Wei20} (dashed line in blue) to the approximation with the $\kappa_{3}$ and $\kappa_{4}$ correction term~\eref{appxk4} (dash-dot line in red). The solid line in black represents simulated true distribution.}
\label{fig:p1}
\end{figure}

In the cases of finite subsystem sizes, approximations to the distribution of von Neumann entropy can be constructed from the exact cumulant expressions by using, for example, the type-A Gram-Charlier series~\cite{Wei20, Wei20BH, Cramer}. 
The simplest form of such an approximation is the standard Gaussian distribution
\begin{equation}
f_{X}(x)\approx\frac{1}{\sqrt{2\pi}}\e^{-\frac{1}{2}x^{2}}.
\end{equation}
The obtained $\kappa_3$ and $\kappa_4$ expressions can now be utilized to provide a refined approximation as~\cite{Cramer}
\begin{equation}\label{appxk4}
f_{X}(x)\approx\frac{1}{\sqrt{2\pi}}\e^{-\frac{1}{2}x^{2}}\left(1+\frac{\kappa_{3}}{6\kappa_{2}^{3/2}}H_{3}(x)+\frac{\kappa_{4}}{24\kappa_{2}^{2}}H_{4}(x)\right),
\end{equation}
where $H_k(x)$ denotes the probabilist's Hermite polynomials with
\begin{equation}
H_{3}(x)=x^{3}-3x, \qquad\qquad H_{4}(x)=x^{4}-6x^2+3.
\end{equation}
Hereinafter, the correction terms ${\kappa_{3}}/(6{\kappa_{2}^{3/2}})H_{3}(x)$ and ${\kappa_{4}}/(24{\kappa_{2}^{2}})H_{4}(x)$ in~\eref{appxk4} are referred to as the $\kappa_{3}$ correction term and the $\kappa_{4}$ correction term, respectively. Correction terms involving higher order cumulants are expected to lead to more accurate approximations to the tail of the distribution~\cite{Cramer}. 
The Gaussian approximation with the $\kappa_3$ correction term \cite{Wei20} is shown to provide improved accuracy in describing the left-skewed nature of the distribution of von Neumann entropy. Here, we further refine the approximation by adding the $\kappa_4$ correction term. In \fref{fig:p1}, we numerically compare the tails of the refined approximation~\eref{appxk4} to that with only the $\kappa_{3}$ correction term\cite{Wei20}, where the dimensions of the subsystems are $m=4$, $n=4$ in \fref{fig:p1}(a) and $m=8$, $n=8$ in \fref{fig:p1}(b). We observe from the figure that before reaching the tails, the accuracy of both approximations are similar. The differences occur when approaching the tails, where the accuracy improvement of the new approximation~\eref{appxk4} becomes more obvious.

The rest of the paper is organized as follows. In \sref{sec:deri}, we derive the main result of the exact fourth cumulant of von Neumann entropy under the Hilbert-Schmidt ensemble. Specifically, in \sref{sec:rela} we relate the computation of the kurtosis of von Neumann entropy to that of an induced entropy over the Wishart-Laguerre ensemble, which boils down to computing five integrals over the correlation functions. In \sref{sec:CoIns}, we outline the procedure of evaluating the five integrals into summations involving gamma and polygamma functions. In~\sref{sec:EvSum}, we discuss the simplification procedure of the resulting summations. In particular, we introduce the unsimplifiable bases as summarized in table 2 and describe the  new cancellation phenomena. The relevant summation identities utilized in the simplification are listed in the appendices.
\section{Derivation of the fourth cumulant}\label{sec:deri}
\subsection{Cumulant relation}\label{sec:rela}
In this subsection, we perform the transformation from the fourth cumulant of von Neumann entropy over the fixed-trace ensemble to that of an induced entropy over the Laguerre ensemble. 

First, we discuss the relation of the fourth moments. By construction, the random coefficient matrix $\mathbf{X}$ has a natural relation with a Wishart matrix $\mathbf{YY}^{\dag}$ as
\begin{equation}\label{eq:wf}
\mathbf{XX}^{\dag}=\frac{\mathbf{YY}^{\dag}}{\tr\left(\mathbf{YY}^{\dag}\right)},
\end{equation}
where $\mathbf{Y}$ is an $m\times n$ ($m\leq n$) matrix of independently and identically distributed complex Gaussian entries. The density of the eigenvalues $0<\theta_{m}<\dots<\theta_{1}<\infty$ of $\mathbf{YY}^{\dag}$ is given by~\cite{Forrester}
\begin{equation}\label{eq:we}
g\left(\bm{\theta}\right)=\frac{1}{C}\prod_{1\leq i<j\leq m}\left(\theta_{i}-\theta_{j}\right)^{2}\prod_{i=1}^{m}\theta_{i}^{n-m}\e^{-\theta_{i}},
\end{equation}
where $C$ is given by~(\ref{eq:con}) and the above ensemble is known as the Wishart-Laguerre ensemble. The trace of the Wishart matrix
\begin{equation}
r=\tr\left(\mathbf{YY}^{\dag}\right)=\sum_{i=1}^{m}\theta_{i}
\end{equation}
follows a gamma distribution with the density
\begin{equation}\label{eq:r}
h_{mn}(r)=\frac{1}{\Gamma(mn)}\e^{-r}r^{mn-1},~~~~~~r\in[0,\infty).
\end{equation}
The relation~(\ref{eq:wf}) induces the change of variables
\begin{equation}\label{eq:cv}
\lambda_{i}=\frac{\theta_{i}}{r},~~~~~~i=1,\ldots,m,
\end{equation}
that leads to a well-known relation (see, e.g.~\cite{Page93}) among the densities~(\ref{eq:fte}),~(\ref{eq:we}), and~(\ref{eq:r}) as
\begin{equation}\label{eq:relation}
f\left(\bm{\lambda}\right)h_{mn}(r)\dd r\prod_{i=1}^{m}\dd\lambda_{i}=g\left(\bm{\theta}\right)\prod_{i=1}^{m}\dd\theta_{i}.
\end{equation}
This relation implies the fact that $r$ is independent of each $\lambda_{i}$, $i=1,\ldots,m$, since their densities factorize. We now define 
\begin{equation}\label{eq:T}
\mathrm T=\sum_{i=1}^{m}\theta_{i}\ln\theta_{i},
\end{equation}
which is referred to as an induced entropy over the Wishart-Laguerre ensemble.
By using the relation (\ref{eq:cv}), the von Neumann entropy $S$ can be converted to the induced entropy $T$ as
\begin{equation}\label{eq:StT}
{\mathrm S}=-\sum_{i=1}^{m}\frac{\theta_i}{r}\ln\frac{\theta_i}{r}=r^{-1}(r\ln{r}-{\mathrm T}).
\end{equation}
Accordingly, one could rewrite the fourth moment of von Neumann entropy as 
\begin{equation}\label{eq:MST1} 
\fl\mathbb{E}_{f}\!\left[{\mathrm S}^{4}\right]=\int_{\bm{\lambda}}r^{-4}\left({\mathrm T}^{4}+4{\mathrm S}^{3}r^{4}\ln r-6{\mathrm S}^2r^{4}\ln^{2}r+4{\mathrm S}r^{4}\ln^{3}r-r^{4}\ln^{4}r\right)f\left(\bm{\lambda}\right)\prod_{i=1}^{m}\dd\lambda_{i}.
\end{equation}
By multiplying an appropriate constant
\begin{equation}
1=\int_{r}h_{mn+4}(r)\dd{r},
\end{equation}
we are able to change the measures of (\ref{eq:MST1}) using the relation  (\ref{eq:relation}) and considering the fact that $r$ and $\lambda$ are independent, the fourth moment of von Neumann entropy is converted to
\begin{eqnarray}\label{eq:MST2} 
\fl\mathbb{E}_{f}\!\left[{\mathrm S}^{4}\right]=&\frac{1}{(mn)_{4}}\left(\mathbb{E}_{g}\!\left[{\mathrm T}^{4}\right]+4\mathbb{E}_{h}\!\left[r^{4}\ln r\right]\mathbb{E}_{f}\!\left[{\mathrm S}^{3}\right]-6\mathbb{E}_{h}\!\left[r^{4}\ln^{2}r\right]\mathbb{E}_{f}\!\left[{\mathrm S}^2\right]+\right.\nonumber\\
\fl&4\mathbb{E}_{h}\!\left[r^{4}\ln^{3}r\right]\mathbb{E}_{f}\!\left[{\mathrm S}\right]-\mathbb{E}_{h}\!\!\left.\left[r^{4}\ln^{4}r\right]\right),
\end{eqnarray}
 with $(a)_{n}$ denoting the Pochhammer's symbol. In (\ref{eq:MST2}), the expected values of the form $r^{4}\ln^{l}r$ over the density $h_{mn}(r)$ can be obtained by 
\begin{equation}
\int_{0}^{\infty}\!\e^{-r}r^{a-1}\ln^n{r}\dd{r}=\frac{\partial^n}{\partial a}\Gamma(a),\qquad\Re(a)>0,
\end{equation}
\begin{eqnarray}
\fl\mathbb{E}_{h}\!\left[r^{4}\ln r\right]= (mn)_{4}\psi_{0}(mn+4) 
\end{eqnarray}
\begin{eqnarray}
\fl\mathbb{E}_{h}\!\left[r^{4}\ln^{2}r\right]= (mn)_{4}\left(\psi_{0}^{2}(mn+4)+\psi_{1}(mn+4)\right)
\end{eqnarray}
\begin{eqnarray}
\fl\mathbb{E}_{h}\!\left[r^{4}\ln^{3}r\right]= (mn)_{4}\left(\psi_{0}^{3}(mn+4)+3\psi_{0}(mn+4)\psi_{1}(mn+4)+\psi_{2}(mn+4)\right)
\end{eqnarray}
\begin{eqnarray}
\fl\mathbb{E}_{h}\!\left[r^{4}\ln^{4}r\right] =& (mn)_{4}\left(\psi_0^4 (mn+4)+6 \psi_0^2 (mn+4)\psi _1(mn+4) +4 \psi_0(mn+4)\times\nonumber\right.\\
&\psi _2(mn+4) +\left.3 \psi _1^2(mn+4)+\psi _3(mn+4)\right).
\end{eqnarray}
From the above results along with the first three moments relations between ${\mathrm S}$ and ${\mathrm T}$ derived in \cite{Ruiz95, Wei17, Wei20}, we obtain the fourth cumulant relation 
\begin{eqnarray}\label{eq:k4ST}
\fl\kappa_4=&\frac{1}{ (mn)_{4}}\left(\kappa_4^{\mathrm T}-\frac{12 \kappa _1^{\mathrm T} \kappa _3^{\mathrm T}}{m n}-\frac{4 \left(3 m^2 n^2+12 m n+11\right) \kappa _3^{\mathrm T}}{(m n+1) (m n+2)}-\frac{6 (2 m n+3) \left(\kappa _2^{\mathrm T}\right){}^2}{m n (m n+1)}+\right.\nonumber\\
\fl&\frac{12 (5 m n+6) \left(\kappa _1^{\mathrm T}\right){}^2 \kappa _2^{\mathrm T}}{m^2 n^2 (m n+1)}+\frac{24 (2 m n+3) (2 m n+5) \kappa _1^{\mathrm T} \kappa _2^{\mathrm T}}{m n (m n+1) (m n+2)}+\frac{12 (m n+3) }{(m n+1)^2}\times\nonumber\\
\fl&\frac{\left(3 m^2 n^2+9 m n+7\right) \kappa _2^{\mathrm T}}{ (m n+2)}-\frac{6 (5 m n+6) \left(\kappa _1^{\mathrm T}\right){}^4}{m^3 n^3 (m n+1)}-\frac{8 (2 m n+3) (5 m n+12) \left(\kappa _1^{\mathrm T}\right){}^3}{m^2 n^2 (m n+1) (m n+2)}-\nonumber\\
\fl&\!\left.\frac{12 (m n+3) (2 m n+3) (3 m n+4) \left(\kappa _1^{\mathrm T}\right){}^2}{m n (m n+1)^2 (m n+2)}-\frac{24 (m n+2) (m n+3) \kappa _1^{\mathrm T}}{(m n+1)^2}\right)-\nonumber\\
\fl&\psi _3(m n+1),
\end{eqnarray}
where $\kappa _i^{\mathrm T}$ denotes the $i$-th cumulant of the induced von Neumann entropy ${\mathrm T}$. The first three cumulants are obtained in \cite{Wei20} as
\begin{eqnarray}\label{eq:kT1} 
\fl\kappa_{1}^{{\mathrm T}}=&mn\psi_{0}(n)+\frac{1}{2}m(m+1) 
\end{eqnarray}
\begin{eqnarray} \label{eq:kT2}
\fl\kappa_{2}^{{\mathrm T}}=&mn(m+n)\psi_{1}(n)+mn\psi_{0}^{2}(n)+m\left(m+2n+1\right)\psi_{0}(n)+\frac{1}{2}m(m+1)
\end{eqnarray}
\begin{eqnarray}
\fl\kappa_{3}^{{\mathrm T}}=&mn\left(m^2+3mn+n^2+1\right)\psi_{2}(n)+6mn(m+n)\psi_{0}(n)\psi_{1}(n)+m\left(2m^2+12mn+\right.\nonumber\\
\fl&3m+\!\left.6n^{2}+3n+1\right)\psi_{1}(n)+2mn\psi_{0}^{3}(n)+3m(m+3n+1)\psi_{0}^{2}(n)+6m(m+\nonumber\\
\fl&n+1)\psi_{0}(n)+m(m+1). \label{eq:kT3}
\end{eqnarray}
The remaining task of this work is to compute the fourth cumulant of the induced von Neumann entropy $\kappa_4^\mathrm{T}$. By definition, one has
\begin{eqnarray}
\fl {\mathrm T}^4=&\sum_{i=1}^{m}\theta_i^4\ln{\theta_i^4}+\sum_{1\leq i\neq j\leq m}\!\!6\theta_i^2\theta_j^2\ln^2{\theta_i}\ln^2{\theta_j}+\sum_{1\leq i\neq j\leq m}\!\!8\theta_i^3\theta_j\ln^3{\theta_{i}}\ln{\theta_j}+\nonumber\\
\fl&\sum_{1\leq i\neq j\neq k\leq m}\!\!36\theta_i^2\theta_j\theta_k\ln^2{\theta_{i}}\ln{\theta_j}\ln{\theta_k}+\sum_{1\leq i\neq j\neq k\neq l\leq m}\!\!24\theta_i\theta_j\theta_k\theta_l\ln{\theta_{i}}\ln{\theta_j}\ln{\theta_k}\ln{\theta_l}.
\end{eqnarray}
It then follows that
\begin{eqnarray*}\label{eq:MT4}
\fl\mathbb{E}_{g}[{\mathrm T}^4]=&{m\choose1}\mathbb{E}_g[\theta_1^4\ln{\theta_1^4}]+6{m\choose2}\mathbb{E}_g[\theta_1^2\theta_2^2\ln^2{\theta_1}\ln^2{\theta_2}]+8{m\choose2}\mathbb{E}_g[\theta_1^3\theta_2\ln^3{\theta_1}\ln{\theta_2}]+\\
\fl&36{m\choose3}\mathbb{E}_g[\theta_1^2\theta_2\theta_3\ln^2{\theta_1}\ln{\theta_2}\ln{\theta_3}]+24{m\choose4}\mathbb{E}_g[\theta_1\theta_2\theta_3\theta_4\ln{\theta_1}\ln{\theta_2}\ln{\theta_3}\ln{\theta_4}],
\end{eqnarray*}
where $g_{1}(x_{1})$, $g_{2}(x_{1},x_{2})$, $g_{3}(x_{1},x_{2},x_{3})$, and $g_{4}(x_{1},x_{2},x_{3},x_{4})$ respectively denote one to four arbitrary eigenvalue densities of the Wishart-Laguerre ensemble~(\ref{eq:we}).
It is a well-known result that the joint density $g_{N}(x_{1},\dots,x_{N})$ of $N$ (out of $m$) arbitrary eigenvalues of various matrix models, including the Wishart-Laguerre ensemble, can be written in terms of a determinant of a correlation kernel $K(x_{i},x_{j})$ as~\cite{Forrester}
\begin{equation}\label{eq:Nei}
g_{N}(x_{1},\dots,x_{N})=\frac{(m-N)!}{m!}\det\left(K\left(x_{i},x_{j}\right)\right)_{i,j=1}^{N}.
\end{equation}
By using (\ref{eq:Nei}) as well as the cumulant and moment relation~(\ref{eq:4mk}), we obtain
\begin{equation}\label{kti}
\kappa_4^\mathrm{T}=\mathrm{I_A}-3\mathrm{I_{B_1}}-4\mathrm{I_{B_2}}+12\mathrm{I_C}-6\mathrm{I_D},
\end{equation}
where we denote 
\begin{eqnarray}
\fl \mathrm {I_A}=&\int_0^\infty\!\! x_1^4\ln^4\!x_1\:K(x_1,x_1)\dd x_1\label{eq:IA}
\\\fl \mathrm{I_{B_1}}=&\int_{(0,\infty)^2} x_1^2x_2^2\ln^2\!x_1\ln^2\!x_2\:K^2(x_1,x_2)\dd x_1\dd x_2\label{eq:IB1}
\\\fl \mathrm{I_{B_2}}=&\int_{(0,\infty)^2} x_1^3x_2\ln^3\!x_1\ln x_2\:K^2(x_1,x_2)\dd x_1\dd x_2\label{eq:IB2}
\\\fl \mathrm{I_C}=&\int_{(0,\infty)^3} x_1^2x_2x_3\ln^2\!x_1\ln x_2\ln x_3\:K(x_1,x_2)K(x_1,x_3)K(x_2,x_3)\dd x_1\dd x_2\dd x_3\label{eq:IC}
\\\fl \mathrm{I_D}=&\int_{(0,\infty)^4} x_1x_2x_3x_4\ln x_1\ln x_2\ln x_3\ln x_4\:K(x_1,x_3)K(x_1,x_4)K(x_2,x_3)K(x_2,x_4)\times\nonumber\\
\fl&\dd x_1\dd x_2\dd x_3\dd x_4\label{eq:ID}.
\end{eqnarray}
The computation of the above integrals are provided in the next two subsections, where we will eventually show that
\begin{eqnarray}\label{eq:kT4}
\fl\kappa_4^\mathrm{T}=&m n (m+n) \left(m^2+5 m n+n^2+5\right) \psi _3(n)+12 m n \left(m^2+3 m n+n^2+1\right) \psi _0(n) \times\nonumber\\
\fl&\psi _2(n)+\!\left(3 m^3+36 m^2 n+6 m^2+54 m n^2+18 m n+9 m+12 n^3+6 n^2+26 n+6\right)\times \nonumber\\
\fl&m \psi _2(n)+6 m n \left(2 m^2+5 m n+2 n^2+1\right) \psi _1^2(n)+36 m n (m+n) \psi _0^2(n)\psi _1(n)+12 \times\nonumber\\
\fl&m \left(2 m^2+14 m n+3 m+8 n^2+3 n+1\right) \psi _0(n) \psi _1(n)+18 m \left(2 m^2+6 m n+3 m+\right.\nonumber\\
\fl&2 \!\left.n^2+2 n+1\right) \psi _1(n)+6 m n \psi _0^4(n)+4 m (3 m+11 n+3) \psi _0^3(n)+24 m (2 m+3 n+\nonumber\\
\fl&2) \psi _0^2(n)+12 m (3 m+2 n+3) \psi _0(n)+3 m (m+1).
\end{eqnarray}
Inserting the above expression along with~(\ref{eq:kT1}),~(\ref{eq:kT2}), and~(\ref{eq:kT3}) into~(\ref{eq:k4ST}), the claimed main result~(\ref{eq:k4}) is then established.

\subsection{Calculations of integrals}\label{sec:CoIns}
In this subsection, we list the necessary results on the Wishart-Laguerre ensemble for the computation of the integrals~(\ref{eq:IA})-(\ref{eq:ID}). First, the correlation kernel of the Wishart-Laguerre ensemble can be explicitly written as~\cite{Forrester}
\begin{equation}\label{eq:ker}
K(x_{i},x_{j})=\sqrt{\e^{-x_{i}-x_{j}}(x_{i}x_{j})^{n-m}}\sum_{k=0}^{m-1}\frac{C_{k}(x_{i})C_{k}(x_{j})}{k!(n-m+k)!},
\end{equation}
where
\begin{equation}\label{eq:kerc}
C_{k}(x)=(-1)^{k}k!L_{k}^{(n-m)}(x)
\end{equation}
with
\begin{equation}\label{eq:Lar}
L_{k}^{(n-m)}(x)=\sum_{i=0}^{k}(-1)^{i}{n-m+k\choose k-i}\frac{x^i}{i!}
\end{equation}
being the (generalized) Laguerre polynomial of degree $k$. The orthogonality relation of Laguerre polynomials~\cite{Forrester}
\begin{equation}\label{eq:oc}
\int_{0}^{\infty}\!\!x^{n-m}\e^{-x}L_{k}^{(n-m)}(x)L_{l}^{(n-m)}(x)\dd{x}=\frac{(n-m+k)!}{k!}\delta_{kl}
\end{equation}
can be generalized to
\begin{equation}\label{eq:Swm}
\fl\int_{0}^{\infty}\!\!x^{q}\e^{-x}L_{s}^{(\alpha)}(x)L_{t}^{(\beta)}(x)\dd{x}=(-1)^{s+t}\sum_{k=0}^{\min(s,t)}{q-\alpha\choose s-k}{q-\beta\choose t-k}\frac{\Gamma(q+1+k)}{k!},
\end{equation}
where $\Re(q)>-1$, which is due to Schr{\"{o}}dinger~\cite{Schrodinger1926}. 
By taking up to the fourth derivatives of the Schr{\"{o}}dinger's integral~(\ref{eq:Swm}) with respect to $q$, we obtain the following integrals useful in computing $\mathrm{I_A}$, $\mathrm{I_{B_1}}$, $\mathrm{I_{B_2}}$, $\mathrm{I_{C}}$, and $\mathrm{I_{D}}$,

\begin{eqnarray}
\fl\int_{0}^{\infty}\!\!x^{q}\e^{-x}\ln{x}~L_{s}^{(\alpha)}(x)L_{t}^{(\beta)}(x)\dd{x}=(-1)^{s+t}\sum_{k=0}^{\min(s,t)}{q-\alpha\choose s-k}{q-\beta\choose t-k}\frac{\Gamma(q+1+k)}{k!}\Psi_{0}\label{eq:1d}
\end{eqnarray}

\begin{eqnarray}
\fl\int_{0}^{\infty}\!\!x^{q}\e^{-x}\ln^{2}{x}~L_{s}^{(\alpha)}(x)L_{t}^{(\beta)}(x)\dd{x}=&(-1)^{s+t}\sum_{k=0}^{\min(s,t)}{q-\alpha\choose s-k}{q-\beta\choose t-k}\frac{\Gamma(q+1+k)}{k!}\times\nonumber\\
\fl&\left(\Psi_{0}^{2}+\Psi_{1}\right)\label{eq:2d}
\end{eqnarray}

\begin{eqnarray}
\fl\int_{0}^{\infty}\!\!x^{q}\e^{-x}\ln^{3}{x}~L_{s}^{(\alpha)}(x)L_{t}^{(\beta)}(x)\dd{x}=&(-1)^{s+t}\sum_{k=0}^{\min(s,t)}{q-\alpha\choose s-k}{q-\beta\choose t-k}\frac{\Gamma(q+1+k)}{k!}\times\nonumber\\
\fl&\left(\Psi_{0}^{3}+3\Psi_{0}\Psi_{1}+\Psi_{2}\right)\label{eq:3d}
\end{eqnarray}

\begin{eqnarray}
\fl\int_{0}^{\infty}\!\!x^{q}\e^{-x}\ln^{4}{x}~L_{s}^{(\alpha)}(x)L_{t}^{(\beta)}(x)\dd{x}=&(-1)^{s+t}\sum_{k=0}^{\min(s,t)}{q-\alpha\choose s-k}{q-\beta\choose t-k}\frac{\Gamma(q+1+k)}{k!}\times\nonumber\\
\fl&\left(\Psi^4_0+6\Psi^2_0\Psi_1+3\Psi^2_1+4\Psi_0\Psi_2+\Psi_3\right),\label{eq:4d}
\end{eqnarray}
where we denote
\begin{eqnarray}
\!\!\!\!\!\Psi_{j}&=&\psi_{j}(q+1+k)+\psi_{j}(q-\alpha+1)+\psi_{j}(q-\beta+1)- \nonumber \\
\!\!\!\!\!&&\psi_{j}(q-\alpha-s+1+k)-\psi_{j}(q-\beta-t+1+k),\qquad j=0,1,2,3.
\end{eqnarray}
Instead of the summation form~(\ref{eq:ker}), the one arbitrary eigenvalue density
\begin{equation}
g_1(x_1)=\frac{1}{m}K(x_1,x_1)
\end{equation}
admits a more convenient form by the Christoffel-Darboux formula~\cite{Ruiz95,Forrester} as
\begin{equation}\label{eq:one}
\!\!\!\!\!\!\!\!\!\!\!\!\!\!\!\!\!\!\!\!g_{1}(x)=\frac{(m-1)!}{(n-1)!}x^{n-m}\e^{-x}\left(\left(L_{m-1}^{(n-m+1)}(x)\right)^{2}-L_{m-2}^{(n-m+1)}(x)L_{m}^{(n-m+1)}(x)\right).
\end{equation}
The computations of $\mathrm{I_A}$, $\mathrm{I_{B_1}}$, $\mathrm{I_{B_2}}$, $\mathrm{I_{C}}$, and $\mathrm{I_{D}}$ share the same procedure, which has been demonstrated for the lower order moments in \cite{Ruiz95, Wei17, Wei20}. Namely, by using the correlation kernel~(\ref{eq:ker}) and the integral identities~(\ref{eq:1d})-(\ref{eq:4d}), all the integrals in $\mathrm{I_A}$, $\mathrm{I_{B_1}}$, $\mathrm{I_{B_2}}$, $\mathrm{ I_{C}}$, and $\mathrm {I_{D}}$ can be evaluated into summations involving gamma and polygamma functions. We note that in the computation of $\mathrm{I_A}$, we can also directly apply the Christoffel-Darboux form of the one-point density~(\ref{eq:one}). In writing down the summation forms, one will also need the following formulas to resolve indeterminacy induced by gamma and polygamma functions with non-positive arguments, i.e.,
\numparts
\label{eq:pgna}\begin{eqnarray}
\!\!\!\!\!\!\!\!\!\!\!\!\!\!\!\!\!\!\!\!\!\!\!\!\!\!\!\!\!\Gamma(-l+\epsilon)&\!=&\frac{(-1)^{l}}{l!\epsilon}\left(1+\psi_{0}(l+1)\epsilon+o\left(\epsilon^2\right)\right)\label{eq:pgna1}\\
\!\!\!\!\!\!\!\!\!\!\!\!\!\!\!\!\!\!\!\!\!\!\!\!\!\!\!\!\!\psi_{0}(-l+\epsilon)&=&-\frac{1}{\epsilon}+\psi_{0}(l+1)+\left(2\psi_{1}(1)-\psi_{1}(l+1)\right)\epsilon+\frac{1}{2}\psi_{2}(l+1)\epsilon^{2}+o\left(\epsilon^3\right)\label{eq:pgna2}\\
\!\!\!\!\!\!\!\!\!\!\!\!\!\!\!\!\!\!\!\!\!\!\!\!\!\!\!\!\!\psi_{1}(-l+\epsilon)&=&\frac{1}{\epsilon^{2}}-\psi_{1}(l+1)+\psi_{1}(1)+\zeta(2)+o\left(\epsilon\right)\label{eq:pgna3}\\
\!\!\!\!\!\!\!\!\!\!\!\!\!\!\!\!\!\!\!\!\!\!\!\!\!\!\!\!\!\psi_{2}(-l+\epsilon)&=&-\frac{2}{\epsilon^{3}}+\psi_{2}(l+1)+\psi_{2}(1)+2\zeta(3)+o\left(\epsilon\right)\label{eq:pgna4}\\
\!\!\!\!\!\!\!\!\!\!\!\!\!\!\!\!\!\!\!\!\!\!\!\!\!\!\!\!\!\psi_{3}(-l+\epsilon)&=&\frac{6}{\epsilon^4}-\psi _3(l+1)+\psi _3(1)+6 \zeta (3)+o\left(\epsilon\right).\label{eq:pgna5}
\end{eqnarray}
\endnumparts
Using the procedure described above, we obtain the summation forms of all integrals $\mathrm{I_A}$, $\mathrm{I_{B_1}}$, $\mathrm{I_{B_2}}$, $\mathrm{I_{C}}$, and $\mathrm{I_{D}}$ as has been similarly performed in \cite{Ruiz95, Wei17, Wei20}. The strategies in evaluating the corresponding summations will be discussed in the following subsection.

\subsection{Evaluation of summations in $\mathrm{I_A}$, $\mathrm{I_{B_1}}$, $\mathrm{I_{B_2}}$, $\mathrm{I_{C}}$, and $\mathrm{I_{D}}$}\label{sec:EvSum}
The last step in the calculation of $\kappa_{4}^{\mathrm T}$ is to evaluate the finite summations obtained from integrals $\mathrm{I_A}$, $\mathrm{I_{B_1}}$, $\mathrm{I_{B_2}}$, $\mathrm{I_{C}}$, and $\mathrm{I_{D}}$. These summations will be eventually simplified to closed-form expressions along with some unsimplifiable sums, which we refer to as unsimplifiable bases. In the literature of moment computation of von Neumann entropy \cite{Wei17, Wei20}, one observes the complete cancellations of all unsimplifiable sums such that the final results remain closed-form expressions. In the current work of computing the fourth cumulant $\kappa_{4}$, we observe the similar cancellation phenomenon, where unsimplifiable sums with higher order polygamma functions appear as expected. The list of unsimplifiable bases of this work is summarized in~\tref{t:bases}, where the new ones (in addition to the ones in \cite{Wei17, Wei20}) are $\Omega_4$--$\Omega_{17}$.
\renewcommand\thetable{2} 
\begin{longtable}[h!]{ll}
\caption{Unsimplifiable bases in $\mathrm{I_A}$, $\mathrm{I_{B_1}}$, $\mathrm{I_{B_2}}$, $\mathrm {I_{C}}$, and $\mathrm{I_{D}}$}\\

\hline\hline

\bs  $\Omega_{1}=\displaystyle\displaystyle\sum _{k=1}^m \frac{\psi _0(k+n-m)}{k}$ & \qquad  {$\Omega_{2}=\displaystyle\displaystyle\sum _{k=1}^m \frac{\psi _0^2(k+n-m)}{k}$}\\ \bs
\bs {$\Omega_{3}=\displaystyle\sum _{k=1}^m \frac{\psi _1(k+n-m)}{k}$} &\qquad{$\Omega_{4}=\displaystyle\sum _{k=1}^m \frac{\psi _0(k+n-m)}{k^2}$}  \\ \bs

\bs {$\Omega_{5}=\displaystyle\sum _{k=1}^m \frac{\psi _0^2(k+n-m)}{k^2}$} &\qquad{$\Omega_{6}=\displaystyle\sum _{k=1}^m \frac{\psi _0^2(k+n-m)}{k+n-m}$} \\ \bs
\bs {$\Omega_{7}=\displaystyle\sum _{k=1}^m \frac{\psi _0(k+n-m)\psi _0(k) }{k}$}&\qquad{$\Omega_{8}=\displaystyle\sum _{k=1}^m \frac{\psi _0^3(k+n-m)}{k}$}  \\ \bs
\bs {$\Omega_{9}=\displaystyle\sum _{k=1}^m \frac{\psi _0^2(k+n-m)\psi _0(k) }{k}$} & \qquad{$\Omega_{10}=\displaystyle\sum _{k=1}^m \frac{\psi _0^3(k+n-m)}{k+n-m}$} \\ \bs
\bs {$\Omega_{11}=\displaystyle\sum _{k=1}^m \frac{\psi _1(k+n-m)}{k+n-m}$} &\qquad{$\Omega_{12}=\displaystyle\sum _{k=1}^m \frac{\psi _0(k+n-m) \psi _1(k+n-m)}{k}$}\\ \bs
\bs {$\Omega_{13}=\displaystyle\sum _{k=1}^m \frac{ \psi _1(k+n-m)\psi _0(k)}{k}$} & \qquad  {$\Omega_{14}=\displaystyle\sum _{k=1}^m \frac{\psi _0(k+n-m) \psi _1(k+n-m)}{k+n-m}$} \\ \bs

\bs {$\Omega_{15}=\displaystyle\sum _{k=1}^m \frac{\psi _2(k+n-m)}{k}$} & \qquad  {$\Omega_{16}=\displaystyle\sum _{k=1}^m \frac{\psi _2(k+n-m)}{k+n-m}$}\\ \bs

\bs {$\Omega_{17}=\left(\displaystyle\sum _{k=1}^m \frac{\psi _0(k+n-m)}{k}\right)^2$} \\ \bs

\hline\hline\bs

\label{t:bases}
\end{longtable}

In the simplification of sums involved in $\kappa_{4}^{\mathrm{T}}$, some new cancellation phenomena appear.
The first new phenomenon is the cancellations between double summations from $\mathrm{I_{B_1}}$, $\mathrm {I_{C}}$, and $\mathrm{I_{D}}$. Specifically, these summations are not summable individually but will cancel among the obtained combination $\mathrm{I_A}-3\mathrm{I_{B_1}}-4\mathrm{I_{B_2}}+12\mathrm{I_C}-6\mathrm{I_D}$. This cancellation becomes visible by appropriate shifts of summation indices leading to a closed-form expression. More explicitly, there exists three types of double summations with such a cancellation feature, i.e.,
\begin{eqnarray}
\!\!\!\!\!\!\!\!\!\!\sum _{j=1}^{m-r_b-1}\frac{(m-j-v_a)!}{(n-j-r_a)!}\mathrm p(j)\sum _{k=1}^{m-j-r_b} \frac{ (n-j-k-v_b)!}{ (m-j-k-r_b)!(n-j-k-v_b+1)}\label{eq:unS1}\\
\!\!\!\!\!\!\!\!\!\!\sum _{j=1}^{m-r_b-1}\frac{(m-j-v_a)!}{(n-j-r_a)!}\mathrm p(j)\sum _{k=1}^{m-j-r_b} \frac{ (n-j-k-v_b)!\psi_{0}(n-j-v_c)}{ (m-j-k-r_b)!(k+r_c)^d}\label{eq:unS2}\\
\!\!\!\!\!\!\!\!\!\!\sum _{j=1}^{m-r_b-1}\frac{(m-j-v_a)!}{(n-j-r_a)!}\mathrm p(j)\sum _{k=1}^{m-j-r_b} \frac{ (n-j-k-v_b)!\psi_{0}^2(n-j-k-v_b+1)}{(m-j-k-r_b)!(k+r_c)^d}\label{eq:unS3},
\end{eqnarray}
where $\mathrm p(j)$ denotes a polynomial in $j$, and the constants $r_a$, $r_b$, $r_c$, $v_a$, $v_b$, $v_c$, $d$ are non-negative integers. 
The second new phenomenon is the cancellations that occur in quadruple summations from $\mathrm{I_{B_1}}$, $\mathrm {I_{C}}$, and $\mathrm{I_{D}}$ of the form
\begin{eqnarray}\label{eq:unS4}
\sum _{j=1}^{m-6} \frac{(m-j)!}{(n-j)!}\sum _{k=1}^{m-j-5} \frac{ (m-j-k-2)!}{ (n-j-k-2)!}\sum _{i=1}^{m-j-k-4} h_i \sum _{s=1}^{m-j-k-4}g_s.
\end{eqnarray}
Here, we note that the inner summations over $h_r$ and $g_r$ contain the following unsummable term
\begin{equation}\label{eq:unst}
\frac{(n-j-k-r-4)!}{ (m-j-k-r-4)! (r+k+c)}.
\end{equation}
By collecting all the quadruple summations of the type~(\ref{eq:unS4}) from $\mathrm{I_{B_1}}$, $\mathrm {I_{C}}$, and $\mathrm{I_{D}}$, we observe summations in pairs with different signs resulting in a cancellation. We are left with three distinct quadruple sums that can be combined into the form
\begin{eqnarray}
\sum _{j=1}^{m-6}\sum _{k=1}^{m-j-5} \frac{(m-j)! (m-j-k-2)!}{(n-j)! (n-j-k-2)!}\left( \sum _{i=1}^{m-j-k-4}\left(4h_{i}-g_i\right)\right)^2\label{eq:qcom},
\end{eqnarray}
where the inner sum over $i$ now admits a closed-form evaluation due to the complete cancellation of the unsummable term (\ref{eq:unst}).

Other than the summations~(\ref{eq:unS1})-(\ref{eq:unS4}), the rest ones in $\kappa_4^\mathrm{T}$ can be evaluated into closed-form expressions and unsimplifiable bases as listed in~\tref{t:bases} by using the identities in the appendices. After the expected but yet miraculous cancellations, the claimed closed-form identity~(\ref{eq:kT4}) is obtained.

\section{Conclusion}
In this work, we continue the study of exact cumulants of von Neumann entanglement entropy under the Hilbert-Schmidt measure. In particular, our main result is the exact yet explicit formula of the fourth cumulant valid for any subsystem dimensions. The result is derived by integrating over up to the four-point densities of the Wishart-Laguerre ensemble. The obtained formula provides additional information to the tail distribution of von Neumann entropy by constructing finite-size approximations. In computing the result, we have also derived new closed-form and semi closed-form summation identities involving gamma and polygamma functions. 

\section*{Acknowledgments}
The authors thank Xhoendi Collaku for helping with the simplification.

\appendix\

\section{Polygamma summation identities of the first type}\label{App1}
In this appendix, we list finite summation identities of polygamma functions of the type
\begin{equation}\label{eq:A1}
\sum_{k=1}^{n}k^{c}\psi_{i_{1}}^{j_{1}}(k+a_{1})\psi_{i_{2}}^{j_{2}}(k+a_{2}),
\end{equation}
$ i_{1},  i_{2}, j_{1}, j_{1}, c \in \mathbb{Z}_{\geq 0}, \Re(a_1, a_2)>-1$, hereinafter referred to as the first type useful in the simplification process in \sref{sec:EvSum}.  The corresponding identities are listed in appendix\,A.1 and semi-closed ones (containing unsimplifiable terms) are listed in  appendix\,A.2. We also inculde identities involving functions of unsimplifiable terms of the type
\begin{equation}\label{eq:A2}
\sum _{k=1}^n \frac{\psi _i^j(k+a)}{(k+a)^c},\qquad c,i,j \in \mathbb{Z}_{\geq 0}, \Re( a)>-1
\end{equation}
in appendix\,A.2.
Some remarks on derivation and implementation of the listed formulas are provided in appendix\,A.3. We also note that the unsimplifiable bases involved in appendix\, A.2 are $\Omega_1$--$\Omega_4$, $\Omega_6$--$\Omega_7$, $\Omega_{10}$--$\Omega_{11}$, $\Omega_{14}$, and $\Omega_{16}$ of table~\ref{t:bases}.

\label{sec:ap1-c}\subsection{Closed-form expressions}
\begin{equation}\label{eq:A3}
\fl\sum_{k=1}^n\psi_{0}(k+a)=(a+n)\psi_{0}(a+n+1)-a\psi_{0}(a+1)-n
\end{equation}
\begin{eqnarray}\label{eq:A4}
\fl\sum_{k=1}^{n}k\psi_{0}(k+a)&=&\frac{1}{2}\left(-a^2+a+n^2+n\right)\psi_{0}(a+n+1)+\frac{1}{2}(a-1)a\psi_{0}(a+1)+\nonumber \\
&&\frac{1}{4}n(2a-n-3)
\end{eqnarray}
\begin{eqnarray}\label{eq:A5}
\fl\sum_{k=1}^{n}k^{2}\psi_{0}(k+a)&=&\frac{1}{6}\left(2a^3-3a^2+a+2n^3+3n^2+n\right)\psi_{0}(a+n+1)-\frac{1}{6}a\big(2a^2-\nonumber\\
&&3a+1\big)\psi_{0}(a+1)-\frac{1}{36}n\left(12a^2-6an-24a+4n^2+15n+17\right)
\end{eqnarray}
\begin{eqnarray}\label{eq:A6}
\fl\sum_{k=1}^{n}k^3\psi_{0}(k+a)&=&\!-\frac{1}{4} \left(a^4-2a^3+a^2-n^4-2n^3-n^2\right)\psi_{0}(a+n+1)+\frac{1}{4}(a-1)^{2}\times\nonumber\\
&&a^{2}\psi_{0}(a+1)-\frac{1}{48}n\big(\!-12a^3+6a^{2}n+30a^2-4an^2-18an-26a+\nonumber\\
&&3n^3+14n^2+21n+10\big)
\end{eqnarray}
\begin{eqnarray}\label{eq:A7}
\fl\sum_{k=1}^{n}k^4\psi_{0}(k+a)&=&\frac{1}{1800} \Big(60\left(6a^5-15a^4+10a^3-a+6n^5+15n^4+10n^3-n\right)\times\nonumber\\
&&\psi_{0}(a+n+1)-60\left(6a^5-15a^4+10a^3-a\right)\psi_{0}(a+1)-120 a^2 n^3+\nonumber\\
&&180 a^3 n^2-630 a^2 n^2-360 a^4 n+1080 a^3 n-1110 a^2 n+90 a n^4+480\times\nonumber\\
&&a n^3+840 a n^2+450 a n-72 n^5-405 n^4-770 n^3-525 n^2-28 n\Big)
\end{eqnarray}
\begin{eqnarray}\label{eq:A8}
\fl\sum_{k=1}^{n}\psi_{0}^{2}(k+a)&=&(a+n)\psi_{0}^{2}(a+n+1)-(2a+2n+1)\psi_{0}(a+n+1)-a\psi_{0}^{2}(a+1)+\nonumber\\
&&(2a+1)\psi_{0}(a+1)+2n
\end{eqnarray}
\begin{eqnarray}\label{eq:A9}
\fl\sum_{k=1}^{n}k\psi_{0}^{2}(k+a)&=&\frac{1}{2}\left(-a^2+a+n^2+n\right)\psi_{0}^{2}(a+n+1)+\frac{1}{4}\big(6a^2+4an-2a-2n^2-\nonumber\\
&&6n-2\big)\times\psi_{0}(a+n+1)+\frac{1}{2}(a-1)a\psi_{0}^{2}(a+1)+\frac{1}{4}\big(-6a^2+2a+\nonumber\\
&&2\big)\psi_{0}(a+1)+\frac{1}{4}n(-6a+n+5)
\end{eqnarray}
\begin{eqnarray}\label{eq:A10}
\fl\sum_{k=1}^{n}k^2\psi_{0}^{2}(k+a)&=&\frac{1}{6}\left(2a^3-3a^2+a+2n^3+3n^2+n\right)\psi_{0}^{2}(a+n+1)-\frac{1}{18}\left(22a^3+\right.\nonumber\\
&&\!\left.12a^{2}n-21a^2-6an^2-24an-a+4n^3+15n^2+17n+3\right)\times\nonumber\\
&&\psi_{0}(a+n+1)-\frac{1}{6}a\left(2a^2-3a+1\right)\psi_{0}^{2}(a+1)+\frac{1}{18}\big(22a^3-21a^2-\nonumber\\
&&a+3\big)\psi_{0}(a+1)+\frac{1}{108}n\left(132a^2-30an-192a+8n^2+39n+79\right)\nonumber\\
&&
\end{eqnarray}
\begin{eqnarray}\label{eq:A11}
\fl\sum_{k=1}^{n}k^3\psi_{0}^{2}(k+a)&=&-\frac{1}{4}\left(a^4-2a^3+a^2-n^4-2n^3-n^2\right)\psi_{0}^{2}(a+n+1)+\frac{1}{24}\left(25a^4+\right.\nonumber\\
&&12a^{3}n-38a^3-6a^{2}n^2-30a^{2}n+11a^2+4an^3+18an^2+26an+\nonumber\\
&&\!\left.2a-3n^4-14n^3-21n^2-10n\right)\psi_{0}(a+n+1)+\frac{1}{4}(a-1)^{2}a^2\times\nonumber\\
&&\psi_{0}^{2}(a+1)-\frac{1}{24}a\left(25a^3-38a^2+11a+2\right)\psi_{0}(a+1)+\frac{1}{288}n\times\nonumber\\
&&\!\left(-300a^3+78a^{2}n+606a^2-28an^2-162an-410a+9n^3+\right.\nonumber\\
&&\!\left.50n^2+111n+118\right)
\end{eqnarray}
\begin{eqnarray}\label{eq:A12}
\fl\sum_{k=1}^{n}k^4\psi_{0}^{2}(k+a)&=&\frac{1}{54000}\Big(1800\left(6a^5-15a^4+10a^3-a+6n^5+15n^4+10n^3-n\right)\times\nonumber\\
&&\psi_{0}^{2}(a+n+1)-60\left(822a^5+15a^4\left(24n-113\right)-20a^3\big(9n^2+54n-\right.\nonumber\\
&&46\big)+15a^2\left(8n^3+42n^2+74n-5\right)-2a\big(45n^4+240n^3+420n^2+\nonumber\\
&&\!\left.225n-14\big)+72n^5+405n^4+770n^3+525n^2+28n-30\right)\times\nonumber\\
&&\psi_{0}(a+n+1)-1800\left(6a^5-15a^4+10a^3-a\right)\psi_{0}^{2}(a+1)+60\times\nonumber\\
&&\!\left(822a^5-1695a^4+920a^3-75a^2+28a-30\right)\psi_{0}(a+1)+5640\times\nonumber\\
&& a^2 n^3-13860 a^3 n^2+37710 a^2 n^2+49320 a^4 n-126360 a^3 n+114270\times\nonumber\\
&& a^2 n-2430 a n^4-15360 a n^3-39480 a n^2-49050 a n+864 n^5+\nonumber\\
&&5535 n^4+14590 n^3+19725 n^2+11486 n\Big)
\end{eqnarray}
\begin{eqnarray}\label{eq:A13}
\fl\sum_{k=1}^{n}\psi_{0}^{3}(k+a)&=&\!-\frac{1}{2}\psi_{1}(a+n+1)+\frac{1}{2}\psi_{1}(a+1)+(a+n)\psi_{0}^{3}(a+n+1)-\frac{3}{2}(2a+\nonumber\\
&&2n+1)\psi_{0}^{2}(a+n+1)+3(2a+2n+1)\psi_{0}(a+n+1)-a\times\nonumber\\
&&\psi_{0}^{3}(a+1)+\frac{3}{2}(2a+1)\psi_{0}^{2}(a+1)-3(2a+1)\psi_{0}(a+1)-6n
\end{eqnarray}
\begin{eqnarray}\label{eq:A14}
\fl\sum_{k=1}^{n}k\psi_{0}^{3}(k+a)&=&\frac{1}{4}\left(2a-1\right)\psi_{1}(a+n+1)+\frac{1}{4}\left(-2a+1\right)\psi_{1}(a+1)+\frac{1}{2}\big(\!-a^2+\nonumber\\
&&a+n^2+n\big)\psi_{0}^{3}(a+n+1)+\frac{3}{4}\left(3a^2+2an-a-n^2-3n-1\right)\times\nonumber\\
&&\psi_{0}^{2}(a+n+1)+\frac{1}{8}\left(-42a^2-36an+6a+6n^2+30n+14\right)\times\nonumber\\
&&\psi_{0}(a+n+1)+\frac{1}{2}(a-1)a\psi_{0}^{3}(a+1)+\frac{3}{4}\left(-3a^2+a+1\right)\times\nonumber\\
&&\psi_{0}^{2}(a+1)+\frac{1}{4}\left(21a^2-3a-7\right)\psi_{0}(a+1)+\frac{1}{8}\left(42an-3n^2-27n\right)\nonumber\\
&&
\end{eqnarray}
\begin{eqnarray}\label{eq:A15}
\fl\sum_{k=1}^{n}k^{2}\psi_{0}^{3}(k+a)&=&\frac{1}{12}\left(-6a^2+6a-1\right)\psi_{1}(a+n+1)+\frac{1}{12}\left(6a^2-6a+1\right)\psi_{1}(a+1)+\nonumber\\
&&\frac{1}{6}\left(2a^3-3a^2+a+2n^3+3n^2+n\right)\psi_{0}^{3}(a+n+1)-\frac{1}{12}\big(22a^3+\nonumber\\
&&12a^{2}n-21a^2-6an^2-24an-a+4n^3+15n^2+17n+3\big)\times\nonumber\\
&&\psi_{0}^{2}(a+n+1)+\frac{1}{36}\left(170a^3+132a^{2}n-123a^2-30an^2-192an-\right.\nonumber\\
&&\!\left.47a+8n^3+39n^2+79n+33\right)\psi_{0}(a+n+1)-\frac{1}{6}a\left(2a^2-3a+1\right)\times\nonumber\\
&&\psi_{0}^{3}(a+1)+\frac{1}{12}\left(22a^3-21a^2-a+3\right)\psi_{0}^{2}(a+1)-\frac{1}{36}\left(170a^3-\right.\nonumber\\
&&\!\left.123a^2-47a+33\right)\psi_{0}(a+1)+\frac{1}{216}\left(-1020a^{2}n+114an^2+\right.\nonumber\\
&&\!\left.1248an-16n^3-105n^2-365n\right)
\end{eqnarray}
\begin{eqnarray}\label{eq:A16}
\fl\sum_{k=1}^{n}k^{3}\psi_{0}^{3}(k+a)&=&\frac{1}{4}a\left(2a^2-3a+1\right)\psi_{1}(a+n+1)+\frac{1}{4}a\left(-2a^2+3a-1\right)\psi_{1}(a+1)-\nonumber\\
&&\frac{1}{4}\left(a^4-2a^3+a^2-n^4-2n^3-n^2\right)\psi_{0}^{3}(a+n+1)+\frac{1}{16}\left(25a^4+\right.\nonumber\\
&&12a^{3}n-38a^3-6a^{2}n^2-30a^{2}n+11a^{2}+4an^{3}+18an^2+26an+\nonumber\\
&&\!\left.2a-3n^4-14n^3-21n^2-10n\right)\psi_{0}^{2}(a+n+1)-\frac{1}{96}\left(415a^4+\right.\nonumber\\
&&300a^{3}n-530a^3-78a^{2}n^{2}-606a^{2}n+17a^2+28an^3+162an^2+\nonumber\\
&&\!\left.410an+146a-9n^4-50n^3-111n^2-118n-36\right)\psi_{0}(a+n+1)+\nonumber\\
&&\frac{1}{4}(a-1)^{2}a^{2}\psi_{0}^{3}(a+1)-\frac{1}{16}a\left(25a^3-38a^2+11a+2\right)\psi_{0}^{2}(a+1)+\nonumber\\
&&\frac{1}{96}\left(415a^4-530a^3+17a^2+146a-36\right)\psi_{0}(a+1)+\frac{1}{1152}\times\nonumber\\
&&\!\left(4980a^{3}n-690a^{2}n^{2}-8850a^{2}n+148an^3+1134an^2+4790an-\right.\nonumber\\
&&\!\left.27n^4-182n^3-525n^2-850n\right)
\end{eqnarray}
\begin{eqnarray}\label{eq:A17}
\fl\sum_{k=1}^{n}k^{4}\psi_{0}^{3}(k+a)&=&\frac{1}{1080000}\Big(\!-18000\left(30a^4-60a^3+30a^2-1\right)\psi_{1}(a+n+1)+\nonumber\\
&&18000\left(30a^4-60a^3+30a^2-1\right)\psi_{1}(a+1)+36000\left(6a^5-15a^4+\right.\nonumber\\
&&10a^3-a+\!\left.6n^5+15n^4+10n^3-n\right)\psi_{0}^{3}(a+n+1)-1800\left(822a^5+\right.\nonumber\\
&&15a^4\left(24n-113\right)-20a^3\left(9n^2+54n-46\right)+15a^2\big(8n^3+42n^2+\nonumber\\
&&74n-5\big)-2a\left(45n^4+240n^3+420n^2+225n-14\right)+72n^5+\nonumber\\
&&405n^4+\left.770n^3+525n^2+28n-30\right)\psi_{0}^{2}(a+n+1)+60\left(72114a^5+\right.\nonumber\\
&&15a^4(3288n-8731)-20a^3\left(693n^2+6318n-2117\right)+15a^2\times\nonumber\\
&&(376n^3+2514n^2+7618n+2015)-2a\big(1215n^4+7680n^3+\nonumber\\
&&19740n^2+24525n+7757\big)+864n^5+5535n^4+14590n^3+\nonumber\\
&&\!\left.19725n^2+11486n+840\right)\psi_{0}(a+n+1)-36000\left(6a^5-15a^4+\right.\nonumber\\
&&\!\left.10a^3-a\right)\psi_{0}^{3}(a+1)+1800\left(822a^5-1695a^4+920a^3+28a-\right.\nonumber\\
&&\!\left.75a^2-30\right)\psi_{0}^{2}(a+1)-60\left(72114a^5-130965a^4+42340a^3+\right.\nonumber\\
&&\!\left.30225a^2-15514a+840\right)\psi_0(a+1)-178680 a^2 n^3+683820 a^3 n^2-\nonumber\\
&&1561770 a^2 n^2-4326840 a^4 n+10021320 a^3 n-7730490 a^2 n+\nonumber\\
&&49410 a n^4+376320 a n^3+1250760 a n^2+2386350 a n-10368 n^5-\nonumber\\
&&76545 n^4-239330 n^3-400575 n^2-331582 n\Big)
\end{eqnarray}
\begin{equation}\label{eq:A18}
\fl\sum_{k=1}^{n}\psi_{1}(k+a)=(a+n)\psi_{1}(a+n+1)-a\psi_{1}(a+1)+\psi_{0}(a+n+1)-\psi_{0}(a+1)
\end{equation}
\begin{eqnarray}\label{eq:A19}
\fl\sum_{k=1}^{n}k\psi_{1}(k+a)&=&\frac{1}{2}\Big(\!\left(-a^2+a+n^2+n\right)\psi_{1}(a+n+1)+(a-1)a\psi_{1}(a+1)+\nonumber\\
&&(-2a+1)\psi_{0}(a+n+1)+(2a-1)\psi_{0}(a+1)+n\Big)
\end{eqnarray}
\begin{eqnarray}\label{eq:A20}
\fl\sum_{k=1}^{n}k^{2}\psi_{1}(k+a)&=&\frac{1}{6}\Big(\!\left(2a^3-3a^2+a+2n^3+3n^2+n\right)\psi_{1}(a+n+1)+a(a-1)\times\nonumber\\
&&(-2a+1)\psi_{1}(a+1)+\left(6a^2-6a+1\right)\psi_{0}(a+n+1)+\big(\!-6a^2+\nonumber\\
&&6a-1\big)\psi_{0}(a+1)-4an+n^2+4n\Big)
\end{eqnarray}
\begin{eqnarray}\label{eq:A21}
\fl\sum_{k=1}^{n}k^{3}\psi_{1}(k+a)&=&\frac{1}{24}\Big(6\left(-a^4+2a^3-a^2+n^4+2n^3+n^2\right)\psi_{1}(a+n+1)+6\left(a^4-\right.\nonumber\\
&&2a^3+\!\left.a^2\right)\psi_{1}(a+1)-12a\left(2a^2-3a+1\right)\psi_{0}(a+n+1)+12a\times\nonumber\\
&&\!\left(2a^2-3a+1\right)\psi_{0}(a+1)+18a^{2}n-6an^2-30an+2n^3+9n^2+\nonumber\\
&&13n\Big)
\end{eqnarray}
\begin{eqnarray}\label{eq:A22}
\fl\sum_{k=1}^{n}k^{4}\psi_{1}(k+a)&=&\frac{1}{60}\Big(\!\left(12a^5-30a^4+20a^3-2a+12n^5+30n^4+20n^3-2n\right)\times\nonumber\\
&&\psi_{1}(a+n+1)+\left(-12a^5+30a^4-20a^3+2a\right)\psi_{1}(a+1)+\nonumber\\
&&\!\left(60a^4-120a^3+60a^2-2\right)\psi_{0}(a+n+1)+\left(-60a^4+120a^3-\right.\nonumber\\
&&\!\left.60a^2+2\right)\psi_{0}(a+1)+18 a^2 n^2-48 a^3 n+108 a^2 n-8 a n^3-42 a n^2-\nonumber\\
&&74 a n+3 n^4+16 n^3+28 n^2+15 n\Big)
\end{eqnarray}
\begin{eqnarray}\label{eq:A23}
\fl\sum_{k=1}^{n}\psi_{0}(k+a)\psi_{1}(k+a)&=&(a+n)\psi_{0}(a+n+1)\psi_{1}(a+n+1)-a\psi_{0}(a+1)\psi_{1}(a+1)-\nonumber\\
&&\frac{1}{2}(2a+2n+1)\psi_{1}(a+n+1)+\frac{1}{2}(2a+1)\psi_{1}(a+1)+\nonumber\\
&&\frac{1}{2}\psi_{0}^{2}(a+n+1)-\psi_{0}(a+n+1)-\frac{1}{2}\psi_{0}^{2}(a+1)+\nonumber\\
&&\psi_{0}(a+1)
\end{eqnarray}
\begin{eqnarray}\label{eq:A24}
\fl\sum_{k=1}^{n}k\psi_{0}(k+a)\psi_{1}(k+a)&=&\frac{1}{4}\Big(2\left(-a^2+a+n^2+n\right)\psi_{0}(a+n+1)\psi_{1}(a+n+1)+\nonumber\\
&&2a(a-1)\psi_{0}(a+1)\psi_{1}(a+1)-\big(\!-3a^2-2an+a+n^2+\nonumber\\
&&3n+1\big)\psi_{1}(a+n+1)+\left(-3a^2+a+1\right)\psi_{1}(a+1)+\nonumber\\
&&(1-2a)\psi_{0}^{2}(a+n+1)+(6a+2n-1)\psi_{0}(a+n+1)+\nonumber\\
&&(2a-1)\psi_{0}^{2}(a+1)+(1-6a)\psi_{0}(a+1)-3n\Big)
\end{eqnarray}
\begin{eqnarray}\label{eq:A25}
\fl\sum_{k=1}^{n}k^{2}\psi_{0}(k+a)\psi_{1}(k+a)&=&\frac{1}{36}\Big(6\left(2a^3-3a^2+a+2n^3+3n^2+n\right)\psi_{0}(a+n+1)\times\nonumber\\
&&\psi_1(a+n+1)-6a\left(2a^2-3a+1\right)\psi_{0}(a+1)\psi_{1}(a+1)+\nonumber\\
&&\!\left(-22a^3-12a^{2}n+21a^2+6an^2+24an+a-4n^3-\right.\nonumber\\
&&\!\left.15n^2-17n-3\right)\psi_{1}(a+n+1)+\left(22a^3-21a^2-a+3\right)\times\nonumber\\
&&\psi_{1}(a+1)+3\left(6a^2-6a+1\right)\psi_{0}^{2}(a+n+1)+\left(-66a^2-\right.\nonumber\\
&&24an+42a+\left.6n^2+24n+1\right)\psi_{0}(a+n+1)-3\big(6a^2-\nonumber\\
&&6a+1\big)\psi_{0}^{2}(a+1)-\left(-66a^2+42a+1\right)\psi_{0}(a+1)+\nonumber\\
&&44an-5n^2-32n\Big)
\end{eqnarray}
\begin{eqnarray}\label{eq:A26}
\fl\sum_{k=1}^{n}k^{3}\psi_{0}(k+a)\psi_{1}(k+a)&=&\frac{1}{288}\Big(\!-72\left(a^4-2a^3+a^2-n^4-2n^3-n^2\right)\psi_{0}(a+n+1)\times\nonumber\\
&&\psi_{1}(a+n+1)+72(a-1)^{2}a^2\psi_{0}(a+1)\psi_{1}(a+1)+\nonumber\\
&&\big(150a^4+72a^{3}n-228a^3-36a^{2}n^{2}-180a^{2}n+66a^2+\nonumber\\
&&24an^3+108an^2+156an+12a-18n^4-84n^3-126n^2-\nonumber\\
&&60n\big)\psi_{1}(a+n+1)+\left(-150a^4+228a^3-66a^2-12a\right)\times\nonumber\\
&&\psi_{1}(a+1)-72a\left(2a^2-3a+1\right)\psi_{0}^{2}(a+n+1)-12\big(\!-50\times\nonumber\\
&&a^3-18a^{2}n+57a^2+6an^2+30an-11a-2n^3-9n^2-\nonumber\\
&&13n-1\big)\psi_{0}(a+n+1)+72a\left(2a^2-3a+1\right)\psi_{0}^{2}(a+1)+\nonumber\\
&&12\left(-50a^3+57a^2-11a-1\right)\psi_{0}(a+1)-450a^{2}n+\nonumber\\
&&78an^2+606an-14n^3-81n^2-205n\Big)
\end{eqnarray}
\begin{eqnarray}\label{eq:A27}
\fl\sum_{k=1}^{n}k^{4}\psi_{0}(k+a)\psi_{1}(k+a)&=&\frac{1}{3600}\Big(120\left(6a^5-15a^4+10a^3-a+6n^5+15n^4+\right.\nonumber\\
&&\!\left.10n^3-n\right)\psi_{0}(a+n+1)\psi_{1}(a+n+1)-120\left(6a^5-\right.\nonumber\\
&&15a^4+\left.10a^3-a\right)\psi_{0}(a+1)\psi_{1}(a+1)-2\left(822a^5+15\times\right.\nonumber\\
&&(24n-113)a^4-20a^3\left(9n^2+54n-46\right)+15a^2\big(8n^3+\nonumber\\
&&42n^2+74n-5\big)-2a\big(45n^4+240n^3+420n^2+225n-\nonumber\\
&&14\big)+\left.72n^5+405n^4+770n^3+525n^2+28n-30\right)\times\nonumber\\
&&\psi_{1}(a+n+1)+2\big(822a^5-1695a^4+920a^3-75a^2+\nonumber\\
&&28a-30\big)\psi_{1}(a+1)+60\left(30a^4-60a^3+30a^2-1\right)\times\nonumber\\
&&\psi_{0}^{2}(a+n+1)-4\left(2055a^4+30a^3\left(24n-113\right)-30a^2\times\right.\nonumber\\
&&(9n^2+54n-46)+15a\left(8n^3+42n^2+74n-5\right)-45n^4-\nonumber\\
&&240n^3-\!\left.420n^2-225n+14\right)\psi_{0}(a+n+1)-60\left(30a^4-\right.\nonumber\\
&&60a^3+\!\left.30a^2-1\right)\psi_{0}^{2}(a+1)+4\left(2055a^4-3390a^3+\right.\nonumber\\
&&\!\left.1380a^2-75a+14\right)\psi_{0}(a+1)-1386 a^2 n^2+6576 a^3 n-\nonumber\\
&&12636 a^2 n+376 a n^3+2514 a n^2+7618 a n-81 n^4-\nonumber\\
&&512 n^3-1316 n^2-1635 n\Big)
\end{eqnarray}
\begin{eqnarray}\label{eq:A28}
\fl\sum_{k=1}^{n}\psi_{0}^{2}(k+a)\psi_{1}(k+a)&=&\frac{1}{6}\Big(\!-\psi_{2}(a+n+1)+\psi_{2}(a+1)+6\left(a+n\right)\psi_{0}^{2}(a+n+1)\times\nonumber\\
&&\psi_{1}(a+n+1)-6a\psi_{0}^{2}(a+1)\psi_{1}(a+1)-6\left(2a+2n+1\right)\times\nonumber\\
&&\psi_{0}(a+n+1)\psi_{1}(a+n+1)+6\left(2a+1\right)\psi_{0}(a+1)\times\nonumber\\
&&\psi_{1}(a+1)+6\left(2a+2n+1\right)\psi_{1}(a+n+1)-6\left(2a+1\right)\times\nonumber\\
&&\psi_{1}(a+1)+2\psi_{0}^{3}(a+n+1)-6\psi_{0}^{2}(a+n+1)+12\times\nonumber\\
&&\psi_{0}(a+n+1)-2\psi_{0}^{3}(a+1)+6\psi_{0}^{2}(a+1)-12\psi_{0}(a+1)\Big)\nonumber\\
&&
\end{eqnarray}
\begin{eqnarray}\label{eq:A29}
\fl\sum_{k=1}^{n}k\psi_{0}^{2}(k+a)\psi_{1}(k+a)&=&\frac{1}{12}\Big(\!\left(2a-1\right)\psi_{2}(a+n+1)-\left(2a-1\right)\psi_{2}(a+1)+6\times\nonumber\\
&&\!\left(-a^2+a+n^2+n\right)\psi_{0}^{2}(a+n+1)\psi_{1}(a+n+1)+6a(a-\nonumber\\
&&1)\psi_{0}^{2}(a+1)\psi_{1}(a+1)+6\big(3a^2+a\left(2n-1\right)-n^2-3n-\nonumber\\
&&1\big)\psi_{0}(a+n+1)\psi_{1}(a+n+1)+6\left(-3a^2+a+1\right)\times\nonumber\\
&&\psi_{0}(a+1)\psi_{1}(a+1)+3\left(-7a^2-6an+a+n^2+5n+3\right)\times\nonumber\\
&&\psi_{1}(a+n+1)+3\left(7a^2-a-3\right)\psi_{1}(a+1)-2\left(2a-1\right)\times\nonumber\\
&&\psi_{0}^{3}(a+n+1)+3\left(6a+2n-1\right)\psi_{0}^{2}(a+n+1)-3\big(14a+\nonumber\\
&&6n-1\big)\psi_{0}(a+n+1)+2\left(2a-1\right)\psi_{0}^{3}(a+1)-3\left(6a-1\right)\times\nonumber\\
&&\psi_{0}^{2}(a+1)+3\left(14a-1\right)\psi_{0}(a+1)+21n\Big)
\end{eqnarray}
\begin{eqnarray}\label{eq:A30}
\fl\sum_{k=1}^{n}k^2\psi_{0}^{2}(k+a)\psi_{1}(k+a)&=&\frac{1}{108}\Big(\!-3\left(6a^2-6a+1\right)\psi_{2}(a+n+1)+3\big(6a^2-6a\nonumber\\
&&+1\big)\psi_{2}(a+1)+18\left(2a^3-3a^2+a+2n^3+3n^2+n\right)\times\nonumber\\
&&\psi_{0}^{2}(a+n+1)\psi_{1}(a+n+1)-18\left(2a^3-3a^2+a\right)\times\nonumber\\
&&\psi_{0}^{2}(a+1)\psi_{1}(a+1)-6\left(12 a^2 n+22 a^3-21 a^2-6 a n^2-\right.\nonumber\\
&&24 a n-a+4 n^3+\!\left.15 n^2+17 n+3\right)\psi_{0}(a+n+1)\times\nonumber\\
&&\psi_{1}(a+n+1)+6\left(22a^3-21a^2-a+3\right)\psi_{0}(a+1)\times\nonumber\\
&&\psi_{1}(a+1)+\left(132 a^2 n+170 a^3-123 a^2-30 a n^2-\right.\nonumber\\
&&192 a n-83 a+8 n^3+\!\left.39 n^2+79 n+51\right)\psi_{1}(a+n+1)+\nonumber\\
&&\!\left(-170a^3+123a^2+83a-51\right)\psi_{1}(a+1)+6\big(6a^2-6a+\nonumber\\
&&1\big)\psi_{0}^{3}(a+n+1)+3\big(\!-66 a^2-24 a n+42 a+6 n^2+24 n+\nonumber\\
&&1\big)\psi_{0}^{2}(a+n+1)+\big(510 a^2+264 a n-246 a-30 n^2-\nonumber\\
&&192 n-47\big)\psi_{0}(a+n+1)-6\left(6a^2-6a+1\right)\psi_{0}^{3}(a+1)+\nonumber\\
&&3\!\left(66a^2-42a-1\right)\psi_{0}^{2}(a+1)+\left(-510a^2+246a+47\right)\times\nonumber\\
&&\psi_{0}(a+1)-340 a n+19 n^2+208 n\Big)
\end{eqnarray}
\begin{eqnarray}\label{eq:A31}
\fl\sum_{k=1}^{n}k^3\psi_{0}^{2}(k+a)\psi_{1}(k+a)&=&\frac{1}{1728}\Big(144\left(2a^3-3a^2+a\right)\psi_{2}(a+n+1)-144\left(2a^3-\right.\nonumber\\
&&\!\left.3a^2+a\right)\psi_{2}(a+1)-432\left(a^4-2a^3+a^2-n^4-2n^3-\right.\nonumber\\
&&\!\left.n^2\right)\psi_{0}^{2}(a+n+1)\psi_{1}(a+n+1)+432\left(a^4-2a^3+a^2\right)\times\nonumber\\
&&\psi_{0}^{2}(a+1)\psi_{1}(a+1)+72\left(-6 a^2 n^2+12 a^3 n-30 a^2 n+\right.\nonumber\\
&& 25a^4-38 a^3+11 a^2+4 a n^3+18 a n^2+26 a n+2 a-\nonumber\\
&&3 n^4-14 n^3-\!\left.21 n^2-10 n\right)\psi_{0}(a+n+1)\psi_{1}(a+n+1)-\nonumber\\
&&72\left(25a^4-38a^3+11a^2+2a\right)\psi_{0}(a+1)\psi_{1}(a+1)-6\times\nonumber\\
&&\big(\!-78a^2 n^2+300 a^3 n-606 a^2 n+415 a^4-530 a^3-127 a^2+\nonumber\\
&&28a n^3+162 a n^2+410 a n+290 a-9 n^4-50 n^3-111 n^2-\nonumber\\
&&118 n-60\big)\psi_{1}(a+n+1)+6\big(415a^4-530a^3-127a^2+\nonumber\\
&&290a-60\big)\psi_{1}(a+1)-288\left(2a^3-3a^2+a\right)\times\nonumber\\
&&\psi_{0}^{3}(a+n+1)+72\left(18 a^2 n+50 a^3-57 a^2-6 a n^2-30\right.\times\nonumber\\
&& a n+11 a+\!\left.2 n^3+9 n^2+13 n+1\right)\psi_{0}^{2}(a+n+1)-12\times\nonumber\\
&&\!\left(450 a^2 n+830 a^3-795 a^2-78 a n^2-606 a n+17 a+14 \right.\times\nonumber\\
&&n^3+\!\left.81 n^2+205 n+73\right)\psi_{0}(a+n+1)+288\big(2a^3-3a^2+\nonumber\\
&&a\big)\psi_{0}^{3}(a+1)-72\left(50a^3-57a^2+11a+1\right)\psi_{0}^{2}(a+1)+\nonumber\\
&&12\left(830a^3-795a^2+17a+73\right)\psi_{0}(a+1)+7470 a^2 n-\nonumber\\
&&690 a n^2-8850 a n+74 n^3+567 n^2+2395 n\Big)
\end{eqnarray}
\begin{eqnarray}\label{eq:A32}
\fl\sum_{k=1}^{n}k^4\psi_{0}^{2}(k+a)\psi_{1}(k+a)&=&\frac{1}{108000}\Big(\!-600\left(30a^4-60a^3+30a^2-1\right)\psi_{2}(a+n+1)+\nonumber\\
&&600\left(30a^4-60a^3+30a^2-1\right)\psi_{2}(a+1)+3600\big(6a^5-\nonumber\\
&&15a^4+10a^3-a+6n^5+15n^4+10n^3-n\big)\psi_{0}^{2}(a+n+1)\times\nonumber\\
&&\psi_{1}(a+n+1)-3600\left(6a^5-15a^4+10a^3-a\right)\psi_{0}^{2}(a+1)\times\nonumber\\
&&\psi_{1}(a+1)-120\left(822a^5+15a^4\left(24n-113\right)-20a^3\big(9n^2+\right.\nonumber\\
&&54n-46\big)+15a^2\left(8n^3+42n^2+74n-5\right)-2a\big(45n^4+\nonumber\\
&&240n^3+420n^2+225n-14\big)+72n^5+405n^4+770n^3+\nonumber\\
&&\!\left.525n^2+28n-30\right)\psi_{0}(a+n+1)\psi_{1}(a+n+1)+120\times\nonumber\\
&&\!\left(822a^5-1695a^4+920a^3-75a^2+28a-30\right)\psi_{0}(a+1)\times\nonumber\\
&&\psi_{1}(a+1)+2\big(72114a^5+15a^4\left(3288n-8731\right)-20a^3\times\nonumber\\
&&(693n^2+6318n-317)+15a^2\big(376n^3+2514n^2+7618n+\nonumber\\
&&5615\big)-2a\big(1215n^4+7680n^3+19740n^2+24525n+\nonumber\\
&&16757\big)+864n^5+5535n^4+14590n^3+19725n^2+\nonumber\\
&&11486n+840\big)\psi_{1}(a+n+1)-2\left(72114a^5-130965\times\right.\nonumber\\
&&a^4+\!\left.6340a^3+84225a^2-33514a+840\right)\psi_{1}(a+1)+\nonumber\\
&&1200\left(30a^4-60a^3+30a^2-1\right)\psi_{0}^{3}(a+n+1)-120\times\nonumber\\
&&\big(2055a^4+30a^3\left(24n-113\right)-30a^2\left(9n^2+54n-46\right)+\nonumber\\
&&15a\left(8n^3+42n^2+74n-5\right)-45n^4-240n^3-420n^2-\nonumber\\
&&225n+14\big)\psi_{0}^{2}(a+n+1)+4\big(180285a^4+30a^3\big(3288n-\nonumber\\
&&8731\big)-30a^2\left(693n^2+6318n-2117\right)+15a\big(376n^3+\nonumber\\
&&2514n^2+7618n+2015\big)-1215n^4-7680n^3-19740\times\nonumber\\
&&n^2-24525n-7757\big)\psi_{0}(a+n+1)-1200\big(30a^4-60a^3+\nonumber\\
&&30a^2-1\big)\psi_{0}^{3}(a+1)+120\big(2055a^4-3390a^3+1380a^2-\nonumber\\
&&75a+14\big)\psi_{0}^{2}(a+1)-4\left(180285a^4-261930a^3+63510\times\right.\nonumber\\
&&\!\left.a^2+30225a-7757\right)\psi_{0}(a+1)+68382 a^2 n^2-576912\times\nonumber\\
&& a^3 n+1002132 a^2 n-11912 a n^3-104118 a n^2-515366 a n+\nonumber\\
&&1647 n^4+12544 n^3+41692 n^2+79545 n\Big)
\end{eqnarray}
\begin{eqnarray}\label{eq:A33}
\fl\sum_{k=1}^{n}\psi_{2}(k+a)&=&(a+n)\psi_{2}(a+n+1)-a\psi_{2}(a+1)+2\psi_{1}(a+n+1)-\nonumber\\
&&2\psi_{1}(a+1)
\end{eqnarray}
\begin{eqnarray}\label{eq:A34}
\fl\sum_{k=1}^{n}k\psi_{2}(k+a)&=&\frac{1}{2}\left(-a^2+a+n^2+n\right)\psi_{2}(a+n+1)+\frac{1}{2}a(a-1)\psi_{2}(a+1)+\nonumber\\
&&(-2a+1)\psi_{1}(a+n+1)+(2a-1)\psi_{1}(a+1)-\psi_{0}(a+n+1)+\nonumber\\
&&\psi_{0}(a+1)
\end{eqnarray}
\begin{eqnarray}\label{eq:A35}
\fl\sum_{k=1}^{n}k^{2}\psi_{2}(k+a)&=&\frac{1}{6}\Big(\!\left(2a^3-3a^2+a+2n^3+3n^2+n\right)\psi_{2}(a+n+1)+a\big(\!-2a^2+\nonumber\\
&&3a-1\big)\psi_{2}(a+1)+2\left(6a^2-6a+1\right)\psi_{1}(a+n+1)+2\big(\!-6a^2+\nonumber\\
&&6a-1\big)\psi_{1}(a+1)+6(2a-1)\psi_{0}(a+n+1)+6(-2a+1)\times\nonumber\\
&&\psi_{0}(a+1)-4n\Big)
\end{eqnarray}
\begin{eqnarray}\label{eq:A36}
\fl\sum_{k=1}^{n}k^{3}\psi_{2}(k+a)&=&\frac{1}{4}\Big(\!\left(-a^4+2a^3-a^2+n^4+2n^3+n^2\right)\psi_{2}(a+n+1)+(a-1)^{2}a^{2}\times\nonumber\\
&&\psi_{2}(a+1)+4a\left(-2a^2+3a-1\right)\psi_{1}(a+n+1)+4a\left(2a^2-3a+1\right)\times\nonumber\\
&&\psi_{1}(a+1)-2\left(6a^2-6a+1\right)\psi_{0}(a+n+1)+2\left(6a^2-6a+1\right)\times\nonumber\\
&&\psi_{0}(a+1)+6an-n^2-5n\Big)
\end{eqnarray}
\begin{eqnarray}\label{eq:A37}
\fl\sum_{k=1}^{n}k^{4}\psi_{2}(k+a)&=&\frac{1}{30}\Big(\!\left(6a^5-15a^4+10a^3-a+6n^5+15n^4+10n^3-n\right)\times\nonumber\\
&&\psi_{2}(a+n+1)+\left(-6a^5+15a^4-10a^3+a\right)\psi_{2}(a+1)+\left(60a^4-\right.\nonumber\\
&&120a^3+\!\left.60a^2-2\right)\psi_{1}(a+n+1)+\left(-60a^4+120a^3-60a^2+2\right)\times\nonumber\\
&&\psi_{1}(a+1)+\left(120a^3-180a^2+60a\right)\psi_{0}(a+n+1)+\left(-120a^3+\right.\nonumber\\
&&\!\left.180a^2-60a\right)\psi_{0}(a+1)-n\big(72a^2-18a\left(n+6\right)+4n^2+21n+\nonumber\\
&&37\big)\Big)
\end{eqnarray}
\subsection{Semi closed-form expressions}\label{sec:ap1-sc}
\begin{eqnarray}\label{eq:A38}
\fl\sum_{k=1}^{n}\psi_{0}(k+a)\psi_{0}(k)&=&\!-a \sum _{k=1}^n \frac{\psi (a+k)}{k}+(a+n) \psi _0(n+1) \psi _0(a+n+1)-(a+n)\times\nonumber\\
&& \psi _0(a+n+1)-a \psi _0(1) \psi _0(a+1)+a \psi _0(a+1)-(n+1)\times \nonumber\\
&&\psi _0(n+1)+\psi _0(1)+2 n
\end{eqnarray}
\begin{eqnarray}\label{eq:A39}
\fl\sum_{k=1}^{n}k\psi_{0}(k+a)\psi_{0}(k)&=&\frac{1}{2}a (a-1) \sum _{k=1}^n \frac{\psi (a+k)}{k}+\frac{1}{4}\Big(\!\left(-2 a^2+2 a+2 n^2+2 n\right)\times\nonumber\\
&& \psi _0(n+1) \psi _0(a+n+1)+\left(a^2-3 a-n^2-3 n\right) \psi _0(a+n+1)+\nonumber\\
&&\!\left(2 a^2-2 a\right) \psi _0(1) \psi _0(a+1)+\left(2 a n+2 a-n^2-3 n-2\right)\times \nonumber\\
&&\psi _0(n+1)-\left(a^2-3a\right) \psi _0(a+1)-(2a-2) \psi _0(1)-3 a n+\nonumber\\
&&n^2+5 n\Big)
\end{eqnarray}
\begin{eqnarray}\label{eq:A40}
\fl\sum_{k=1}^{n}k^{2}\psi_{0}(k+a)\psi_{0}(k)&=&\!-\frac{1}{6} a \left(2 a^2-3 a+1\right) \sum _{k=1}^n \frac{\psi (a+k)}{k}+\frac{1}{108}\Big(\!\left(36 a^3-54 a^2+\right.\nonumber\\
&&18 a+36 n^3+\!\left.54 n^2+18 n\right) \psi _0(n+1) \psi _0(a+n+1)+\left(-12 a^3+\right.\nonumber\\
&&45 a^2-51 a-12 n^3-\!\left.45 n^2-51 n\right) \psi _0(a+n+1)+\left(-36 a^3+\right.\nonumber\\
&&\!\left.54 a^2-18 a\right) \psi _0(1) \psi _0(a+1)+\left(12 a^3-45 a^2+51 a\right) \psi _0(a+1)+\nonumber\\
&&\!\left(-36 a^2 n-36 a^2+18 a n^2+72 a n+54 a-12 n^3-51 n-\right.\nonumber\\
&&\!\left.45 n^2-18\right) \psi _0(n+1)+\left(36 a^2-54 a+18\right) \psi _0(1)+48 a^2 n-\nonumber\\
&&15 a n^2-96 a n+8 n^3+39 n^2+79 n\Big)
\end{eqnarray}
\begin{eqnarray}\label{eq:A41}
\fl\sum_{k=1}^{n}k^{3}\psi_{0}(k+a)\psi_{0}(k)&=&\frac{1}{4}a^2 (a-1)^2  \sum _{k=1}^n \frac{\psi (a+k)}{k}+\frac{1}{288}\Big(\!\left(-72 a^4+144 a^3-72 a^2+\right.\nonumber\\
&&72 n^4+144 n^3+\!\left.72 n^2\right) \psi _0(n+1) \psi _0(a+n+1)+\left(18 a^4-\right.\nonumber\\
&&84 a^3+126 a^2-60 a-18 n^4-84 n^3-\!\left.126 n^2-60 n\right)\times\nonumber\\
&& \psi _0(a+n+1)+72 (a-1)^2 a^2 \psi _0(1) \psi _0(a+1)+6 a \left(-3 a^3+\right.\nonumber\\
&&14\!\left. a^2-21 a+10\right) \psi _0(a+1)+\left(-36 a^2 n^2+72 a^3 n-180 a^2 n+\right.\nonumber\\
&&72 a^3-144 a^2+24 a n^3+108 a n^2+156 a n+72 a-18 n^4-\nonumber\\
&&84 n^3-\left.126 n^2-60 n\right) \psi _0(n+1)+\left(-72 a^3+144 a^2-72 a\right)\times\nonumber\\
&& \psi _0(1)+27 a^2 n^2-90 a^3 n+219 a^2 n-14 a n^3-81 a n^2-205\times\nonumber\\
&&a n+9 n^4+50 n^3+111 n^2+118 n\Big)
\end{eqnarray}
\begin{eqnarray}\label{eq:A42}
\fl\sum_{k=1}^{n}k^{4}\psi_{0}(k+a)\psi_{0}(k)&=&\!-\frac{1}{30} a \left(6 a^4-15 a^3+10 a^2-1\right) \sum _{k=1}^n \frac{\psi (a+k)}{k}+\frac{1}{1800}\times\nonumber\\
&&\Big(\!\left(360 a^5-900 a^4+600 a^3-60 a+360 n^5+900 n^4-60 n+\right.\nonumber\\
&&\!\left.600 n^3\right) \psi _0(n+1) \psi _0(a+n+1)+\left(-72 a^5+405 a^4-770 a^3+\right.\nonumber\\
&&525 a^2-28 a-72 n^5-405 n^4-770 n^3-\!\left.525 n^2-28 n\right)\times\nonumber\\
&& \psi _0(a+n+1)+\left(-360 a^5+900 a^4-600 a^3+60 a\right) \psi _0(1)\times\nonumber\\
&& \psi _0(a+1)+\left(72 a^5-405 a^4+770 a^3-525 a^2+28 a\right) \psi _0(a+1)+\nonumber\\
&&\!\left(-120 a^2 n^3+180 a^3 n^2-630 a^2 n^2-360 a^4 n+1080 a^3 n-1110 \times\right.\nonumber\\
&&a^2 n-360 a^4+900 a^3-600 a^2+90 a n^4+480 a n^3+840 a n^2+\nonumber\\
&&450 a n-72 n^5-405 n^4-770 n^3-\!\left.525 n^2-28 n+60\right)\times\nonumber\\
&& \psi _0(n+1)+\left(360 a^4-900 a^3+600 a^2-60\right) \psi _0(1)\Big)+\frac{1}{54000}\times\nonumber\\
&&n \Big(1920 a^2 n^2-3780 a^3 n+13005 a^2 n+12960 a^4-37530 a^3+\nonumber\\
&&40485 a^2-1215 a n^3-7680 a n^2-19740 a n-24525 a+864 n^4+\nonumber\\
&&5535 n^3+14590 n^2+19725 n+11486\Big)
\end{eqnarray}
\begin{eqnarray}\label{eq:A43}
\fl\sum_{k=1}^{n}\psi_{0}(k+a)\psi_{0}^2(k)&=&\!-2 a \sum _{k=1}^n \frac{\psi_0 (k+a)\psi_0 (k) }{k}-a \sum _{k=1}^n \frac{\psi_0 (k+a)}{k^2}+(2 a-1)\times\nonumber\\
&& \sum _{k=1}^n \frac{\psi_0 (k+a)}{k}+(a+n) \psi _0^2(n+1) \psi _0(a+n+1)-(2 a+2 n)\times\nonumber\\
&& \psi _0(n+1) \psi _0(a+n+1)+(2 a+2 n) \psi _0(a+n+1)-a \psi _0^2(1)\times\nonumber\\
&& \psi _0(a+1)+2 a \psi _0(1) \psi _0(a+1)-2 a \psi _0(a+1)-(n+1) \times\nonumber\\
&&\psi _0^2(n+1)+(4 n+3) \psi _0(n+1)+\psi _0^2(1)-3 \psi _0(1)-6 n
\end{eqnarray}
\begin{eqnarray}\label{eq:A44}
\fl\sum_{k=1}^{n}k\psi_{0}(k+a)\psi_{0}^2(k)&=&\!\left(a^2-a\right) \sum _{k=1}^n \frac{\psi_0 (k+a)\psi_0 (k) }{k}+\frac{1}{2} \left(a^2-a\right) \sum _{k=1}^n \frac{\psi_0 (k+a)}{k^2}+\frac{1}{2}\times\nonumber\\
&&\!\left(-a^2+3 a-1\right) \sum _{k=1}^n \frac{\psi_0 (k+a)}{k}+\frac{1}{8}\Big(\!\left(-4 a^2+4 a+4 n^2+4 n\right)\times\nonumber\\
&& \psi _0^2(n+1) \psi _0(a+n+1)+\left(4 a^2-12 a-4 n^2-12 n\right) \times\nonumber\\
&&\psi _0(n+1) \psi _0(a+n+1)+\left(4 a^2-4 a\right) \psi _0^2(1) \psi _0(a+1)+\nonumber\\
&&\!\left(-2 a^2+10 a+2 n^2+10 n\right) \psi _0(a+n+1)+\left(12 a-4 a^2\right) \times\nonumber\\
&&\psi _0(1) \psi _0(a+1)+\left(2 a^2-10 a\right) \psi _0(a+1)+\big(\!-12 a n-8 a+\nonumber\\
&&4 n^2+20 n+14\big) \psi _0(n+1)+\left(4 a n+4 a-2 n^2-6 n-4\right)\times\nonumber\\
&& \psi _0^2(n+1)+(4-4 a) \psi _0^2(1)+(8 a-14) \psi _0(1)+14 a n-3 n^2-\nonumber\\
&&27 n\Big)
\end{eqnarray}
\begin{eqnarray}\label{eq:A45}
\fl\sum_{k=1}^{n}k^2\psi_{0}(k+a)\psi_{0}^2(k)&=&\!-\frac{1}{3} \left(2 a^3-3 a^2+a\right) \sum _{k=1}^n \frac{ \psi_0 (k+a)\psi_0(k)}{k}-\frac{1}{6} \left(2 a^3-3 a^2+a\right)\times\nonumber\\
&&\sum _{k=1}^n \frac{\psi_0 (k+a)}{k^2}+\frac{1}{18} \left(4 a^3-15 a^2+17 a-3\right) \sum _{k=1}^n \frac{\psi_0 (k+a)}{k}+\nonumber\\
&&\frac{1}{216}\Big(\!\left(72 a^3-108 a^2+36 a+72 n^3+108 n^2+36 n\right)\times\nonumber\\
&& \psi _0^2(n+1)\psi _0(a+n+1)+\left(-48 a^3+180 a^2-204 a-48 n^3-\right.\nonumber\\
&&180\!\left. n^2-204 n\right) \psi _0(n+1) \psi _0(a+n+1)+\left(16 a^3-78 a^2+\right.\nonumber\\
&&158 a+16 \!\left.n^3+78 n^2+158 n\right) \psi _0(a+n+1)+\left(-72 a^3+108\times\right.\nonumber\\
&&\!\left.a^2-36 a\right) \psi _0^2(1) \psi _0(a+1)+\left(48 a^3-180 a^2+204 a\right) \psi _0(1) \times\nonumber\\
&&\psi _0(a+1)+\left(-16 a^3+78 a^2-158 a\right) \psi _0(a+1)+\big(\!-72a^2 \times\nonumber\\
&& n-72 a^2+36 a n^2+144 a n+108 a-24 n^3-90 n^2-102\times\nonumber\\
&& n-36\big)\psi _0^2(n+1)+\left(192 a^2 n+120 a^2-60 a n^2-384 a n-\right.\nonumber\\
&&252 a+32 n^3+156 \!\left.n^2+316 n+198\right) \psi _0(n+1)+\big(72 a^2-\nonumber\\
&&108 a+36\big) \psi _0^2(1)+\left(-120 a^2+252 a-198\right) \psi _0(1)-208 a^2 n+\nonumber\\
&&38 a n^2+416 a n-16 n^3-105 n^2-365 n\Big)
\end{eqnarray}
\begin{eqnarray}\label{eq:A46}
\fl\sum_{k=1}^{n}k^3\psi_{0}(k+a)\psi_{0}^2(k)&=&\frac{1}{2} \left(a^4-2 a^3+a^2\right) \sum _{k=1}^n \frac{\psi_0 (k+a)\psi_0 (k) }{k}+\frac{1}{4} \left(a^4-2 a^3+a^2\right)\times\nonumber\\
&&\!\sum _{k=1}^n \frac{\psi_0 (k+a)}{k^2}-\frac{a}{24} \left(3 a^3-14 a^2+21 a-10\right) \sum _{k=1}^n \frac{\psi_0 (k+a)}{k}+\nonumber\\
&&\frac{1}{3456}\Big(\!\left(-864 a^4+1728 a^3-864 a^2+864 n^4+1728 n^3+\right.\nonumber\\
&&864\!\left.n^2\right)\psi _0^2(n+1) \psi _0(a+n+1)+\left(432 a^4-2016 a^3+3024 a^2-\right.\nonumber\\
&&1440 a-432 n^4-2016 n^3-3024\!\left. n^2-1440 n\right) \psi _0(n+1) \times\nonumber\\
&&\psi _0(a+n+1)+\left(-108 a^4+600 a^3-1332 a^2+1416 a+108\times \right.\nonumber\\
&&n^4+600 n^3+1332 \!\left.n^2+1416 n\right) \psi _0(a+n+1)+\left(864 a^4-\right. \nonumber\\
&&1728 a^3+864\!\left. a^2\right) \psi _0^2(1) \psi _0(a+1)+\left(-432 a^4+2016 a^3-3024 \times\right. \nonumber\\
&&\!\left.a^2+1440 a\right) \psi _0(1) \psi _0(a+1)+\big(108 a^4-600 a^3+1332 a^2- \nonumber\\
&&1416 a\big) \psi _0(a+1)+\left(-432 a^2 n^2+864 a^3 n-2160 a^2 n+864 \times\right.\nonumber\\
&&a^3-1728 a^2+288 a n^3+1296 a n^2+1872 a n+864 a-216 n^4-\nonumber\\
&&1008 n^3-1512 \!\left.n^2-720 n\right) \psi _0^2(n+1)+\left(648 a^2 n^2-2160 a^3 n+\right.\nonumber\\
&&5256 a^2 n-1296 a^3+3312 a^2-336 a n^3-1944 a n^2-4920 a n-\nonumber\\
&&3168 a+216 n^4+1200 n^3+2664 \!\left.n^2+2832 n+1296\right)\times\nonumber\\
&& \psi _0(n+1)+\left(-864 a^3+1728 a^2-864 a\right) \psi _0^2(1)+\left(1296 a^3-\right.\nonumber\\
&&3312\!\left. a^2+3168 a-1296\right) \psi _0(1)-378 a^2 n^2+2268 a^3 n-\nonumber\\
&& 5370a^2 n+148 a n^3+1134 a n^2+4790 a n-81 n^4-546 n^3-\nonumber\\
&&1575 n^2-2550 n\Big)
\end{eqnarray}
\begin{eqnarray}\label{eq:A47}
\fl\sum_{k=1}^{n}k^4\psi_{0}(k+a)\psi_{0}^2(k)&=&\frac{1}{15} \left(-6 a^5+15 a^4-10 a^3+a\right) \sum _{k=1}^n \frac{ \psi_0 (k+a)\psi_0 (k)}{k}-\frac{1}{30} \big(6 a^5-\nonumber\\
&&15 a^4+10 a^3-a\big) \sum _{k=1}^n \frac{\psi_0 (k+a)}{k^2}+\frac{1}{900} \big(72 a^5-405 a^4+770 a^3-\nonumber\\
&&525 a^2+28 a+30\big) \sum _{k=1}^n \frac{\psi_0 (k+a)}{k}+\frac{1}{54000}\Big(\big(10800 a^5-27000 \times\nonumber\\
&&a^4+18000 a^3-1800 a+10800 n^5+27000 n^4+18000 n^3-\nonumber\\
&&1800 n\big) \psi _0^2(n+1) \psi _0(a+n+1)+\big(\!-4320 a^5+24300 a^4-\nonumber\\
&&46200 a^3+31500 a^2-1680 a-4320 n^5-24300 n^4-46200 n^3-\nonumber\\
&&31500 n^2-1680 n\big) \psi _0(n+1) \psi _0(a+n+1)+\big(864 a^5-5535 a^4+\nonumber\\
&&14590 a^3-19725 a^2+11486 a+864 n^5+5535 n^4+14590 n^3+\nonumber\\
&&19725 n^2+11486 n\big) \psi _0(a+n+1)+\big(\!-10800 a^5+27000 a^4-\nonumber\\
&&18000 a^3+1800 a\big) \psi _0^2(1) \psi _0(a+1)+\big(4320 a^5-24300 a^4+\nonumber\\
&&46200 a^3-31500 a^2+1680 a\big) \psi _0(1) \psi _0(a+1)+\left(-864 a^5+\right.\nonumber\\
&&5535 a^4-14590 a^3+\!\left.19725 a^2-11486 a\right) \psi _0(a+1)+\big(\!-3600\times\nonumber\\
&& a^2 n^3+5400 a^3 n^2-18900 a^2 n^2-10800 a^4 n+32400 a^3 n-\nonumber\\
&&33300 a^2 n-10800 a^4+27000 a^3-18000 a^2+2700 a n^4+\nonumber\\
&&14400 a n^3+25200 a n^2+13500 a n-2160 n^5-12150 n^4-\nonumber\\
&&23100 n^3-15750 n^2-840 n+1800\big) \psi _0^2(n+1)+\left(3840 a^2 n^3-\right.\nonumber\\
&&7560 a^3 n^2+26010 a^2 n^2+25920 a^4 n-75060 a^3 n+80970 \times\nonumber\\
&&a^2 n+15120 a^4-45900 a^3+52500 a^2-2430 a n^4-15360\times\nonumber\\
&& a n^3-39480 a n^2-49050 a n-27000 a+1728 n^5+11070 n^4+\nonumber\\
&&29180\!\left. n^3+39450 n^2+22972 n+2520\right) \psi _0(n+1)+\left(10800 a^4-\right.\nonumber\\
&&27000 a^3+\!\left.18000 a^2-1800\right) \psi _0^2(1)+\left(-15120 a^4+45900 a^3-\right.\nonumber\\
&&\left.52500 a^2+27000 a-2520\right) \psi _0(1)\Big)+\frac{1}{1080000} \Big(\!-31360 a^2 n^3+\nonumber\\
&&84240 a^3 n^2-283290 a^2 n^2-535680 a^4 n+1501740 a^3 n-\nonumber\\
&&1515130 a^2 n+16470 a n^4+125440 a n^3+416920 a n^2+\nonumber\\
&& 795450a n-10368 n^5-76545 n^4-239330 n^3-400575 n^2-\nonumber\\
&&331582 n\Big)
\end{eqnarray}
\begin{eqnarray}\label{eq:A48}
\fl\sum _{k=1}^n \psi_0^2 (k+a)\psi_0(k)&=&\!-a \sum _{k=1}^n \frac{\psi _0^2(k+a)}{k}+(2 a-1) \sum _{k=1}^n \frac{\psi _0(k+a)}{k}+(a+n) \times\nonumber\\
&& \psi _0(n+1)\psi _0^2(a+n+1)-(a+n) \psi _0^2(a+n+1)-(2 a+2 n+\nonumber\\
&&1) \psi _0(n+1) \psi _0(a+n+1)+(4 a+4 n+1) \psi _0(a+n+1)-a \times\nonumber\\
&&\psi _0(1) \psi _0^2(a+1)+a \psi _0^2(a+1)+(2 a+1) \psi _0(1) \psi _0(a+1)-(4 a+\nonumber\\
&&1) \psi _0(a+1)+(2 n+2) \psi _0(n+1)-2 \psi _0(1)-6 n
\end{eqnarray}
\begin{eqnarray}\label{eq:A49}
\fl\sum _{k=1}^n k \psi_0^2 (k+a)\psi_0(k)&=&\frac{1}{2} a (a-1) \sum _{k=1}^n \frac{\psi _0^2(k+a)}{k}-\frac{1}{2} \left(3 a^2-3 a+1\right) \sum _{k=1}^n \frac{\psi _0(k+a)}{k}+\nonumber\\
&&\frac{1}{8}\Big(\!\left(-4 a^2+4 a+4 n^2+4 n\right) \psi _0(n+1) \psi _0^2(a+n+1)+\left(2 a^2\right.-\nonumber\\
&&6 a-2\!\left.n^2-6 n\right) \psi _0^2(a+n+1)+\big(12 a^2+8 a n-4 a-4 n^2-12\times\nonumber\\
&& n-4\big) \psi _0(n+1) \psi _0(a+n+1)+\big(\!-16 a^2-12 a n+16 a+4 n^2+\nonumber\\
&&20 n+6\big) \psi _0(a+n+1)+4\left( a^2- a\right) \psi _0(1) \psi _0^2(a+1)+2a (3-\nonumber\\
&&a) \psi _0^2(a+1)-\big(\!12 a^2-4 a-4\big) \psi _0(1) \psi _0(a+1)+\big(16 a^2-16\times\nonumber\\
&& a-6\big) \psi _0(a+1)+\left(-12 a n-12 a+2 n^2+10 n+8\right) \psi _0(n+1)+\nonumber\\
&&(12 a-8) \psi _0(1)+28 a n-3 n^2-27 n\Big)
\end{eqnarray}
\begin{eqnarray}\label{eq:A50}
\fl\sum _{k=1}^n k^2 \psi_0^2 (k+a)\psi_0(k)&=&\!-\frac{1}{6} a \left(2 a^2-3 a+1\right) \sum _{k=1}^n \frac{\psi _0^2(k+a)}{k}+\frac{1}{18} \big(22 a^3-33 a^2+17 \times\nonumber\\
&&a-3\big) \sum _{k=1}^n \frac{\psi _0(k+a)}{k}+\frac{1}{216}\Big(\!\left(72 a^3-108 a^2+36 a+72 n^3+\right.\nonumber\\
&&108\!\left.n^2+36 n\right) \psi _0(n+1) \psi _0^2(a+n+1)+\left(-24 a^3+90 a^2-\right.\nonumber\\
&&102 a-24 \!\left.n^3-90 n^2-102 n\right) \psi _0^2(a+n+1)+\big(\!-144 a^2 n-\nonumber\\
&&264 a^3+252 a^2+72 a n^2+288 a n+12 a-48 n^3-180 n^2-\nonumber\\
&&204 n-36\big) \psi _0(n+1) \psi _0(a+n+1)+\left(192 a^2 n+284 a^3-468\times\right.\nonumber\\
&&a^2-60 a n^2-384 a n+136 a+32 n^3+156\!\left.n^2+316 n+102\right)\times\nonumber\\
&&\psi _0(a+n+1)+\left(-72 a^3+108 a^2-36 a\right) \psi _0(1) \psi _0^2(a+1)+\nonumber\\
&&\left(24 a^3-90 a^2+102 a\right) \psi _0^2(a+1)+\big(264 a^3-252 a^2-12 a+\nonumber\\
&&36\big) \psi _0(1) \psi _0(a+1)+\left(-284 a^3+468 a^2-136 a-102\right)\times\nonumber\\
&& \psi _0(a+1)+\left(264 a^2 n+264 a^2-60 a n^2-384 a n-324 a+16\times\right.\nonumber\\
&& n^3+78\!\left.n^2+158 n+96\right) \psi _0(n+1)+\left(-264 a^2+324 a-96\right) \times\nonumber\\
&&\psi _0(1)-548 a^2 n+76 a n^2+832 a n-16 n^3-105 n^2-365 n\Big)\nonumber\\
\end{eqnarray}
\begin{eqnarray}\label{eq:A51}
\fl\sum _{k=1}^n k^3 \psi_0^2 (k+a)\psi_0(k)&=&\frac{1}{4}  a^2(a-1)^2 \sum _{k=1}^n \frac{\psi _0^2(k+a)}{k}-\frac{5}{24} a \left(5 a^3-10 a^2+7 a-2\right)\times\nonumber\\
&&\sum _{k=1}^n \frac{\psi _0(k+a)}{k}+\frac{1}{288}\Big(\!\left(-72 a^4+144 a^3-72 a^2+72 n^4+144\right.\times\nonumber\\
&& \!\left.n^3+72 n^2\right) \psi _0(n+1) \psi _0^2(a+n+1)+\left(18 a^4-84 a^3+126 a^2-\right.\nonumber\\
&&\!\left.60 a-18 n^4-84 n^3-126 n^2-60 n\right) \psi _0^2(a+n+1)+\left(-72 a^2\right.\times\nonumber\\
&&n^2+144 a^3 n-360 a^2 n+300 a^4-456 a^3+132 a^2+48 a n^3+\nonumber\\
&&\!\left.216 a n^2+312 a n+24 a-36 n^4-168 n^3-252 n^2-120 n\right) \times\nonumber\\
&&\psi _0(n+1) \psi _0(a+n+1)+\left(54 a^2 n^2-180 a^3 n+438 a^2 n-280\right. \times\nonumber\\
&&a^4+628 a^3-380 a^2-28 a n^3-162 a n^2-410 a n-16 a+18 n^4+\nonumber\\
&&\!\left.100 n^3+222 n^2+236 n+60\right) \psi _0(a+n+1)+\left(72 a^4-144 a^3+\right.\nonumber\\
&&\!\left.72 a^2\right) \psi _0(1) \psi _0^2(a+1)+\left(-18 a^4+84 a^3-126 a^2+60 a\right)\times\nonumber\\
&& \psi _0^2(a+1)-\left(300 a^4-456 a^3+132 a^2+24 a\right) \psi _0(1) \psi _0(a+1)+\nonumber\\
&&\!\left(280 a^4-628 a^3+380 a^2+16 a-60\right) \psi _0(a+1)+\left(78 a^2 n^2-\right.\nonumber\\
&&300 a^3 n+606 a^2 n-300 a^3+528 a^2-28 a n^3-162 a n^2-410\times\nonumber\\
&&\!\left.a n-276 a+9 n^4+50 n^3+111 n^2+118 n+48\right) \psi _0(n+1)+\nonumber\\
&&\!\left(300 a^3-528 a^2+276 a-48\right) \psi _0(1)\Big)+\frac{1}{3456} \left(\!-1068 a^2 n^2+\right.\nonumber\\
&&6960 a^3 n-14220 a^2 n+296 a n^3+2268 a n^2+9580 a n-81 n^4-\nonumber\\
&&\!\left.546 n^3-1575 n^2-2550 n\right)
\end{eqnarray}
\begin{eqnarray}\label{eq:A52}
\fl\sum _{k=1}^n k^4 \psi_0^2 (k+a)\psi_0(k)&=&\frac{1}{30} \left(-6 a^5+15 a^4-10 a^3+a\right) \sum _{k=1}^n \frac{\psi _0^2(k+a)}{k}+\frac{1}{900} \left(822 a^5-\right.\nonumber\\
&&\!\left.2055 a^4+1820 a^3-675 a^2+28 a+30\right) \sum _{k=1}^n \frac{\psi _0(k+a)}{k}+\nonumber\\
&&\frac{1}{54000}\Big(\!\left(10800 a^5-27000 a^4+18000 a^3-1800 a+10800 n^5+\right.\nonumber\\
&&\!\left.27000 n^4+18000 n^3-1800 n\right) \psi _0(n+1) \psi _0^2(a+n+1)+\nonumber\\
&&\!\left(-2160 a^5+12150 a^4-23100 a^3+15750 a^2-840 a-2160 n^5-\right.\nonumber\\
&&\!\left.12150 n^4-23100 n^3-15750 n^2-840 n\right) \psi _0^2(a+n+1)+\nonumber\\
&&\!\left(10800 a^3 n^2-7200 a^2 n^3-37800 a^2 n^2-21600 a^4 n+64800\times\right.\nonumber\\
&&a^3 n-66600 a^2 n-49320 a^5+101700 a^4-55200 a^3+4500 a^2+\nonumber\\
&&5400 a n^4+28800 a n^3+50400 a n^2+27000 a n-1680 a-4320\times\nonumber\\ 
&&\!\left.n^5-24300 n^4-46200 n^3-31500 n^2-1680 n+1800\right) \times\nonumber\\ 
&&\psi _0(n+1) \psi _0(a+n+1)+\left(-7560 a^3 n^2+3840 a^2 n^3+26010\times\right.\nonumber\\
&& a^2 n^2+25920 a^4 n-75060 a^3 n+80970 a^2 n+41478 a^5-\nonumber\\
&&116700 a^4+101030 a^3-19200 a^2-2430 a n^4-15360 a n^3-\nonumber\\
&&39480 a n^2-49050 a n-8528 a+1728 n^5+11070 n^4+29180\times\nonumber\\
&&\!\left.n^3+39450 n^2+22972 n+840\right) \psi _0(a+n+1)+\left(-10800 a^5+\right.\nonumber\\
&&\!\left.27000 a^4-18000 a^3+1800 a\right) \psi _0(1) \psi _0^2(a+1)+\left(2160 a^5-\right.\nonumber\\
&&\!\left.12150 a^4+23100 a^3-15750 a^2+840 a\right) \psi _0^2(a+1)+\left(49320 a^5-\right.\nonumber\\
&&\!\left.101700 a^4+55200 a^3-4500 a^2+1680 a-1800\right) \psi _0(1) \times\nonumber\\
&&\psi _0(a+1)+\left(-41478 a^5+116700 a^4-101030 a^3+19200 a^2+\right.\nonumber\\
&&\!\left.8528 a-840\right) \psi _0(a+1)+\left(5640 a^2 n^3-13860 a^3 n^2+37710\times\right.\nonumber\\
&& a^2 n^2+49320 a^4 n-126360 a^3 n+114270 a^2 n+49320 a^4-\nonumber\\
&&112500 a^3+82200 a^2-2430 a n^4-15360 a n^3-39480 a n^2-\nonumber\\
&&49050 a n-22500 a+864 n^5+5535 n^4+14590 n^3+11486 n+\nonumber\\
&&\!\left.19725 n^2+1680\right) \psi _0(n+1)+\left(-49320 a^4+112500 a^3-\right.\nonumber\\
&&\!\left.82200 a^2+22500 a-1680\right) \psi _0(1)\Big)+\frac{1}{1080000} \Big(\!-90920 a^2 n^3+\nonumber\\
&&294180 a^3 n^2-803880 a^2 n^2-1815960 a^4 n+4644180 a^3 n-\nonumber\\
&&4091960 a^2 n+32940 a n^4+250880 a n^3+833840 a n^2+\nonumber\\
&& 1590900a n-10368 n^5-76545 n^4-239330 n^3-400575 n^2-\nonumber\\
&&331582 n\Big)
\end{eqnarray}
\begin{eqnarray}\label{eq:A53}
\fl\sum _{k=1}^n \psi_0^4 (k+a)&=&2 \sum _{k=1}^n \frac{\psi _1(k+a)}{k+a}-\frac{1}{2} \psi _2(a+n+1)+\frac{1}{2} \psi _2(a+1)-2 \psi_{0}(a+n+1)\times\nonumber\\
&& \psi _1(a+n+1)+2 \psi_{0} (a+1) \psi _1(a+1)+2 \psi _1(a+n+1)-2 \psi _1(a+1)+\nonumber\\
&&(a+n) \psi_{0}^4(a+n+1)+(-4 a-4 n-2) \psi_{0}^3(a+n+1)+\big(12 a+12 n+\nonumber\\
&&6\big) \psi_{0}^2(a+n+1)+(-24 a-24 n-12) \psi_{0}(a+n+1)-a\psi_{0}^4(a+1)+ \nonumber\\
&&(4 a+2) \psi_{0} ^{3}(a+1)-(12 a+6) \psi_{0} ^{2}(a+1)+(24 a+12)\psi_{0} (a+1)+24 n \nonumber\\
\end{eqnarray}
\begin{eqnarray}\label{eq:A54}
\fl\sum _{k=1}^n k\psi_0^4 (k+a)&=&(-2 a+1) \sum _{k=1}^n \frac{\psi _1(k+a)}{k+a}+\frac{1}{4}\Big((2 a-1) \psi _2(a+n+1)+(-2a+1) \times\nonumber\\
&&\psi _2(a+1)+(8 a-4) \psi_{0}(a+n+1) \psi _1(a+n+1)+(-8a+4) \times\nonumber\\
&&\psi_{0} (a+1) \psi _1(a+1)+(-8 a+4) \psi _1(a+n+1)+(8 a-4) \psi _1(a+1)+\nonumber\\
&&\!\left(-2 a^2+2 a+2 n^2+2 n\right) \psi_{0}^4(a+n+1)+\big(12 a^2+8 a n-4 a-4 n^2-\nonumber\\
&&12 n-4\big) \psi_{0}^3(a+n+1)+\left(-42 a^2-36 a n+6 a+6 n^2+30 n+14\right) \times\nonumber\\
&&\psi_{0}^2(a+n+1)+\left(90 a^2+84 a n-6 a-6 n^2-54 n-26\right)\times\nonumber\\
&& \psi_{0}(a+n+1)+\left(2 a^2-2 a\right) \psi_{0}^4(a+1)+\left(-12 a^2+4 a+4\right)\times \nonumber\\
&&\psi_{0} ^{3}(a+1)+\left(42 a^2-6 a-14\right) \psi_{0} ^{2}(a+1)+\left(-90 a^2+6 a+26\right)\times \nonumber\\
&&\psi_{0} (a+1)-90 a n+3 n^2+51 n\Big)
\end{eqnarray}
\begin{eqnarray}\label{eq:A55}
\fl\sum _{k=1}^n k^2\psi_0^4 (k+a)&=&\frac{1}{3} \left(6 a^2-6 a+1\right) \sum _{k=1}^n \frac{\psi _1(k+a)}{k+a}+\frac{1}{324}\Big(\!\left(-162 a^2+162 a-27\right)\times\nonumber\\
&& \psi _2(a+n+1)+\left(162 a^2-162 a+27\right) \psi _2(a+1)+\big(\!-648 a^2+\nonumber\\
&& 648a-108\big) \psi_{0}(a+n+1) \psi _1(a+n+1)+\left(648 a^2-648 a+108\right) \times\nonumber\\
&&\psi_{0} (a+1) \psi _1(a+1)+\left(648 a^2-648 a+144\right) \psi _1(a+n+1)+\nonumber\\
&&\!\left(-648 a^2+648 a-144\right) \psi _1(a+1)+\left(108 a^3-162 a^2+54 a+108\times\right. \nonumber\\
&&n^3+162\!\left. n^2+54 n\right) \psi_{0}^4(a+n+1)+\left(-432 a^2 n-792 a^3+756 a^2+\right. \nonumber\\
&&216a n^2+864 a n+36 a-144\!\left.n^3-540 n^2-612 n-108\right)\times\nonumber\\
&& \psi_{0}^3(a+n+1)+\left(2376 a^2 n+3060 a^3-2214 a^2-540 a n^2-3456\times\right.\nonumber\\
&& a n-846 a+144 n^3+702\!\left.n^2+1422 n+594\right) \psi_{0}^2(a+n+1)+\nonumber\\
&&\left(-6120 a^2 n-6900 a^3+4230 a^2+684 a n^2+7488 a n+2022 a-\right.\nonumber\\
&&96\!\left.n^3-630 n^2-2190 n-990\right)\psi_{0}(a+n+1)+\left(-108 a^3+162\times\right.\nonumber\\
&&\!\left.a^2-54 a\right) \psi_{0}^4(a+1)+\left(792a^3-756 a^2-36 a+108\right)\psi_{0} ^{3}(a+1)+\nonumber\\
&&\!\left(-3060 a^3+2214 a^2+846 a-594\right) \psi_{0} ^{2}(a+1)+\left(6900 a^3-4230 \times\right.\nonumber\\
&&\!\left.a^2-2022 a+990\right) \psi_{0} (a+1)+6900 a^2 n-390 a n^2-7680 a n+\nonumber\\
&&32 n^3+291 n^2+1891 n\Big)
\end{eqnarray}
\begin{eqnarray}\label{eq:A56}
\fl\sum _{k=1}^n k^3\psi_0^4 (k+a)&=&\!\left(-2 a^3+3 a^2-a\right) \sum _{k=1}^n \frac{\psi _1(k+a)}{k+a}+\frac{1}{3456}\Big(\big(1728 a^3-2592 a^2+864 \times \nonumber\\
&&a\big) \psi _2(a+n+1)+\left(-1728 a^3+2592 a^2-864 a\right) \psi _2(a+1)+\big(6912 \times\nonumber\\
&&a^3-10368 a^2+3456 a\big) \psi_{0}(a+n+1) \psi _1(a+n+1)+\left(-6912 a^3+\right.\nonumber\\
&&10368\!\left.a^2-3456 a\right) \psi_{0} (a+1) \psi _1(a+1)+\big(\!-6912 a^3+10368 a^2-\nonumber\\
&& 4608a+576\big) \psi _1(a+n+1)+\left(6912 a^3-10368 a^2+4608 a-576\right)\times\nonumber\\
&& \psi _1(a+1)-864 \left(a^4-2 a^3+a^2-n^4-2 n^3-n^2\right)\psi_{0}^4(a+n+1)+\nonumber\\
&& 288 \left(-6 a^2 n^2+12 a^3 n-30 a^2 n+25 a^4-38 a^3+11 a^2+4 a n^3+18\times\right.\nonumber\\
&&a n^2+26 a n+2 a-3 n^4-14 n^3-21\!\left.n^2-10 n\right) \psi_{0}^3(a+n+1)-\nonumber\\
&&72 \left(-78 a^2 n^2+300 a^3 n-606 a^2 n+415 a^4-530 a^3+17 a^2+28 a n^3+\right.\nonumber\\
&&162 a n^2+410 a n+146 a-9 n^4-50\!\left. n^3-111 n^2-118 n-36\right)\times\nonumber\\
&& \psi_{0}^2(a+n+1)+12 \left(-690 a^2 n^2+4980 a^3 n-8850 a^2 n+5845 a^4-\right.\nonumber\\
&&6710 a^3-301 a^2+148 a n^3+1134 a n^2+4790 a n+1790 a-27 n^4-\nonumber\\
&&182 n^3-525\!\left.n^2-850 n-348\right) \psi_{0}(a+n+1)+\left(864a^4-1728 a^3+\right.\nonumber\\
&&864\!\left.a^2\right) \psi_{0} ^4(a+1)+\left(-7200 a^4+10944 a^3-3168 a^2- 576 a\right)\times\nonumber\\
&&\psi_{0} ^{3}(a+1)+\left(29880 a^4-38160 a^3+1224 a^2+10512 a-2592\right)\times\nonumber\\
&& \psi_{0} ^{2}(a+1)+\left(-70140 a^4+80520 a^3+3612 a^2-21480 a+4176\right)\times\nonumber\\
&& \psi_{0} (a+1)+5190 a^2 n^2-70140 a^3 n+115590 a^2 n-700 a n^3-7290\times\nonumber\\
&&a n^2-55250 a n+81 n^4+674 n^3+2631 n^2+7414 n\Big)
\end{eqnarray}
\begin{eqnarray}\label{eq:A57}
\fl\sum _{k=1}^n k^4\psi_0^4 (k+a)&=&\!\left(2 a^4-4 a^3+2 a^2-\frac{1}{15}\right) \sum _{k=1}^n \frac{\psi _1(k+a)}{k+a}+\frac{1}{270000}\Big(\!-4500 \big(30 a^4-\nonumber\\
&&60 a^3+30 a^2-1\big) \psi _2(a+n+1)+4500 \left(30 a^4-60 a^3+30 a^2-1\right)\times\nonumber\\
&&\psi _2(a+1)-18000 \left(30 a^4-60 a^3+30 a^2-1\right) \psi_{0}(a+n+1) \times\nonumber\\
&&\psi _1(a+n+1)+18000 \left(30 a^4-60 a^3+30 a^2-1\right) \psi_{0} (a+1) \times\nonumber\\
&&\psi _1(a+1)+1200 \left(450 a^4-900 a^3+600 a^2-150 a+7\right) \times\nonumber\\
&&\psi _1(a+n+1)-1200\left(450 a^4-900 a^3+600 a^2-150 a+7\right)\times \nonumber\\
&&\psi _1(a+1)+9000 \big(6 a^5-15 a^4+10 a^3-a+6 n^5+15 n^4+10 n^3-\nonumber\\
&&n\big) \psi_{0}^4(a+n+1)-600 \left(-180 a^3 n^2+120 a^2 n^3+630 a^2 n^2+360\times\right.\nonumber\\
&& a^4 n-1080 a^3 n+1110 a^2 n+822 a^5-1695 a^4+920 a^3-75 a^2-90\times \nonumber\\
&&a n^4-480 a n^3-840 a n^2-450 a n+28 a+72 n^5+405 n^4+770 n^3+\nonumber\\
&& 525\!\left. n^2+28 n-30\right) \psi_{0}^3(a+n+1)+30\left(-13860 a^3 n^2+5640 a^2 n^3+\right.\nonumber\\
&&37710 a^2 n^2+49320 a^4 n-126360 a^3 n+114270 a^2 n+72114 a^5-\nonumber\\
&&130965 a^4+42340 a^3+30225 a^2-2430 a n^4-15360 a n^3-39480\times\nonumber\\
&&a n^2-49050 a n-15514 a+864 n^5+5535 n^4+14590 n^3+11486 n+\nonumber\\
&& 19725\!\left. n^2+840\right) \psi_{0}^2(a+n+1)-\left(-683820a^3 n^2+178680 a^2 n^3+\right.\nonumber\\
&&1561770 a^2 n^2+4326840 a^4 n-10021320 a^3 n+7730490 a^2 n+\nonumber\\
&&5249118 a^5-8795955 a^4+2200580 a^3+2163075 a^2-49410 a n^4-\nonumber\\
&&376320 a n^3-1250760 a n^2-2386350 a n-973418 a+10368 n^5+\nonumber\\
&& 76545 n^4+239330 n^3+\!\left.400575 n^2+331582 n+119580\right)\times\nonumber\\
&&\psi_{0}(a+n+1)-9000 \left(6 a^5-15 a^4+10 a^3-a\right) \psi_{0}^4 (a+1)+600\times\nonumber\\
&&\!\left(822 a^5-1695 a^4+920 a^3-75 a^2+28 a-30\right) \psi_{0} ^{3}(a+1)-30\times\nonumber\\
&&\!\left(72114 a^5-130965 a^4+42340 a^3+30225 a^2-15514 a+840\right) \times\nonumber\\
&&\psi_{0} ^{2}(a+1)+\big(5249118 a^5-8795955 a^4+2200580 a^3+2163075 a^2-\nonumber\\
&&973418 a+119580\big) \psi_{0} (a+1)\big)+\frac{1}{16200000}\Big(4769160 a^2 n^3-\nonumber\\
&&27668340 a^3 n^2+55809990 a^2 n^2+314947080 a^4 n-685230840 a^3 n+\nonumber\\
&&480804630 a^2 n-896670 a n^4-8355840 a n^3-37164120 a n^2-\nonumber\\
&&121392450 a n+124416 n^5+1070415 n^4+4055710 n^3+8810025 n^2+\nonumber\\
&&9710234 n\Big)
\end{eqnarray}
\begin{eqnarray}\label{eq:A58}
\fl\sum _{k=1}^n \psi_0 (k+a) \psi _1(k)&=&a \sum _{k=1}^n \frac{\psi _0(k+a)}{k^2}+\sum _{k=1}^n \frac{\psi _0(k+a)}{k}+(a+n)  \psi _0(a+n+1)\times\nonumber\\
&&\psi _1(n+1)-(n+1) \psi _1(n+1)-a \psi _1(1) \psi _0(a+1)-\psi _0(n+1)+\nonumber\\
&&\psi _1(1)+\psi _0(1)
\end{eqnarray}
\begin{eqnarray}\label{eq:A59}
\fl\sum _{k=1}^n k \psi_0 (k+a) \psi _1(k)&=&\frac{1}{2} \left(a-a^2\right) \sum _{k=1}^n \frac{\psi _0(k+a)}{k^2}+\frac{1}{2} \sum _{k=1}^n \frac{\psi _0(k+a)}{k}+\frac{1}{4}\Big(\!\left(-2 a^2+2 a+\right.\nonumber\\
&&2\!\left.n^2+2 n\right) \psi _0(a+n+1) \psi _1(n+1)+(2 a n+2 a-n^2-\nonumber\\
&&3 n-2)\psi _1(n+1)+\left(2 a^2-2 a\right) \psi _1(1) \psi _0(a+1)+(2 a+2 n)\times\nonumber\\
&& \psi _0(a+n+1)-2 a \psi _0(a+1)+(2 a-1) \psi _0(n+1)+(2-2 a) \times\nonumber\\
&&\psi _1(1)+(1-2 a) \psi _0(1)-3 n\Big)
\end{eqnarray}
\begin{eqnarray}\label{eq:A60}
\fl\sum _{k=1}^n k^2 \psi_0 (k+a) \psi _1(k)&=&\frac{1}{6} a \left(2 a^2-3 a+1\right) \sum _{k=1}^n \frac{\psi _0(k+a)}{k^2}+\frac{1}{6} \sum _{k=1}^n \frac{\psi _0(k+a)}{k}+\frac{1}{36}\times\nonumber\\
&&\Big(\!\left(12 a^3-18 a^2+6 a+12 n^3+18 n^2+6 n\right)  \psi _0(a+n+1)\times\nonumber\\
&&\psi _1(n+1)+\left(-12 a^2 n-12 a^2+6 a n^2+24 a n+18 a-4 n^3-\right.\nonumber\\
&&15\!\left.n^2-17 n-6\right) \psi _1(n+1)+\left(-12 a^3+18 a^2-6 a\right) \psi _1(1)\times\nonumber\\
&& \psi _0(a+1)+\left(-6 a^2+24 a+6 n^2+24 n\right) \psi _0(a+n+1)+6a\times\nonumber\\
&&(a-4) \psi _0(a+1)+\left(-12 a^2+12 a+1\right) \psi _0(n+1)+6(2 a^2-\nonumber\\
&&3a+1) \psi _1(1)+\left(12 a^2-12 a-1\right) \psi _0(1)+12 a n-5 n^2-32 n\Big)\nonumber\\
&&
\end{eqnarray}
\begin{eqnarray}\label{eq:A61}
\fl\sum _{k=1}^n k^3 \psi_0 (k+a) \psi _1(k)&=&\!-\frac{1}{4}a^2 (a-1)^2  \sum _{k=1}^n \frac{\psi _0(k+a)}{k^2}+\frac{1}{288}\Big(\!\left(-72 a^4+144 a^3-72 a^2+\right.\nonumber\\
&&72 n^4+\!\left.144 n^3+72 n^2\right) \psi _0(a+n+1)\psi _1(n+1)+\left(-36 a^2 n^2+\right.\nonumber\\
&&72 a^3 n-180 a^2 n+72 a^3-144 a^2+24 a n^3+108 a n^2+156 a n+\nonumber\\
&&72 a-18 n^4-84 n^3-\!\left.126 n^2-60 n\right) \psi _1(n+1)+\left(72 a^4-144\times\right.\nonumber\\
&& a^3+72 \!\left.a^2\right) \psi _1(1) \psi _0(a+1)+\left(24 a^3-108 a^2+156 a+24 n^3+\right.\nonumber\\
&&108\!\left. n^2+156 n\right) \psi _0(a+n+1)+\left(-24 a^3+108 a^2-156 a\right)\times\nonumber\\
&& \psi _0(a+1)+\left(72 a^3-108 a^2+12 a+12\right) \psi _0(n+1)+\left(-72 a^3+\right.\nonumber\\
&&144 \!\left.a^2-72 a\right) \psi _1(1)+\left(-72 a^3+108 a^2-12 a-12\right) \psi _0(1)-\nonumber\\
&&60 a^2 n+24 a n^2+168 a n-14 n^3-81 n^2-205 n\Big)
\end{eqnarray}
\begin{eqnarray}\label{eq:A62}
\fl\sum _{k=1}^n k^4 \psi_0 (k+a) \psi _1(k)&=&\!-\frac{1}{30} a \left(-6 a^4+15 a^3-10 a^2+1\right) \sum _{k=1}^n \frac{\psi _0(k+a)}{k^2}-\frac{1}{30}\times\nonumber\\
&& \sum _{k=1}^n \frac{\psi _0(k+a)}{k}+\frac{1}{3600}\Big(\!\left(720 a^5-1800 a^4+1200 a^3-120 a+\right.\nonumber\\
&&720 n^5+1800 n^4+\!\left.1200 n^3-120 n\right) \psi _1(n+1) \psi _0(a+n+1)+\nonumber\\
&&\big(\!-240 a^2 n^3+360 a^3 n^2-1260 a^2 n^2-720 a^4 n+2160 a^3 n-\nonumber\\
&&2220 a^2 n-720 a^4+1800 a^3-1200 a^2+180 a n^4+960 a n^3+\nonumber\\
&&1680 a n^2+900 a n-144 n^5-810 n^4-1540 n^3-1050 n^2-56 n+\nonumber\\
&&120\big) \psi _1(n+1)+\left(-720 a^5+1800 a^4-1200 a^3+120 a\right) \psi _1(1)\times\nonumber\\
&& \psi _0(a+1)+\left(-180 a^4+960 a^3-1680 a^2+900 a+180 n^4+\right.\nonumber\\
&&960 n^3+1680\!\left.n^2+900 n\right) \psi _0(a+n+1)+\left(180 a^4-960 a^3+\right.\nonumber\\
&&1680\!\left.a^2-900 a\right) \psi _0(a+1)+\big(\!-720 a^4+1440 a^3-420 a^2-\nonumber\\
&&300 a-56\big) \psi _0(n+1)+\left(720 a^4-1800 a^3+1200 a^2-120\right) \times\nonumber\\
&&\psi _1(1)+\left(720 a^4-1440 a^3+420 a^2+300 a+56\right) \psi _0(1)-\nonumber\\
&&210 a^2 n^2+540 a^3 n-1710 a^2 n+120 a n^3+780 a n^2+2220 a n-\nonumber\\
&&81 n^4-512 n^3-1316 n^2-1635 n\Big)
\end{eqnarray}
\begin{eqnarray}\label{eq:A63}
\fl\sum _{k=1}^n\psi _1(k+a) \psi _0(k) &=&\!-a \sum _{k=1}^n \frac{\psi _1(k+a)}{k}-\sum _{k=1}^n \frac{\psi _0(k+a)}{k}+(a+n) \psi _0(n+1)\times\nonumber\\
&& \psi _1(a+n+1)-(a+n) \psi _1(a+n+1)-a \psi _0(1) \psi _1(a+1)+\nonumber\\
&&a \psi _1(a+1)+\psi _0(n+1) \psi _0(a+n+1)-\psi _0(a+n+1)-\psi _0(1) \times\nonumber\\
&&\psi _0(a+1)+\psi _0(a+1)
\end{eqnarray}
\begin{eqnarray}\label{eq:A64}
\fl\sum _{k=1}^n k \psi _1(k+a) \psi _0(k) &=&\frac{1}{2} a(a-1)  \sum _{k=1}^n \frac{\psi _1(k+a)}{k}+\left(a-\frac{1}{2}\right) \sum _{k=1}^n \frac{\psi _0(k+a)}{k}+\frac{1}{4}\times\nonumber\\
&&\Big(\!\left(-2 a^2+2 a+2 n^2+2 n\right) \psi _0(n+1) \psi _1(a+n+1)+\left(a^2-3 a-\right.\nonumber\\
&&\!\left.n^2-3 n\right) \psi _1(a+n+1)+\left(2 a^2-2 a\right) \psi _0(1) \psi _1(a+1)+\big(3 a-\nonumber\\
&&a^2\big) \psi _1(a+1)-(4a-2) \psi _0(n+1) \psi _0(a+n+1)+(2 a-3)\times\nonumber\\
&& \psi _0(a+n+1)+(4 a-2) \psi _0(1) \psi _0(a+1)-(2 a-3) \psi _0(a+1)+\nonumber\\
&&(2 n+2) \psi _0(n+1)-2 \psi _0(1)-3 n\Big)
\end{eqnarray}
\begin{eqnarray}\label{eq:A65}
\fl\sum _{k=1}^n k^2 \psi _1(k+a) \psi _0(k) &=&\!-\frac{1}{6} a \left(2 a^2-3 a+1\right) \sum _{k=1}^n \frac{\psi _1(k+a)}{k}-\left(a^2-a+\frac{1}{6}\right)\times\nonumber\\
&& \sum _{k=1}^n \frac{\psi _0(k+a)}{k}+\frac{1}{36}\Big(6 \left(2 a^3-3 a^2+a+2 n^3+3 n^2+n\right)\times\nonumber\\
&& \psi _0(n+1) \psi _1(a+n+1)+\left(-4 a^3+15 a^2-17 a-4 n^3-15 \times\right.\nonumber\\
&&\!\left.n^2-17 n\right) \psi _1(a+n+1)+\left(36 a^2-36 a+6\right) \psi _0(n+1)\times\nonumber\\
&& \psi _0(a+n+1)+\left(-12 a^2+30 a-17\right) \psi _0(a+n+1)-\left(12 a^3-\right.\nonumber\\
&&\!\left.18 a^2+6 a\right) \psi _0(1) \psi _1(a+1)+\left(4 a^3-15 a^2+17 a\right) \psi _1(a+1)+\nonumber\\
&&\left(-36 a^2+36 a-6\right) \psi _0(1) \psi _0(a+1)+\left(12 a^2-30 a+17\right)\times\nonumber\\
&& \psi _0(a+1)+\left(-24 a n-24 a+6 n^2+24 n+18\right) \psi _0(n+1)+\nonumber\\
&&(24 a-18) \psi _0(1)+32 a n-5 n^2-32 n\Big)
\end{eqnarray}
\begin{eqnarray}\label{eq:A66}
\fl\sum _{k=1}^n k^3 \psi _1(k+a) \psi _0(k) &=&\frac{1}{4} (a-1)^2 a^2 \sum _{k=1}^n \frac{\psi _1(k+a)}{k}+\frac{1}{2} a \left(2 a^2-3 a+1\right) \sum _{k=1}^n \frac{\psi _0(k+a)}{k}+\nonumber\\
&&+\frac{1}{288}\Big(\!\left(-72 a^4+144 a^3-72 a^2+72 n^4+144 n^3+72 n^2\right)\times\nonumber\\
&& \psi _0(n+1) \psi _1(a+n+1)+\left(18 a^4-84 a^3+126 a^2-60 a-18\times\right.\nonumber\\
&&\!\left. n^4-84 n^3-126 n^2-60 n\right) \psi _1(a+n+1)+\left(72 a^4-144 a^3+\right.\nonumber\\
&&\!\left.72 a^2\right) \psi _0(1) \psi _1(a+1)+\left(-18 a^4+84 a^3-126 a^2+60 a\right)\times\nonumber\\
&& \psi _1(a+1)+\left(-288 a^3+432 a^2-144 a\right) \psi _0(n+1) \times\nonumber\\
&&\psi _0(a+n+1)+\left(72 a^3-252 a^2+252 a-60\right) \psi _0(a+n+1)+\nonumber\\
&&\!\left(288 a^3-432 a^2+144 a\right) \psi _0(1) \psi _0(a+1)+\left(-72 a^3-252 a+\right.\nonumber\\
&&\!\left.252 a^2+60\right) \psi _0(a+1)+\left(216 a^2 n+216 a^2-72 a n^2-360 a n-\right.\nonumber\\
&&\!\left.288 a+24 n^3+108 n^2+156 n+72\right) \psi _0(n+1)+\big(\!-216 a^2+\nonumber\\
&&288 a-72\big) \psi _0(1)-270 a^2 n+54 a n^2+438 a n-14 n^3-81 n^2-\nonumber\\
&&205 n\Big)
\end{eqnarray}
\begin{eqnarray}\label{eq:A67}
\fl\sum _{k=1}^n k^4 \psi _1(k+a) \psi _0(k) &=&\left(-a^4+2 a^3-a^2+\frac{1}{30}\right) \sum _{k=1}^n \frac{\psi _0(k+a)}{k}-\frac{1}{30} a \left(6 a^4-15 a^3+\right.\nonumber\\
&&\!\left.10 a^2-1\right) \sum _{k=1}^n \frac{\psi _1(k+a)}{k}+\frac{1}{1800}\Big(\!\left(360 a^5-900 a^4+600 a^3-\right.\nonumber\\
&&\!\left.60 a+360 n^5+900 n^4+600 n^3-60 n\right) \psi _0(n+1)\times\nonumber\\
&&\psi _1(a+n+1)+\left(-72 a^5+405 a^4-770 a^3+525 a^2-28 a-\right.\nonumber\\
&&72\left. n^5-405 n^4-770 n^3-525 n^2-28 n\right) \psi _1(a+n+1)+\nonumber\\
&&\!\left(-360 a^5+900 a^4-600 a^3+60 a\right) \psi _0(1)\psi _1(a+1)+\left(72 a^5-\right.\nonumber\\
&&405\!\left.a^4+770 a^3-525 a^2+28 a\right) \psi _1(a+1)+\left(1800 a^4-3600 a^3+\right.\nonumber\\
&&1800\!\left.a^2-60\right) \psi _0(n+1) \psi _0(a+n+1)+\left(-360 a^4+1620 a^3-\right.\nonumber\\
&&2310\!\left.a^2+1050 a-28\right) \psi _0(a+n+1)+\left(-1800 a^4+3600 a^3-\right.\nonumber\\
&&1800\!\left.a^2+60\right) \psi _0(1) \psi _0(a+1)+\left(360 a^4-1620 a^3-1050 a+\right.\nonumber\\
&&2310\!\left.a^2+28\right) \psi _0(a+1)+\left(540 a^2 n^2-1440a^3 n+3240 a^2 n-\right.\nonumber\\
&&1440 a^3+2700 a^2-240 a n^3-1260 a n^2-2220 a n-1200 a+\nonumber\\
&&90 n^4+\!\left.480 n^3+840 n^2+450 n\right) \psi _0(n+1)+\left(1440 a^3-2700\times\right.\nonumber\\
&&\!\left.a^2+1200 a\right) \psi _0(1)\Big)+\frac{1}{3600} \Big(\!-756 a^2 n^2+3456 a^3 n-7506\times \nonumber\\
&&a^2 n+256a n^3+1734 a n^2+5398 a n-81 n^4-512 n^3-1316 \times\nonumber\\
&&n^2-1635 n\Big)
\end{eqnarray}
\begin{eqnarray}\label{eq:A68}
\fl\sum _{k=1}^n  \psi _1^2(k+a)&=&2 \sum _{k=1}^n \frac{\psi _1(k+a)}{k+a}+\frac{1}{2}\Big(\!-\psi _2(a+n+1)+\psi _2(a+1)+2 (a+n)\times\nonumber\\
&& \psi _1^2(a+n+1)-2 a \psi _1^2(a+1)\Big)
\end{eqnarray}
\begin{eqnarray}\label{eq:A69}
\fl\sum _{k=1}^n k \psi _1^2(k+a)&=&(-2 a+1) \sum _{k=1}^n \frac{\psi _1(k+a)}{k+a}+\frac{1}{4}\Big((2 a-1) \psi _2(a+n+1)+(-2 a+1)\times\nonumber\\
&& \psi _2(a+1)+\left(-2 a^2+2 a+2 n^2+2 n\right) \psi _1^2(a+n+1)+2a\left(a-1\right)\times\nonumber\\
&&\psi _1^2(a+1)+(4 a+4 n+2) \psi _1(a+n+1)-(4a+2) \psi _1(a+1)+4 \times\nonumber\\
&&\psi _0(a+n+1)-4 \psi _0(a+1)\Big)
\end{eqnarray}
\begin{eqnarray}\label{eq:A70}
\fl\sum _{k=1}^n k^2 \psi _1^2(k+a)&=&\!-\frac{1}{3} \left(-6 a^2+6 a-1\right) \sum _{k=1}^n \frac{\psi _1(k+a)}{k+a}+\frac{1}{12}\Big(\!\left(-6 a^2+6 a-1\right)\times\nonumber\\
&& \psi _2(a+n+1)+\left(6 a^2-6 a+1\right) \psi _2(a+1)+\left(4 a^3-6 a^2+2 a+4 n^3+\right.\nonumber\\
&&6\!\left.n^2+2 n\right) \psi _1^2(a+n+1)+\left(-4 a^3+6 a^2-2 a\right) \psi _1^2(a+1)+\nonumber\\
&&\!\left(-20 a^2-16 a n+4 a+4 n^2+16 n+6\right) \psi _1(a+n+1)+\big(20 a^2-\nonumber\\
&&4 a-6\big) \psi _1(a+1)+(-24 a+12) \psi _0(a+n+1)+(24 a-12)\times\nonumber\\
&& \psi _0(a+1)+4 n\Big)
\end{eqnarray}
\begin{eqnarray}\label{eq:A71}
\fl\sum _{k=1}^n k^3 \psi _1^2(k+a)&=&\left(-2 a^3+3 a^2-a\right) \sum _{k=1}^n \frac{\psi _1(k+a)}{k+a}+\frac{1}{12}\Big(\!\left(6 a^3-9 a^2+3 a\right) \times\nonumber\\
&&\psi _2(a+n+1)+\left(-6 a^3+9 a^2-3 a\right) \psi _2(a+1)+\left(-3 a^4+6 a^3-\right.\nonumber\\
&&3 a^2+3 n^4+6 n^3+3\!\left. n^2\right) \psi _1^2(a+n+1)+\left(3 a^4-6 a^3+3 a^2\right) \times\nonumber\\
&&\psi _1^2(a+1)+\left(18 a^2 n+26 a^3-21 a^2-6 a n^2-30 a n-5 a+2 n^3+\right.\nonumber\\
&&9\!\left.n^2+13 n+3\right) \psi _1(a+n+1)+\left(-26 a^3+21 a^2+5 a-3\right)\times\nonumber\\
&& \psi _1(a+1)+\left(36 a^2-36 a+7\right) \psi _0(a+n+1)+\left(-36 a^2+36 a-7\right)\times\nonumber\\
&& \psi _0(a+1)-10 a n+n^2+7 n\Big)
\end{eqnarray}
\begin{eqnarray}\label{eq:A72}
\fl\sum _{k=1}^n k^4 \psi _1^2(k+a)&=&\frac{1}{15} \left(30 a^4-60 a^3+30 a^2-1\right) \sum _{k=1}^n \frac{\psi _1(k+a)}{k+a}+\frac{1}{60}\Big(\!\left(-30 a^4+60 a^3-\right.\nonumber\\
&&30\!\left. a^2+1\right) \psi _2(a+n+1)+\left(30 a^4-60 a^3+30 a^2-1\right) \psi _2(a+1)+\nonumber\\
&&\!\left(12 a^5-30 a^4+20 a^3-2 a+12 n^5+30 n^4+20 n^3-2 n\right) \times\nonumber\\
&&\psi _1^2(a+n+1)+\left(-12 a^5+30 a^4-20 a^3+2 a\right) \psi _1^2(a+1)+\left(36a^2\times\right.\nonumber\\
&&  n^2-96 a^3 n+216 a^2 n-154 a^4+212 a^3-24 a^2-16 a n^3-84 a n^2-\nonumber\\
&&148 a n-30 a+6 n^4+32 n^3+\!\left.56 n^2+30 n\right) \psi _1(a+n+1)+\left(154 a^4-\right.\nonumber\\
&&212 a^3+24\!\left. a^2+30 a\right) \psi _1(a+1)+\left(-240 a^3+360 a^2-140 a+10\right) \times\nonumber\\
&&\psi _0(a+n+1)+\left(240 a^3-360 a^2+140 a-10\right) \psi _0(a+1)+86 a^2 n-\nonumber\\
&&14 a n^2-114 a n+2 n^3+13 n^2+37 n\Big)
\end{eqnarray}
\begin{eqnarray}\label{eq:A73}
\fl\sum _{k=1}^n \psi _0(k+a) \psi _2(k+a)&=&\!-2 \sum _{k=1}^n \frac{\psi _1(k+a)}{k+a}+(a+n) \psi _0(a+n+1) \psi _2(a+n+1)-\nonumber\\
&&a \psi _0(a+1) \psi _2(a+1)-(a+n) \psi _2(a+n+1)+a \psi _2(a+1)+\nonumber\\
&&2 \psi _0(a+n+1) \psi _1(a+n+1)-2 \psi _1(a+n+1)-2\times\nonumber\\
&& \psi _0(a+1) \psi _1(a+1)+2 \psi _1(a+1)
\end{eqnarray}
\begin{eqnarray}\label{eq:A74}
\fl\sum _{k=1}^n k\psi _0(k+a) \psi _2(k+a)&=&(2 a-1) \sum _{k=1}^n \frac{\psi _1(k+a)}{k+a}+\frac{1}{4}\Big(\!\left(-2 a^2+2 a+2 n^2+2 n\right)\times\nonumber\\
&& \psi _0(a+n+1) \psi _2(a+n+1)+\left(2 a^2-2 a\right) \psi _0(a+1) \times\nonumber\\
&&\psi _2(a+1)+\left(3 a^2+2 a n-3 a-n^2-3 n\right) \psi _2(a+n+1)+\nonumber\\
&&\!\left(3 a-3 a^2\right) \psi _2(a+1)-(8 a-4) \psi _0(a+n+1)\times\nonumber\\
&& \psi _1(a+n+1)+(8 a-4) \psi _0(a+1) \psi _1(a+1)+(8 a-4)\times\nonumber\\
&&\psi _1(a+n+1)-(8 a-4) \psi _1(a+1)-2 \psi _0^2(a+n+1)+\nonumber\\
&&2 \psi _0(a+n+1)+2 \psi _0^2(a+1)-2 \psi _0(a+1)\Big)
\end{eqnarray}
\begin{eqnarray}\label{eq:A75}
\fl\sum _{k=1}^n k^2\psi _0(k+a) \psi _2(k+a)&=&\!-\frac{1}{3} \left(6 a^2-6 a+1\right) \sum _{k=1}^n \frac{\psi _1(k+a)}{k+a}+\frac{1}{36}\Big(\!\left(12 a^3-18 a^2+\right.\nonumber\\
&&6 a+12 n^3+\!\left.18 n^2+6 n\right)\psi _0(a+n+1) \psi _2(a+n+1)+\nonumber\\
&&\!\left(-12 a^3+18 a^2-6 a\right) \psi _0(a+1) \psi _2(a+1)+\left(-12 a^2 n-\right.\nonumber\\
&&22 a^3+39 a^2+6 a n^2+24 a n-17 a-4 n^3-\!\left.15 n^2-17 n\right)\times\nonumber\\
&&\psi _2(a+n+1)+\left(22 a^3-39 a^2+17 a\right) \psi _2(a+1)+\big(72 a^2-\nonumber\\
&&72 a+12\big)\psi _0(a+n+1) \psi _1(a+n+1)-\big(72 a^2-72 a+\nonumber\\
&&12\big) \psi _0(a+1) \psi _1(a+1)+\left(-72 a^2+72 a-16\right) \times\nonumber\\
&&\psi _1(a+n+1)+\left(72 a^2-72 a+16\right) \psi _1(a+1)+\big(36 a-\nonumber\\
&&18\big) \psi _0^2(a+n+1)+(-60 a-24 n+6) \psi _0(a+n+1)+\nonumber\\
&&(18-36 a) \psi _0^2(a+1)+(60 a-6) \psi _0(a+1)+32 n\Big)
\end{eqnarray}
\begin{eqnarray}\label{eq:A76}
\fl\sum _{k=1}^n k^3\psi _0(k+a) \psi _2(k+a)&=&\!\left(2 a^3-3 a^2+a\right) \sum _{k=1}^n \frac{\psi _1(k+a)}{k+a}+\frac{1}{48} \Big(\!\left(-12 a^2 + 24 a^3 -\right.\nonumber\\
&& 12 a^4 +\!\left.12 n^2 + 24 n^3 + 12 n^4\right)\psi _0(a+n+1) \psi _2(a+n+1)+\nonumber\\
&&\left(12 a^4-24 a^3+12 a^2\right) \psi _0(a+1) \psi _2(a+1)+\left(-6 a^2 n^2+\right.\nonumber\\
&&12 a^3 n-30 a^2 n+25 a^4-62 a^3+47 a^2+4 a n^3+18 a n^2+\nonumber\\
&&26 a n-10 a-3 n^4-14 n^3-\!\left.21 n^2-10 n\right)\psi _2(a+n+1)+\nonumber\\
&&\left(-25 a^4+62 a^3-47 a^2+10 a\right) \psi _2(a+1)+\left(-96 a^3+\right.\nonumber\\
&&144\!\left.a^2-48 a\right)\psi _0(a+n+1) \psi _1(a+n+1)+\big(96 a^3-144\times\nonumber\\
&&a^2+48 a\big)\psi _0(a+1) \psi _1(a+1)+\big(96 a^3-144 a^2+64 a-\nonumber\\
&& 8\big)\psi _1(a+n+1)+\left(-96 a^3+144 a^2-64 a+8\right) \psi _1(a+1)+\nonumber\\
&&\!\left(-72 a^2+72 a-12\right) \psi _0^2(a+n+1)+\left(156 a^2+72 a n-84\times\right.\nonumber\\
&&a-\!\left.12 n^2-60 n-6\right) \psi _0(a+n+1)+\left(72 a^2-72 a+12\right)\times\nonumber\\
&& \psi _0^2(a+1)+\left(-156 a^2+84 a+6\right) \psi _0(a+1)-110 a n+\nonumber\\
&&9 n^2+73 n\Big)
\end{eqnarray}
\begin{eqnarray}\label{eq:A77}
\fl\sum _{k=1}^n k^4\psi _0(k+a) \psi _2(k+a)&=&\!-\frac{1}{15} \left(30 a^4-60 a^3+30 a^2-1\right) \sum _{k=1}^n \frac{\psi _1(k+a)}{k+a}+\frac{1}{1800}\times\nonumber\\
&&\Big(\big(360 a^5-900 a^4+600 a^3-60 a+360 n^5+900 n^4+ \nonumber\\
&&600n^3-60 n\big) \psi _0(a+n+1) \psi _2(a+n+1)+\left(-360 a^5+\right.\nonumber\\
&&900\!\left. a^4-600 a^3+60 a\right) \psi _0(a+1) \psi _2(a+1)+\big(180 a^3 n^2-\nonumber\\
&&120 a^2 n^3-630 a^2 n^2-360 a^4 n+1080 a^3 n-1110 a^2 n-\nonumber\\
&&822 a^5+2595 a^4-2720 a^3+975 a^2+90 a n^4+480 a n^3+\nonumber\\
&&840 a n^2+450 a n-28 a-72 n^5-405 n^4-770 n^3-525 n^2-\nonumber\\
&&28 n\big) \psi _2(a+n+1)+\big(822 a^5-2595 a^4+2720 a^3-975 a^2+\nonumber\\
&&28 a\big)\psi _2(a+1)+\left(3600 a^4-7200 a^3+3600 a^2-120\right) \times\nonumber\\
&&\psi _0(a+n+1) \psi _1(a+n+1)+\big(\!-3600 a^4+7200 a^3- \nonumber\\
&&3600 a^2+120\big)\psi _0(a+1) \psi _1(a+1)+\left(-3600 a^4+7200 a^3-\right.\nonumber\\
&&4800\!\left. a^2+1200 a-56\right) \psi _1(a+n+1)+\left(3600 a^4-7200 a^3+\right.\nonumber\\
&&4800 \!\left.a^2-1200 a+56\right) \psi _1(a+1)+\big(3600 a^3-5400 a^2+\nonumber\\
&&1800 a\big) \psi _0^2(a+n+1)+\big(\!-4320 a^2 n-9240 a^3+9540 a^2+\nonumber\\
&&1080 a n^2+6480 a n-1320 a-240 n^3-1260 n^2-2220 n-\nonumber\\
&&150\big) \psi _0(a+n+1)+\left(-3600 a^3+5400 a^2-1800 a\right)\times\nonumber\\
&& \psi _0^2(a+1)+\left(9240 a^3-9540 a^2+1320 a+150\right) \psi _0(a+1)+\nonumber\\
&&7284 a^2 n-966 a n^2-9216 a n+128 n^3+867 n^2+2699 n\Big)\nonumber\\
\end{eqnarray}
\begin{eqnarray}\label{eq:A78}
\fl\sum _{k=1}^n \frac{\psi_0 (k+a)^2+\psi _1(k+a)}{k+a}&=&\frac{1}{3}\Big(\psi _2(a+n+1)-\psi _2(a+1)+3 \psi_0 (a+n+1)\times\nonumber\\
&& \psi _1(a+n+1)-3 \psi_0 (a+1) \psi _1(a+1)+\psi_0^3 (a+n+1)-\nonumber\\
&&\psi_0^3 (a+1)\Big)
\end{eqnarray}
\begin{eqnarray}\label{eq:A79}
\fl\sum _{k=1}^n \left(\frac{\psi _0^2(k+a)}{k+a}+\frac{\psi _0(k+a)}{(k+a)^2}\right)&=&\frac{1}{6}\Big(\!-\psi _2(a+n+1)+\psi _2(a+1)+2 \psi _0^3(a+n+1)-\nonumber\\
&&2 \psi _0^3(a+1)\Big)
\end{eqnarray}
\begin{eqnarray}\label{eq:A80}
&\!\fl\sum _{k=1}^n \frac{\psi _0^3(k+a)+3 \psi _0(k+a) \psi _1(k+a)+\psi _2(k+a)}{k+a}=\frac{1}{4}\Big(\psi _3(a+n+1)-\psi _3(a+1)+\nonumber\\
&4 \psi _0(a+n+1) \psi _2(a+n+1) -4\psi _0(a+1)\psi _2(a+1)+3 \psi _1^2(a+n+1)-\nonumber\\
&3 \psi _1^2(a+1)+6 \psi _0^2(a+n+1)\psi _1(a+n+1)-6 \psi _0^2(a+1)\psi _1(a+1)+\nonumber\\
&\psi _0^4(a+n+1)-\psi _0^4(a+1)\Big).
\end{eqnarray}

\subsection{Remarks on the first type summation}\label{sec:ap1-re}
The identities (\ref{eq:A3})-(\ref{eq:A6}), (\ref{eq:A8})-(\ref{eq:A11}), (\ref{eq:A18})-(\ref{eq:A21}), (\ref{eq:A28})-(\ref{eq:A31}), and (\ref{eq:A38})-(\ref{eq:A41}) are available in \cite{Wei20}. The approach in obtaining the first type summation (\ref{eq:A1}) is based on changing the order of sums and making use of the obtained lower order
summation formulas in a recursive manner, i.e.,
\begin{eqnarray}\label{eq:A81}
\fl\sum_{k=1}^{n}f(k)\psi_{i_1}(k+a_1)&=\sum_{k=1}^{n}f(k)\left(\psi_{i_1}(a_1)+\sum_{l=1}^{k}\frac{(-1)^{i_1}i_1!}{\left(a_1+l-1\right)^{i_1+1}}\right)\nonumber\\
&=\psi_{i_1}^{j_1}(a_1)\sum_{k=1}^{n}f(k)+\sum_{l=1}^{n}\frac{(-1)^{i_1}i_1!}{\left(a_1+l-1\right)^{i_1+1}}\sum_{k=l}^{n}f(k),
\end{eqnarray}
where 
\begin{equation*}
f(k)=\psi_{i_{1}}^{j_{1}}(k+a_{1})\psi_{i_{2}}^{j_{2}}(k+a_{2}).
\end{equation*}
By using this approach, one can start building the pyramid of the finite sums described in (\ref{eq:A1}). The closed-form identities (\ref{eq:A78})-(\ref{eq:A81}) that represent the relation between the summations (\ref{eq:A2}) are derived by the same principle. Alternatively, some of the results can also be obtained by the partial differentiation relations, i.e.,
\begin{equation}
\fl\frac{\partial }{\partial a_1}\sum_{k=1}^{n}\psi_{i_1}^{j_1}(k+a_1)\psi_{i_2}^{j_2}(k+a_2)=\sum_{k=1}^{n}j_1\psi_{i_1}^{j_1-1}(k+a_1)\psi_{i_1+1}(k+a_1)\psi_{i_2}^{j_2}(k+a_2),
\end{equation}
or
\begin{eqnarray}
\fl\frac{\partial }{\partial a_1}\sum_{k=1}^{n}\psi_{i_1}^{j_1}(k+a_1)\psi_{i_2}^{j_2}(k+a_1)&=&\sum_{k=1}^{n}j_1\psi_{i_1}^{j_1-1}(k+a_1)\psi_{i_1+1}(k+a_1)\psi_{i_2}^{j_2}(k+a_1)+\nonumber\\
&&\sum_{k=1}^{n}j_2\psi_{i_1}^{j_1}(k+a_1)\psi_{i_2}^{j_2-1}(k+a_1)\psi_{i_2+1}(k+a_1).
\end{eqnarray}
 For examples, by taking the derivatives, one obtains (\ref{eq:A18})-(\ref{eq:A22}) from the corresponding ones in (\ref{eq:A3})-(\ref{eq:A7}). Likeswise, the indentities~(\ref{eq:A23})-(\ref{eq:A27}), (\ref{eq:A28})-(\ref{eq:A32}), (\ref{eq:A33})-(\ref{eq:A37}), (\ref{eq:A63})-(\ref{eq:A67}), (\ref{eq:A73})-(\ref{eq:A77}) can be obtained by the same principle. Finally, we note the interesting fact that the identities (\ref{eq:A53})-(\ref{eq:A57}) only admit semi closed-from expressions whereas their lower order ones (\ref{eq:A3})-(\ref{eq:A17}) admit closed-form expressions.
\section{Polygamma summation identities of the second type}\label{App2}
In this appendix, we list finite summation identities of the type
\begin{equation}\label{eq:B1}
S_{f}(m,n)=\sum_{k=1}^{m}\frac{(n-k)!}{(m-k)!}f(k),~~~~~~m\leq n,
\end{equation}
hereinafter referred to as the second type useful in the simplification process in~\sref{sec:EvSum}. Here, $f(k)$ is referred to as the test functions that may involve polygamma functions and $(n-k)!/(m-k)!$ is referred to as the summation kernel.

The corresponding closed-form and semi closed-form identities are listed in appendix\,B.1 and appendix\,B.2, respectively. The strategy in deriving the second type identities is discussed in appendix\,B.3. Note that the unsimplifiable bases involved in appendix\,B.2 are $\Omega_1$--$\Omega_9$, $\Omega_{12}$, $\Omega_{15}$, and $\Omega_{17}$ of~\tref{t:bases}.

\subsection{Closed-form expressions}\label{sec:ap2-c}
\begin{equation}\label{eq:B2}
\fl\sum_{k=1}^{m}\frac{(n-k)!}{(m-k)!}=\frac{n!}{(m-1)!}\frac{1}{n-m+1}
\end{equation}
\begin{equation}\label{eq:B3}
\fl\sum_{k=1}^{m}\frac{(n-k)!}{(m-k)!}\frac{1}{k}=\frac{n!}{m!}\left(\psi_{0}\left(n+1\right)-\psi_{0}\left(n-m+1\right)\right)
\end{equation}
\begin{eqnarray}\label{eq:B4}
\fl\sum_{k=1}^{m}\frac{(n-k)!}{(m-k)!}\psi_{0}(k)&=&\frac{n!}{(m-1)!(n-m+1)}\Bigg(\psi_{0}(n+1)-\psi_{0}(n-m+1)+\psi_{0}(1)-\nonumber\\
&&\frac{1}{n-m+1}\Bigg)
\end{eqnarray}
\begin{eqnarray}\label{eq:B5}
\fl\sum_{k=1}^{m}\frac{(n-k)!}{(m-k)!}\frac{\psi_{0}(k)}{k}&=&\frac{n!}{m!}\Bigg(\frac{1}{2}\left(\psi_{1}(n+1)-\psi_{1}(n-m+1)+\psi_{0}^{2}(n+1)+\right.\nonumber\\
&&\!\left.\psi_{0}^{2}(n-m+1)\right)+\psi_{0}(1)\left(\psi_{0}(n+1)-\psi_{0}(n-m+1)\right)-\nonumber\\
&&\psi_{0}(n+1)\psi_{0}(n-m+1)\Bigg)
\end{eqnarray}
\begin{eqnarray}\label{eq:B6}
\fl\sum_{k=1}^{m}\frac{(n-k)! }{(m-k)!}\psi_0(n+1-k)&=&\frac{n!}{(m-1)! (n-m+1)}\left(\psi_0(n+1)-\frac{1}{n-m+1}\right)
\end{eqnarray}
\begin{eqnarray}\label{eq:B7}
\fl\sum_{k=1}^{m}\frac{(n-k)!}{(m-k)!}\frac{\psi_{0}(n+1-k)}{k}&=&\frac{n!}{m!}(\psi_{1}(n+1)-\psi_{1}(n-m+1)+\psi_{0}(n+1)(\psi_{0}(n+1)-\nonumber\\
&&\psi_{0}(n-m+1)))
\end{eqnarray}
\begin{eqnarray}\label{eq:B8}
&\fl\sum _{k=1}^m \frac{ (n-k)!}{(m-k)!} \psi _0(k) \psi _0(n-k+1)=\frac{n!}{(m-1)! (n-m+1)^2}\Bigg(\!\!-(n-m+1)\times\nonumber\\
& \psi _1(n-m+1)+(n-m+1) \psi _1(n+1)-(n-m+1) \psi _0(n+1)\times\nonumber\\
& \psi_0(n-m+1)+\psi_0(n-m+1)+(n-m+1) \psi _0^2(n+1) +(n-\nonumber\\
&m+1)\psi_0 (1) \psi _0(n+1)-2\psi _0(n+1)-\psi_0 (1) +\frac{2}{n-m+1}\Bigg)
\end{eqnarray}
\begin{eqnarray}\label{eq:B9}
&\fl\sum _{k=1}^m \frac{ (n-k)!}{(m-k)!}\frac{ \psi _0(k) \psi _0(n-k+1)}{k}=\frac{n!}{2 m!}\Big(\!-\psi _2(n-m+1)+\psi _2(n+1)+\psi _0^3(n+1)-\nonumber\\
&2 \psi _0^2(n+1) (\psi_0(n-m+1)-\psi_0 (1) )+\psi _0(n+1)\big(\psi_0^2(n-m+1)-2 \times\nonumber\\
&\psi_0 (1)  \psi_0(n-m+1)-3 \psi _1(n-m+1)+3\psi _1(n+1)\big)+2 (-\psi_0 (1) +\nonumber\\
&\psi_0 (n-m+1))(\psi _1(n-m+1)-\psi _1(n+1))\Big)
\end{eqnarray}
\begin{eqnarray}\label{eq:B10}
&\fl\sum _{k=1}^m \frac{ (n-k)!}{(m-k)!}\left(\psi _0^2(n-k+1)+\psi _1(n-k+1)\right)=\frac{n!}{(m-1)! (n-m+1)^3}\times\nonumber\\
&\Big((n-m+1)^2 \psi _1(n+1)-2 (n-m+1) \psi_0(n+1)+(n-m+1)^2 \times\nonumber\\
&\psi_0^2(n+1)+2\Big)
\end{eqnarray}
\begin{eqnarray}\label{eq:B11}
&\fl\sum _{k=1}^m \frac{ (n-k)!}{(m-k)!}\frac{\psi _0^2(n-k+1)+\psi _1(n-k+1)}{k}=\frac{n!}{m!}\Big(\!-\psi _2(n-m+1)+\psi _2(n+1)-\nonumber\\
&\psi_0(n-m+1)\psi _1(n+1)+\psi _0(n+1) \left(3 \psi _1(n+1)-2 \psi _1(n-m+1)\right)+\nonumber\\
&\psi _0^3(n+1)-\psi _0^2(n+1) \psi_0(n-m+1)\Big)
\end{eqnarray}
\subsection{Semi closed-form expressions}\label{sec:ap2-sc}
\begin{eqnarray}\label{eq:B12}
\fl\sum_{k=1}^{m}\frac{(n-k)!}{(m-k)!}\frac{1}{k+a}&=&\frac{(a+n)!}{(a+m)!}\sum_{k=1}^{m}\frac{(k+n-m-1)!(k+a-1)!}{(k-1)!(k+a+n-m)!}
\end{eqnarray}
\begin{eqnarray}\label{eq:B13}
\fl\sum_{k=1}^{m}\frac{(n-k)!}{(m-k)!}\frac{1}{k^{2}}&=&\frac{n!}{m!}\sum_{k=1}^{m}\frac{\psi_{0}(k+n-m)}{k}+\frac{n!}{m!}\Bigg(\frac{1}{2}\Big(\psi_{1}(n-m+1)-\psi_{1}(n+1)+\nonumber\\
&&\psi_{0}^{2}(n-m+1)-\psi_{0}^{2}(n+1)\Big)+\psi_{0}(n-m)(\psi_{0}(n+1)-\psi_{0}(m+1)-\nonumber\\
&&\psi_{0}(n-m+1)+\psi_{0}(1))\Bigg)
\end{eqnarray}
\begin{eqnarray}\label{eq:B14}
\fl\sum_{k=1}^{m}\frac{(n-k)!}{(m-k)!}\frac{1}{k^{3}}&=&\frac{n!}{m!}\left(-\frac{1}{2} \sum _{k=1}^m \frac{\psi _1(k+n-m)}{k}+\frac{1}{2} \sum _{k=1}^m \frac{\psi _0^2(k+n-m)}{k}-\right.\nonumber\\
&&\sum _{k=1}^m \frac{ \psi _0(k+n-m)\psi_0 (k)}{k}+\left(\psi _0(m+1)-\psi_0(n+1)\right) \times\nonumber\\
&&\!\left.\sum _{k=1}^m \frac{\psi _0(k+n-m)}{k}\right)+\frac{n!}{12 m!}\Big(\!-2 \psi _2(n-m)+2 \psi _2(n)-6\times\nonumber\\
&& \psi _1(n-m) \left(-\psi _0(m+1)+\psi _0(n+1)+\psi _0(1)\right)+6 \psi _0(n+1)\psi _1(n)-\nonumber\\
&& 2 \psi _0^3(n-m)+6 \left(-\psi _0(m+1)+\psi _0(n+1)+\psi _0(1)\right) \psi _0^2(n-m)-\nonumber\\
&&12 \psi _0(n) \psi _0(n+1) \psi _0(n-m)-4 \psi _0^3(n)+6  \psi _0^2(n)(\psi _0(n-m)+\nonumber\\
&&\psi _0(n+1))-6 \psi _0(n-m) \left(-\psi _1(n-m)+\psi _1(n)+\psi _1(m)-\psi _0^2(m)-\right.\nonumber\\
&& 2 \psi _1(m+1)-2\psi _0(m+1) \psi _0(n+1)-2 \psi _0(1) \psi _0(m+1)+2 \psi _0(m)\times\nonumber\\
&& \psi _0(m+1)+2 \psi _0(1) \psi _0(n+1)+\left.\psi _0^2(1)+\psi _1(1)\right)\!\Big)
\end{eqnarray}
\begin{eqnarray}\label{eq:B15}
\fl\sum_{k=1}^{m}\frac{(n-k)!}{(m-k)!}\frac{\psi_{0}(k)}{k^{2}}&=&\frac{n!}{2m!}\left(\sum_{k=1}^{m}\frac{\psi_{1}(k+n-m)}{k}+\sum_{k=1}^{m}\frac{\psi_{0}^{2}(k+n-m)}{k}\right.-2\times\nonumber\\
&&(\psi_{0}(n-m)-\psi_{0}(1))\!\left.\sum_{k=1}^{m}\frac{\psi_{0}(k+n-m)}{k}\right)+\frac{n!}{2m!}\Bigg(\!-\frac{1}{3}\times\nonumber\\&&\bigg(\psi_{2}(n+1)-\psi_{2}(n-m+1)+\psi_{0}^{3}(n+1)-\psi_{0}^{3}(n-m+1)+\nonumber\\
&&3\psi_{0}(n+1)\psi_{1}(n+1)-3\psi_{0}(n-m+1)\psi_{1}(n-m+1)\bigg)+\nonumber\\
&&(\psi_{0}(n-m)-\psi_{0}(1))\Big(\psi_{1}(n+1)-\psi_{1}(n-m+1)+\psi_{0}^{2}(n+1)-\nonumber\\
&&\psi_{0}^{2}(n-m+1)\Big)-\Big(\psi_{1}(n-m)-\psi_{0}^{2}(n-m)+2\psi_{0}(1)\times\nonumber\\
&&\psi_{0}(n-m)\Big)\left(\psi_{0}(m+1)-\psi_{0}(n+1)+\psi_{0}(n-m+1)-\right.\nonumber\\
&&\!\left.\psi_{0}(1 )\right)\!\Bigg)
\end{eqnarray}
\begin{eqnarray}\label{eq:B16}
\fl\sum_{k=1}^{m}\frac{(n-k)!}{(m-k)!}\frac{\psi_{0}(k)}{k^{3}}&=&\frac{n!}{m!}\left(-\frac{1}{6} \sum _{k=1}^m \frac{\psi _2(k+n-m)}{k}-\frac{1}{2} \sum _{k=1}^m \frac{ \psi _1(k+n-m)\psi_0 (k)}{k}+\right.\nonumber\\
&&\frac{1}{3} \sum _{k=1}^m \frac{\psi _0^3(k+n-m)}{k}-\frac{1}{2} \sum _{k=1}^m \frac{ \psi _0^2(k+n-m)\psi_0 (k)}{k}+\nonumber\\
&&\frac{1}{2} \left(\psi _0(n-m)-\psi _0(n+1)+\psi _0(m+1)-\psi _0(1)\right)\times\nonumber\\
&&\! \sum _{k=1}^m \frac{\psi _1(k+n-m)}{k}-\frac{1}{2} (\psi _0(n-m)+\psi _0(n+1)-\psi _0(m+1)-\nonumber\\
&&\psi _0(1)) \sum _{k=1}^m \frac{\psi _0^2(k+n-m)}{k}+\left(\psi _0(n-m)-\psi _0(1)\right) \times\nonumber\\
&&\!\sum _{k=1}^m \frac{ \psi _0(k+n-m)\psi_0 (k)}{k}+(\psi _0(n+1) \psi _0(n-m)-\psi _0(m+1) \times\nonumber\\
&&\psi _0(n-m)-\psi _0(1) \psi _0(n+1)+\psi _0(1) \psi _0(m+1)) \times\nonumber\\
&&\!\left.\sum _{k=1}^m \frac{\psi _0(k+n-m)}{k}\right)+\frac{n!}{24 m!}\Big(\!-\psi _3(n-m)+\psi _3(n)+(4\times\nonumber\\
&& \psi _0(n-m)-4 \psi _0(n+1)+4 \psi _0(m+1)-8 \psi _0(1)) \psi _2(n-m)+\nonumber\\
&&4 \left(-\psi _0(n-m)+\psi _0(n+1)+\psi _0(1)\right) \psi _2(n)+3 \psi _1^2(n-m)+\nonumber\\
&&3 \psi _1^2(n)-6 \psi _1(n-m) \left(\psi _1(n)-\psi _0^2(n)+2 \psi _0(n) \psi _0(n+1)-\right.\nonumber\\
&&2 \psi _0(m+1) \psi _0(n+1)+4 \psi _0(1) \psi _0(n+1)+\psi _1(m)-\psi _0^2(m)-\nonumber\\
&&4 \psi _0(1) \psi _0(m+1)+2 \psi _0(m) \psi _0(m+1)-2 \psi _1(m+1)+\psi _1(1)+\nonumber\\
&&3\!\left.\psi _0^2(1)\right)-6  \left(\psi _0^2(n)-2 \psi _0(n+1) \psi _0(n)-2 \psi _0(1) \psi _0(n+1)\right)\times\nonumber\\
&&\psi _1(n)+\psi _0^4(n-m)+4 \left(\psi _0(m+1)-\psi _0(n+1)-2 \psi _0(1)\right) \times\nonumber\\
&&\psi _0^3(n-m)+6 \left(-\psi _1(n-m)+\psi _1(n)-2 \psi _1(m+1)-\psi _0^2(n)-\right.\nonumber\\
&&\psi _0^2(m)-2 \psi _0(m+1) \psi _0(n+1)-4 \psi _0(1) \psi _0(m+1)+2 \psi _0(m)\times\nonumber\\
&& \psi _0(m+1)+\psi _1(m)+4 \psi _0(1) \psi _0(n+1)+2 \psi _0(n) \psi _0(n+1)+\nonumber\\
&&\psi _1(1)+\left.3 \psi _0^2(1)\right) \psi _0^2(n-m)+4 \left(2 \psi _0^3(n)-3 \psi _0^2(n) \psi _0(n+1)+\right.\nonumber\\
&&3 \left(\psi _0(n+1)-\psi _0(m+1)+2 \psi _0(1)\right) \psi _1(n-m)-3 (\psi _0(n+1)+\nonumber\\
&&\psi _0(1)) \psi _1(n)-3 \psi _0(1) \psi _1(m)+6 \psi _0(1) \psi _1(m+1)+6 \psi _0(1)\times\nonumber\\
&& \psi _0(m+1)(\psi _0(n+1)+\psi _0(1))+3 \psi _0(1) \psi _0^2(m)-6 \psi _0(1) \psi _0(m) \times\nonumber\\
&&\psi _0(m+1)-6 \psi _0^2(1) \psi _0(n+1)+3 \psi _0(1) \psi _0^2(n)-6 \psi _0(1) \psi _0(n)\times\nonumber\\
&& \psi _0(n+1)-3 \psi _1(1) \psi _0(1)-\left.3 \psi _0^3(1)\right) \psi _0(n-m)-\psi _0^2 (n)\times\nonumber\\
&&\!\left(3 \psi _0^2(n)+8 \psi _0(1) \psi _0(n)-4 \psi _0(n+1) \psi _0(n)-12 \psi _0(1) \psi _0(n+1)\right)\!\!\Big)\nonumber\\
&&
\end{eqnarray}
\begin{eqnarray}\label{eq:B17}
\fl\sum_{k=1}^{m}\frac{(n-k)!}{(m-k)!}\psi_{0}^2(k)&=&\frac{n!}{(m-1)! (n-m+1)}\sum _{k=1}^m \frac{\psi_0 (k+n-m)}{k}+\frac{n!}{2 (m-1)!}\times\nonumber\\
&&\frac{1}{n-m+1}\Big(4 \left(\psi _1(n-m+1)-\psi _1(n-m+2)\right)-4 \times\nonumber\\
&&\!\left(-\psi _0(n-m+1)+\psi _0(n+1)+\psi _0(1)\right) (\psi _0(n-m+2)-\nonumber\\
&&\psi _0(n-m+1))-2 \left(\psi _0(n-m+1)-\psi _0(n-m)\right) \times\nonumber\\
&&\!\left(-\psi _0(n-m+1)-\psi _0(m+1)+\psi _0(n+1)+\psi _0(1)\right)+2 \times\nonumber\\
&&\!\left(\psi _0(m+1)-\psi _0(m)\right) \left(\psi _0(n-m+1)-\psi _0(n+1)\right)+\big(\!-2\times\nonumber\\
&& \psi _0(1) \psi _0(n-m+1)+\psi _0^2(n-m+1)-2 \psi _0(m+1)\times\nonumber\\
&& \psi _0(n-m+1)-2 \psi _0(n+1) \psi _0(n-m+1)-\psi _1(n-m+1)+\nonumber\\
&&4 \psi _0(1) \psi _0(n+1)+\psi _0^2(n+1)+\psi _1(n+1)+2 \psi _0^2(1)\big)\Big)
\end{eqnarray}
\begin{eqnarray}\label{eq:B18}
\fl\sum_{k=1}^{m}\frac{(n-k)!}{(m-k)!}\frac{\psi_0^2 (k)}{k }&=&\frac{n!}{m!}\left(-\sum _{k=1}^m \frac{\psi_0^2 (k+n-m)}{k}+\Bigg(\psi _0(n-m)+\psi _0(n+1)\right.-\nonumber\\
&&\!\left.\frac{1}{n-m}\Bigg) \sum _{k=1}^m \frac{\psi_0 (k+n-m)}{k}\right)+\frac{n!}{6 m!}\Big(\!-\psi _2(n-m+1)+\nonumber\\
&&\psi _2(n+1)-\psi _0^3(n-m+1)+6 \psi _0(m+1) \psi _0^2(n-m+1)+\nonumber\\
&&\psi _0^3(n+1)+\psi _0^2(n+1)\left(6 \psi _0(1)-3 \psi _0(n-m+1)\right)-3\times\nonumber\\
&& \psi _0(n-m+1)\big(2 \left(\psi _0(m+1)-\psi _0(1)\right) \psi _0(n-m)+\psi _1(n+1)-\nonumber\\
&&3 \psi _1(n-m+1)+2 \big(\psi _1(n-m)+\psi _0^2(1)\big)\big)+3 \psi _0(n+1)\big(2\times\nonumber\\
&&\psi _0(n-m) (\psi _0(1)-\psi _0(m+1)) +\psi _0(n-m+1) (-4 \psi _0(1)+\nonumber\\
&&\psi _0(n-m+1))+2 \left(\psi _1(n-m)+\psi _0^2(1)\right)-3 \psi _1(n-m+1)+\nonumber\\
&&\psi _1(n+1)\big)+6 \psi _0(1) \psi _1(n+1)-6\psi _1(n-m) (\psi _0(m+1)-\nonumber\\
&&\psi _0(1)) +6 \psi _1(n-m+1) \left(\psi _0(m+1)-2 \psi _0(1)\right)\!\Big)
\end{eqnarray}
\begin{eqnarray}\label{eq:B19}
\fl\sum_{k=1}^{m}\frac{(n-k)!}{(m-k)!}\frac{\psi_0^2 (k)}{k^2 }&=&\frac{n!}{m!}\left(\frac{1}{6} \sum _{k=1}^m \frac{\psi _2(k+n-m)}{k}+\sum _{k=1}^m \frac{\psi_0 (k+n-m) \psi _1(k+n-m)}{k}\right.\nonumber\\
&&\!+\frac{1}{2} \left(\sum _{k=1}^m \frac{\psi_0 (k+n-m)}{k}\right)^2-\frac{1}{3} \sum _{k=1}^m \frac{\psi_0^3 (k+n-m)}{k}+\nonumber\\
&&\!\sum _{k=1}^m \frac{ \psi_0^2 (k+n-m)\psi_0 (k)}{k}-\left(\psi _0(n-m)-\psi _0(1)\right) \times\nonumber\\
&&\!\sum _{k=1}^m \frac{\psi _1(k+n-m)}{k}+\frac{1}{2} \sum _{k=1}^m \frac{\psi_0^2 (k+n-m)}{k^2}-\psi _0(n-m) \times\nonumber\\
&&\!\sum _{k=1}^m \frac{\psi_0 (k+n-m)}{k^2}+\left(\psi_0(n+1)-\psi _0(m+1)+\psi _0(1)\right) \times\nonumber\\
&&\!\sum _{k=1}^m \frac{\psi_0^2 (k+n-m)}{k}-2 \psi _0(n-m) \sum _{k=1}^m \frac{\psi_0 (k+n-m)\psi_0 (k) }{k}+\nonumber\\
&&\frac{1}{2} \left(-\psi _1(n-m)+\psi _0^2(n-m)-2\psi _0(n-m) \big(\psi _0(n+1)-\right.\nonumber\\
&&\psi _0(m+1)+\psi _0(1)\big)-\psi _1(n+1)-\psi _0^2(n+1)+\left.2 \psi _0^2(1)\right)\times\nonumber\\
&&\!\left.\sum _{k=1}^m \frac{\psi_0 (k+n-m)}{k}\right)+\frac{n!}{24 m!}\Big(\psi _3(n-m+1)-\psi _3(n+1)-4 \times\nonumber\\
&&\!\left(\psi _0(m+1)-3 \psi _0(1)\right) \psi _2(n-m+1)-8 \psi _0(1) \psi _2(n+1)-3 \times\nonumber\\
&&\psi _1^2(n-m+1)+6 \psi _1(n+1) \psi _1(n-m+1)-6 \psi _0^2(n-m+1)  \times\nonumber\\
&&\psi _1(n-m+1)+36 \psi _0^2(1) \psi _1(n-m+1)-24 \psi _0(1) \psi _0(m+1)\times\nonumber\\
&& \psi _1(n-m+1)-3 \psi _1^2(n+1)+6 \psi _0^2(n-m+1)\psi _1(n+1) -\nonumber\\
&&12 \psi _0^2(1) \psi _1(n+1)+3 \psi _0^4(n-m+1)-24 \psi _0(1) \psi _0^3(n-m+1)+\nonumber\\
&&8 \psi _0(m+1) \psi _0^3(n-m+1)+36 \psi _0^2(1) \psi _0^2(n-m+1)-24 \psi _0(1)\times\nonumber\\
&& \psi _0(m+1) \psi _0^2(n-m+1)-\psi _0^4(n+1)-4 \psi _0^3(n+1)(2 \psi _0(1)-\nonumber\\
&&\psi _0(n-m))+6 \psi _0^2(n+1)\left(\psi _0^2(n-m+1)+\psi _1(n-m+1)-\right.\nonumber\\
&&\psi _1(n+1)-2 \psi _0(n-m) (\psi _0(n-m+1)-\psi _0(m+1)-\nonumber\\
&&\psi _0(1))-\left.2 \psi _0^2(1)\right)+4 \psi _0(n-m)\left(9 \psi _0(1)\psi _0^2(n-m+1)-\right.\nonumber\\
&&\psi _2(n-m+1)+\psi _2(n+1)-3 \psi _0(1)(3 \psi _1(n-m+1)-\nonumber\\
&&\psi _1(n+1))-\psi _0^3(n-m+1)-3 \psi _0(n-m+1)(\psi _1(n+1)-\nonumber\\
&&\psi _1(n-m+1)+6 \psi _0^2(1))+3 \psi _0(m+1) (\psi _1(n-m+1)+\nonumber\\
&&\psi _1(n+1)+4 \psi _0(1) \psi _0(n-m+1)-\psi _0^2(n-m+1)-2 \psi _0^2(1))+\nonumber\\
&&6\left. \psi _0^3(1)\right)+4 \psi _0(n+1)\left(\psi _2(n-m+1)-2 \psi _0^3(n-m+1)-\right.\nonumber\\
&&\psi _2(n+1)+6 \psi _0(1) \psi _0^2(n-m+1)+6 \psi _0(1) (\psi _1(n-m+1)-\nonumber\\
&&\psi _1(n+1))+3 \psi _0(n-m)\left(-\psi _1(n-m+1)+\psi _0^2(n-m+1)+\right.\nonumber\\
&&\psi _1(n+1)- \left.4 \psi _0(1)\psi _0(n-m+1)+2\left.\psi _0^2(1)\right)\right)\!\Big)
\end{eqnarray}
\begin{eqnarray}\label{eq:B20}
\fl\sum_{k=1}^{m}\frac{(n-k)!}{(m-k)!}\frac{\psi_{0}(n+1-k)}{k^{2}}&=&\frac{n!}{m!}\left(\sum_{k=1}^{m}\frac{\psi_{1}(k+n-m)}{k}+\psi_{0}(n+1)\right.\times\nonumber\\
&&\!\left.\sum_{k=1}^{m}\frac{\psi_{0}(k+n-m)}{k}\right)+\frac{n!}{m!}\Bigg(\frac{1}{2}\psi_{2}(n-m+1)-\frac{1}{2}\times\nonumber\\
&&\psi_{2}(n+1)+\psi_{0}(n-m+1)\psi_{1}(n-m+1)+\psi_{0}(n+1)\times\nonumber\\
&&\Bigg(\frac{1}{2}\left(\psi_{1}(n-m+1)-\psi_{1}(n+1)+\psi_{0}^{2}(n-m+1)-\right.\nonumber\\
&&\!\left.\psi_{0}^{2}(n+1)\right)+\psi_{0}(n-m)(\psi_{0}(n+1)-\psi_{0}(n-m+1)-\nonumber\\
&&\psi_{0}(m+1)+\psi_{0}(1))+\psi_{1}(n-m)-\psi_{1}(n+1)\Bigg)-\nonumber\\
&&\psi_{1}(n-m)(\psi_{0}(n-m+1)+\psi_{0}(m+1)-\psi_{0}(1))+\nonumber\\
&&\psi_{0}(n-m)\left(\psi_{1}(n+1)-\psi_{1}(n-m+1)\right)\!\Bigg)
\end{eqnarray}\begin{eqnarray}\label{eq:B21}
\fl\sum_{k=1}^{m}\frac{(n-k)!}{(m-k)!}\frac{\psi_{0}(n+1-k)}{k^{3}}&=&\frac{n!}{m!}\left(\sum _{k=1}^m \frac{\psi _0(k+n-m) \psi _1(k+n-m)}{k}-\frac{1}{2} \times\right.\nonumber\\
&&\!\sum _{k=1}^m \frac{\psi _2(k+n-m)}{k}-\sum _{k=1}^m \frac{\psi _1(k+n-m)\psi_0 (k) }{k}-\frac{1}{2}\times\nonumber\\
&& \left(3 \psi _0(n+1)-2 \psi _0(m+1)\right) \sum _{k=1}^m \frac{\psi _1(k+n-m)}{k}-\nonumber\\
&&\psi _0(n+1) \sum _{k=1}^m \frac{ \psi _0(k+n-m)\psi_0 (k)}{k}+\frac{1}{2} \psi _0(n+1)\times\nonumber\\
&&\!\sum _{k=1}^m \frac{\psi _0^2(k+n-m)}{k}-(\psi _1(n+1)-\psi _0(m+1)\times\nonumber\\
&&\psi _0(n+1)+\psi _0^2(n+1))\left. \sum _{k=1}^m \frac{\psi _0(k+n-m)}{k}\right)+\frac{n!}{6 m!}\times\nonumber\\
&&\!\Big(\!-\psi _3(n-m)+\psi _3(n)+(3 \psi _0(n-m)+3 \psi _0(m+1)-\nonumber\\
&&4 \psi _0(n+1)-3 \psi _0(1)) \psi _2(n-m)+(4 \psi _0(n+1)-3\times\nonumber\\
&& \psi _0(n-m))\psi _2(n)+3 \left(\psi _1(n)-\psi _1(n-m)\right)^2+3 \times\nonumber\\
&&\psi _0^2(n-m) \left(\psi _1(n)-\psi _1(n-m)\right)+9  (\psi _0(n-m)+\nonumber\\
&&\psi _0(m+1)-\psi _0(1))\psi _0(n+1)\psi _1(n-m)-3 \left(-\psi _0^2(m)+\right.\nonumber\\
&&\psi _1(m)-2 \psi _1(m+1)-2 \psi _0(1) \psi _0(m+1)+2 \psi _0(m)\times\nonumber\\
&&\psi _0(m+1)+\left.\psi _0^2(1)+\psi _1(1)\right) \psi _1(n-m)-\psi _0(n+1) \times\nonumber\\
&&\psi _0^3(n-m)-3 \psi _0(n+1) \psi _0(n-m) \left(3 \psi _1(n)-\psi _0^2(m)-\right.\nonumber\\
&&2 \psi _1(m+1)-\psi _0(1) \psi _0(n-m)+\psi _0(m+1) \psi _0(n-m)- \nonumber\\
&&2 \psi _0(1)\psi _0(m+1)+2 \psi _0(m) \psi _0(m+1)+\psi _1(m)+\psi _1(1)+\nonumber\\
&&\!\left.\psi _0^2(1)\right)-6 \psi _0(n-m) \left(\psi _0(1)-\psi _0(m+1)\right) (\psi _1(n)-\nonumber\\
&&\psi _1(n-m))-3 \psi _0^4(n)+\left(6 \psi _0(n-m)+4 \psi _0(n+1)\right)\times\nonumber\\
&&\psi _0^3(n)+3 \psi _0^2(n)(2 \psi _1(n-m)-2 \psi _1(n)-(\psi _0(n-m)+\nonumber\\
&&3 \psi _0(n+1)+2 \psi _0(m+1)-2 \psi _0(1)) \psi _0(n-m))+6 \times\nonumber\\
&&\psi _0(n) \psi _0(n+1)(-2 \psi _1(n-m)+2 \psi _1(n)+\psi _0(n-m) \times\nonumber\\
&&\!\left(\psi _0(n-m)+2 \psi _0(m+1)-2 \psi _0(1)\right))\Big)
\end{eqnarray}
\begin{eqnarray}\label{eq:B22}
&\fl\sum _{k=1}^m \frac{ (n-k)!}{(m-k)!}\frac{ \psi _0(k) \psi _0(n-k+1)}{k^2}=\frac{n!}{m!}\left(\sum _{k=1}^m \frac{\psi_0 (k+n-m) \psi _1(k+n-m)}{k}+\right.\nonumber\\
&\frac{1}{2} \sum _{k=1}^m \frac{\psi _2(k+n-m)}{k}+\frac{1}{2} \left(-2 \psi _0(n-m)+\psi _0(n+1)+2\psi _0(1)\right)\times\nonumber\\
& \sum _{k=1}^m \frac{\psi _1(k+n-m)}{k}+\frac{1}{2} \psi _0(n+1) \sum _{k=1}^m \frac{\psi_0^2(k+n-m)}{k}-\nonumber\\
&\!\left(\psi _0(n+1) \psi _0(n-m)+\psi _1(n-m)-\psi _0(1) \psi _0(n+1)\right)  \times\nonumber\\
&\!\left.\sum _{k=1}^m\frac{\psi_0 (k+n-m)}{k}\right)+\frac{n!}{6 m!}\Big(\psi _3(n-m+1)-\psi _3(n+1)+3\times\nonumber\\
& \psi _0(n-m+1) \psi _2(n-m+1)-3 \psi _0(n+1) \psi _2(n+1)-3 \psi _1^2(n+1)+\nonumber\\
&3 \psi _1^2(n-m+1)+3 \psi _0^2(n-m+1)\psi _1(n-m+1)-3 \psi _0^2(n+1)\times\nonumber\\
&\psi _1(n+1)-3 \psi _2(n-m) \left(\psi _0(n-m+1)+\psi _0(m+1)-\psi _0(n+1)-\right.\nonumber\\
&\!\left.\psi _0(1)\right)+3 \left(\psi _1(n+1)-\psi _1(n-m+1)\right) \big(\!-\psi _0^2(n-m)+2 \psi _0(1)\times\nonumber\\
& \psi _0(n-m)+\psi _1(n-m)\big)+3 \psi _1(n-m) \left(-\psi _0^2(n-m+1)-\right.\nonumber\\
&\!\left.\psi _1(n-m+1)+\psi _0^2(n+1)+\psi _1(n+1)\right)+6 \psi _1(n-m) (\psi _0(n-m)-\nonumber\\
&\psi _0(1))(\psi _0(n-m+1)+\psi _0(m+1)-\psi _0(n+1)-\psi _0(1))+3\times\nonumber\\
&(\psi _0(n-m)-\psi _0(1)) (-\psi _2(n-m+1)-2 \psi _0(n-m+1)\times\nonumber\\
& \psi _1(n-m+1)+2 \psi _0(n+1) \psi _1(n+1)+\psi _2(n+1))+3 \psi _0(n+1) \times\nonumber\\
&\!\left(-\psi _0(n-m+1)-\psi _0(m+1)+\psi _0(n+1)+\psi _0(1)\right) \big(\!-\psi _0^2(n-m)+\nonumber\\
&2 \psi _0(1) \psi _0(n-m)+\psi _1(n-m)\big)+3 \psi _0(n+1) \left(\psi _0(n-m)-\psi _0(1)\right)\times\nonumber\\
&\!\left(-\psi _1(n-m+1)-\psi _0^2(n-m+1)+\psi _0^2(n+1)+\psi _1(n+1)\right)+\nonumber\\
&\psi _0(n+1) \left(\psi _0^3(n-m+1)+3 \psi _0(n-m+1) \psi _1(n-m+1)+\right.\nonumber\\
&\!\left.\psi _2(n-m+1)-\psi _0^3(n+1)-3 \psi _1(n+1) \psi _0(n+1)-\psi _2(n+1)\right)\!\Big)
\end{eqnarray}
\begin{eqnarray}\label{eq:B23}
\fl\sum_{k=1}^{m}\frac{(n-k)!}{(m-k)!}\psi_{1}(k)&=&-\frac{n!}{(m-1)!(n-m+1)}\sum_{k=1}^{m}\frac{\psi_{0}(k+n-m)}{k}-\nonumber\\
&&\frac{n!}{(m-1)!(n-m+1)}\Bigg(\frac{1}{2}\Big(\psi_{1}(n-m+1)-\psi_{1}(n)-\psi_{0}^{2}(n)+\nonumber\\
&&\psi_{0}^{2}(n-m+1)\Big)+\psi_{0}(n-m)(\psi_{0}(n)-\psi_{0}(m)-\psi_{0}(n-m+1)+\nonumber\\
&&\psi_{0}(1))-\psi_{1}(1)-\frac{\psi_{0}(n)-\psi_{0}(n-m+1)}{n}-\frac{\psi_{0}(n)}{m}\Bigg)
\end{eqnarray}
\begin{eqnarray}\label{eq:B24}
\fl\sum_{k=1}^{m}\frac{(n-k)!}{(m-k)!}\frac{\psi_{1}(k)}{k}&=&\frac{n! }{m!}\left(\!\left(-\psi _0(n-m)-\psi _0(n+1)+\frac{1}{n-m}\right) \sum _{k=1}^m \frac{\psi_0 (k+n-m)}{k}+\right.\nonumber\\
&&\!\left.\sum _{k=1}^m \frac{\psi_0^2 (k+n-m)}{k}\right)-\frac{n!}{6 m!}\Big(\psi _2(n-m+1)-\psi _2(n+1)-\nonumber\\
&&3 \psi _0(n-m+1) \psi _1(n-m+1)+3 \psi _0(n+1) \psi _1(n-m+1)+3\times\nonumber\\
&& \psi _1(n+1) \psi _0(n-m+1)-3 \psi _0(n+1) \psi _1(n+1)-3 \psi _0(n+1)\times\nonumber\\ 
&&\psi _0^2(n-m+1)+\psi _0^3(n-m+1)+3 \psi _0^2(n+1) \psi _0(n-m+1)-6 \nonumber\\ 
&&\psi _0(m) \psi _0(n+1) \psi _0(n-m+1)+6 \psi _0(n+1) \psi _0(n-m+1)\times\nonumber\\
&&\psi _0(1) +6 \psi _1(1) \psi _0(n-m+1)-6 \psi _1(1) \psi _0(n+1)-\psi _0^3(n+1)+\nonumber\\
&&6 \left(\psi _0(n-m+1)-\psi _0(n-m)\right)(\psi _0(m) \psi _0(n+1)+\psi _0(n+1)\times\nonumber\\
&&\psi _0(n-m+1)-\psi _0(1) \psi _0(n-m+1)+\psi _0(m) \psi _0(n-m+1)-\nonumber\\
&&\psi _0(1) \psi _0(n+1))-6 \left(\psi _1(n-m+1)-\psi _1(n-m)\right)(n^2 \psi _0(m)\times\nonumber\\
&& \psi _0(n+1) \psi _0(n-m+1)-n \psi _0(m) \psi _0(n+1)-n^2 \psi _0(m+1)\times\nonumber\\
&& \psi _0(n+1) \psi _0(n-m+1)+n \psi _0(m+1) \psi _0(n+1)-n \psi _0(m) \times\nonumber\\
&&\psi _0(n-m+1)+n \psi _0(m+1) \psi _0(n-m+1)+n \psi _0(n+1) \times\nonumber\\
&&\psi _0(n-m+1)-2 \psi _0(n-m+1)-\psi _0(m+1)+\psi _0(1))\Big)
\end{eqnarray}
\begin{eqnarray}\label{eq:B25}
\fl\sum_{k=1}^{m}\frac{(n-k)!}{(m-k)!}\frac{\psi_{1}(k)}{k^2}&=&\frac{n!}{m!}\left(-\frac{1}{2} \Bigg(\sum _{k=1}^m \frac{\psi_0 (k+n-m)}{k}\Bigg)^2+\frac{1}{6} \sum _{k=1}^m \frac{\psi _2(k+n-m)}{k}+\right.\nonumber\\
&&\frac{2}{3}\sum _{k=1}^m \frac{\psi_0^3 (k+n-m)}{k}-\frac{1}{2} \sum _{k=1}^m \frac{\psi_0^2 (k+n-m)}{k^2}-\nonumber\\
&&\!\sum _{k=1}^m \frac{ \psi_0^2 (k+n-m)\psi_0 (k)}{k}-(\psi _0(n-m)+\psi _0(n+1)-\nonumber\\
&&\psi _0(m+1)) \sum _{k=1}^m \frac{\psi_0^2 (k+n-m)}{k}+\psi _0(n-m)\times\nonumber\\
&& \sum _{k=1}^m \frac{\psi_0 (k+n-m)}{k^2}+2 \psi _0(n-m) \sum _{k=1}^m \frac{ \psi_0 (k+n-m)\psi_0 (k)}{k}+\nonumber\\
&&\frac{1}{2} \big(\psi _0^2(n-m)+2\psi _0(n+1) \psi _0(n-m)-2 \psi _0(1) \psi _0(n-m)-\nonumber\\
&&2 \psi _0(m+1)\psi _0(n-m)-\psi _1(n-m)+\psi _0^2(n+1)+\psi _1(n+1)+\nonumber\\
&&2 \psi _1(1)\big)\left.\sum _{k=1}^m \frac{\psi_0 (k+n-m)}{k}\right)+\frac{n!}{24 m!}\Big(\!-\psi _3(n-m+1)+2\times\nonumber\\
&&\psi _3(n-m)+\psi _3(n+1)-2 \psi _3(n)+4\psi _2(n-m+1) ( \psi _0(m+1)-\nonumber\\
&& \psi _0(n+1)-3 \psi _0(1))+8 \psi _2(n-m) \big(\!-\psi _0(m+1)+\psi _0(n+1)+ \nonumber\\
&&2 \psi _0(1)\big)+3\psi _1^2(n-m+1)-3 \psi _0^4(n-m+1)-2 \psi _0^4(n-m)-\nonumber\\
&&8\psi _0^3(n-m)\left( \psi _0(m+1)- \psi _0(n+1)-2 \psi _0(1)\right)+6 \psi _0^4(n)+\nonumber\\
&&\psi _0^4(n+1)+8\psi _0^3(n)\left(2 \psi _0(1)- \psi _0(n+1)\right)+8\psi _0^3(n-m+1)\times\nonumber\\
&&(\psi _0(n+1)-\psi _0(m+1)+3 \psi _0(1))-12 \psi _0^2(1)\psi _1(n)+4\psi _0(1)\times\nonumber\\
&& \psi _0(n+1)\left(2 \psi _0^2(n+1)+3 \psi _0(1) \psi _0(n+1)-6 \psi _1(n)\right)-6 \psi _1(n) \times\nonumber\\
&&(\psi _1(n)+2 \psi _1(1))+\psi _0^2(n-m+1) \left(-6 \psi _0^2(n+1)-24 \psi _0(1)\times\right.\nonumber\\
&& \psi _0(n+1)-\left.6 \psi _1(n+1)+24 \psi _0(1) \psi _0(m+1)-36 \psi _0^2(1)\right)+3 \times\nonumber\\
&& \psi _1(n+1)\left(8 \psi _0(1) \psi _0(n+1)+2 \psi _0^2(n+1)+\psi _1(n+1)+4\times\right.\nonumber\\
&&\!\left.\psi _0^2(1)\right)+12 \psi _0(n-m) \psi _0(n-m+1) \left(\psi _1(n+1)+\psi _0^2(n+1)+ \right.\nonumber\\
&&4 \psi _0(1)\!\left.\psi _0(n+1)-4 \psi _0(1) \psi _0(m+1)+6 \psi _0^2(1)\right)+12 \psi _0^2(n)\times\nonumber\\
&&\!\left(-\psi _1(n-m)-2 \psi _0(1) \psi _0(n+1)+\psi _1(n)+\psi _0^2(1)+\psi _1(1)\right)+\nonumber\\
&&12 \psi _0(1) \psi _1(n-m) \left(-2 \psi _0(m+1)+2 \psi _0(n+1)+3 \psi _0(1)\right)-\nonumber\\
&&24 \psi _0(n-m) \psi _1(n-m) \left(-\psi _0(m+1)+\psi _0(n+1)+2 \psi _0(1)\right)+\nonumber\\
&&6 \psi _1(n-m) \left(-\psi _1(n-m)+2 \psi _1(n)+2 \psi _1(1)\right)-12 \psi _0^2(n-m)\times\nonumber\\
&&\!\left(-\psi _1(n-m)-2 \psi _0(1) \psi _0(m+1)+2 \psi _0(1) \psi _0(n+1)-\psi _0^2(n)+\right.\nonumber\\
&&\!\left.2 \psi _0(n) \psi _0(n+1)+\psi _1(n)+3 \psi _0^2(1)+\psi _1(1)\right)-24 \psi _0(n) \times\nonumber\\
&&\psi _0(n+1) \left(-\psi _1(n-m)+\psi _1(n)+\psi _0^2(1)+\psi _1(1)\right)+12\times\nonumber\\
&& \psi _0(n-m) \psi _1(n-m+1) (-\psi _0(n-m+1)-\psi _0(m+1)+\nonumber\\
&&\psi _0(n+1)+3 \psi _0(1))+\psi _1(n-m+1) \left(6 \psi _0^2(n-m+1)+24 \times\right.\nonumber\\
&& \psi _0(1) \psi _0(m+1)-24\psi _0(1) \psi _0(n+1)-6 \psi _0^2(n+1)-6\times\nonumber\\
&&\!\left. \psi _1(n+1)-36 \psi _0^2(1)\right)-8 \psi _2(n) \left(\psi _0(n+1)+\psi _0(1)\right)+\nonumber\\
&& 4 \psi _2(n+1)\left(\psi _0(n+1)+2 \psi _0(1)\right)-4 \psi _0(n-m) \big(2 \psi _2(n-m)-\nonumber\\
&&\psi _2(n-m+1)+\psi _2(n+1)-2 \psi _2(n)-\psi _0^3(n-m+1)+3 \times\nonumber\\
&&\psi _0(m+1) \psi _0^2(n+1)+9 \psi _0(1) \psi _0^2(n-m+1)-3 \psi _0(m+1)\times\nonumber\\
&& \psi _0^2(n-m+1)+3 \psi _0(n+1) \psi _0^2(n-m+1)+3 \psi _0(m+1) \times\nonumber\\
&&\psi _1(n+1)+6 \psi _1(1) \psi _0(m+1)+4 \psi _0^3(n)+6 \psi _0(1) \psi _0^2(n)-6 \times\nonumber\\
&&\psi _0(n+1) \psi _0^2(n)-12 \psi _0(1) \psi _0(n+1) \psi _0(n)+\psi _0^3(n+1)+3\times\nonumber\\
&& \psi _0(1) \psi _0^2(n+1)-6 \psi _1(1) \psi _0(n+1)-6 \psi _0(1) \psi _1(n)+3 \psi _0(1)\times\nonumber\\
&& \psi _1(n+1)-6 \psi _0(n+1) \psi _1(n)+3 \psi _0(n+1) \psi _1(n+1)-6 \psi _0(1)\times\nonumber\\
&& \psi _1(1)\big)\Big)
\end{eqnarray}
\begin{eqnarray}\label{eq:B26}
&\!\fl\sum _{k=1}^m \frac{ (n-k)!}{(m-k)!}\frac{\psi _0^2(n-k+1)+\psi _1(n-k+1)}{k^2}=\frac{n!}{m!}\left(\sum _{k=1}^m \frac{\psi _2(k+n-m)}{k}+\psi _0(n+1)\times\right.\nonumber\\
&2\sum _{k=1}^m \frac{\psi _1(k+n-m)}{k}+\left(\psi _0^2(n+1)+\psi _1(n+1)\right)\left. \sum _{k=1}^m \frac{\psi_0 (k+n-m)}{k}\right)+\nonumber\\
&\frac{n!}{2 m!}\Big(\psi _3(n-m+1)-\psi _3(n+1)+\psi _2(n-m+1) \left(2 \psi _0(n-m+1)+2\right.\nonumber\\
&\!\left. \psi _0(n+1)\right)+2 \psi _1^2(n-m+1)+\psi _1(n-m+1) \left(-4 \psi _1(n-m)+4\times\right.\nonumber\\
&\!\left. \psi _0(n+1) \psi _0(n-m+1)+\psi _0^2(n+1)+\psi _1(n+1)\right)-\psi _0^4(n+1)+\nonumber\\
&\psi _0^2(n+1) \psi _0^2(n-m+1)-2\psi _0(n-m) \psi _0(n-m+1)( \psi _0^2(n+1)+\nonumber\\
& \psi _1(n+1)) -3 \psi _1^2(n+1)+\psi _1(n+1) \left(\psi _0^2(n-m+1)+4 \psi _1(n-m)-\right.\nonumber\\
&\!\left.6 \psi _0^2(n+1)\right)+4 \psi _0(n+1) \psi _1(n-m) \left(-\psi _0(n-m+1)-\psi _0(m+1)+\right.\nonumber\\
&\!\left.\psi _0(n+1)+\psi _0(1)\right)-2 \psi _2(n-m) \left(\psi _0(n-m+1)+\psi _0(m+1)-\right.\nonumber\\
&\!\left.\psi _0(1)\right)+2 \psi _0(n+1) \left(-2 \psi _0(n-m) \psi _1(n-m+1)+\psi _2(n-m)-\right.\nonumber\\
&\!\left.2 \psi _2(n+1)\right)+\psi _0(n-m)\left(-2 \psi _0(m+1) \psi _0^2(n+1)-2 \psi _0(m+1) \times\right.\nonumber\\
&\psi _1(n+1)-2 \psi _2(n-m+1)+2 \psi _0^3(n+1)+6 \psi _1(n+1) \psi _0(n+1)+2\times\nonumber\\ 
&\!\left. \psi _0(1) \psi _0^2(n+1)+2\psi _0(1) \psi _1(n+1)+2 \psi _2(n+1)\right)\!\Big).
\end{eqnarray}
\subsection{Remarks on the second type summation}\label{sec:ap2-re}
We first point out that the identities  (\ref{eq:B2})-(\ref{eq:B7}), (\ref{eq:B12})-(\ref{eq:B13}), (\ref{eq:B15}), (\ref{eq:B20}), and (\ref{eq:B23}) have been derived in \cite[appendix\,B]{Wei20}. The idea of deriving (\ref{eq:B4}), (\ref{eq:B17}), and (\ref{eq:B23}) is similar to that of the first type, where we expand the test function $f(k)$ by series form of polygamma functions~(\ref{eq:p0})-(\ref{eq:p3}) before changing the order of summations to utilize the established identities.
For example, the identity (\ref{eq:B17}) can be obtained by using (\ref{eq:B2}), (\ref{eq:B5}), and (\ref{eq:B23}) as
\begin{eqnarray}\label{eq:B27}
\fl\sum _{k=1}^m \frac{ (n-k)!}{(m-k)!}\psi_0^2(k)&=&\sum _{k=1}^m \frac{ (n-k)!}{(m-k)!}\left(\sum _{l=1}^{k-1} \frac{1}{l}+\psi _0(1)\right)^2\nonumber\\
&=&\left(\psi_0 (1) ^2+\psi _1(1)\right)\sum _{k=1}^m \frac{(n-k)!}{(m-k)!}+\frac{2 }{n-m+1}\times\nonumber\\
&&\sum _{k=1}^{m-1} \frac{ (n-k)!}{ (m-1-k)!}\frac{\psi_0 (k)}{k}-\sum _{k=1}^m \frac{(n-k)!}{(m-k)!}\psi _1(k).
\end{eqnarray}
Here, we notice that the above approach works when the test function is only consisted of polygamma functions. In the case when $f(k)$ contains a factor $1/k^c$,
\begin{equation}\label{eq:B28}
f(k)=\frac{\psi_i^{j}(k)}{k^c},\qquad i,j\in \mathbb{Z}_{\geq 0},
\end{equation}
where $c$ is a positive integer, the procedure is as follows. 
First, we denote
\begin{equation}
f_0(k)=kf(k)
\end{equation}
as the test function of a known second type identity.
Considering summation (\ref{eq:B1}) associated with the test function of the type (\ref{eq:B28}), one could rewrite (\ref{eq:B1}) as
\begin{equation}
S_{f}(m,n)=\sum_{k=1}^{m}\frac{(n-1-k)!}{(m-1-k)!}\frac{n-k}{(m-k)k}f_0(k).
\end{equation}
Now taking partial fraction decomposition of
\begin{equation}  
\frac{n-k}{(m-k)k},
\end{equation}
followed by the corresponding recurrence relation, one has
\begin{equation}
S_{f}(m,n)=\frac{n}{m}S_{f}(m-1,n-1)+\frac{n-m}{m}\sum_{k=1}^{m}\frac{(n-1-k)!}{(m-k)!}f_0(k).
\end{equation}
After $m$ recursions, we arrive at
\begin{equation}\label{eq:B33}
S_{f}(m,n)=\frac{n! (n-m)}{m!}\sum _{i=0}^{m-1} \frac{(m-i-1)!}{(n-i)!}\sum _{k=1}^{m-i} \frac{ (n-i-1-k)!}{(m-i-k)!}f_0(k),
\end{equation}
where the relation between the desired identity and a known one is established.  Finally, we note that the listed formulas in appendix\,B are analytically continued to any complex number $n$ except for integers smaller than $m$. This fact makes it possible to take derivative of the formulas (\ref{eq:B2})-(\ref{eq:B5}) and (\ref{eq:B13})-(\ref{eq:B15}) with respect to $n$. For example, the identities (\ref{eq:B6})-(\ref{eq:B9}) and  (\ref{eq:B20})-(\ref{eq:B22}) can be obtained as
\begin{equation}
\fl \frac{\partial}{\partial n}\sum_{k=1}^{m}\frac{(n-k)!}{(m-k)!}\frac{\psi_i^j(k)}{k^{c}}=\sum_{k=1}^{m}\frac{(n-k)!}{(m-k)!}\frac{\psi_{0}(n+1-k)\psi_i^j(k)}{k^{c}},\qquad c,i,j\in \mathbb{Z}_{\geq 0}.
\end{equation}
Taking twice derivative on (\ref{eq:B2}), (\ref{eq:B3}), and (\ref{eq:B13}) with respect to $n$,
\begin{equation}
\fl \frac{\partial^2}{\partial n}\sum_{k=1}^{m}\frac{(n-k)!}{(m-k)!}\frac{1}{k^{c}}=\sum_{k=1}^{m}\frac{(n-k)!}{(m-k)!}\frac{\psi_{0}^2(n+1-k)+\psi_{1}(n+1-k)}{k^{c}},\qquad c\in \mathbb{Z}_{\geq 0},
\end{equation}
the identities (\ref{eq:B10}), (\ref{eq:B11}), and (\ref{eq:B26}) can be obtained.
When $n=m$, all the listed identities in appendix\,B are still valid with some indeterminate forms resolved by using the non-positive integer arguments of gamma and polygamma functions~(\ref{eq:pgna1})-(\ref{eq:pgna5}).

\end{document}